\documentclass{aa}  

\usepackage{graphicx}
\usepackage{txfonts}
\usepackage[colorlinks]{hyperref}
\def\aap{A\&A}
\def\apj{ApJ}
\def\apjs{ApJS}

\def \hi {H\,{\sc i}}

\def\kms{km\,s$^{-1}$}

\def\deg{\hbox{$^\circ$}}
\def\arcmin{\hbox{$^\prime$}}

\def\farcm{\hbox{$.\mkern-4mu^\prime$}}

\begin{document}

\title{Properties of cold and warm \hi\ gas phases derived from a
  Gaussian decomposition of HI4PI data }

   \subtitle{}

   \author{P.\ M.\ W.\ Kalberla \inst{1} \and U.\ Haud \inst{2} }

\institute{Argelander-Institut f\"ur Astronomie,
           Auf dem H\"ugel 71, 53121 Bonn, Germany \\
           \email{pkalberla@astro.uni-bonn.de}
           \and
           Tartu Observatory, University of Tartu,
           61602 T\~oravere, Tartumaa, Estonia           
           }

   \authorrunning{P.\,M.\,W. Kalberla \& U.\ Haud } 

   \titlerunning{Distribution of cold and warm \hi\ gas}

   \date{Received 2 April 2018 / Accepted 2 June 2018 }

% \abstract{}{}{}{}{} 
% 5 {} token are mandatory
  \abstract 
  {A large fraction of the interstellar medium can be characterized as a
    multiphase medium.  The neutral hydrogen gas is bistable with a cold
    and warm neutral medium (CNM and WNM respectively) but there is
    evidence for an additional phase at intermediate temperatures, a
    lukewarm neutral medium (LNM) that is thermally unstable. }
  {We use all sky data from the HI4PI survey to separate these neutral
    \hi\ phases with the aim to determine their distribution and phase
    fractions $f$ in the local interstellar medium.  }
  {HI4PI observations, gridded on an nside = 1024 HEALPix grid, were 
    decomposed into Gaussian components. From the frequency distribution
    of the velocity dispersions we infer three separate linewidth
    regimes. Accordingly we extract the \hi\ line emission corresponding
    to the CNM, LNM, and WNM. We generateed all-sky maps of these phases
    in the local \hi\ gas with $ -8 < v_{ \mathrm{LSR} } < 8 $ \kms.  }
  {Each of the \hi\ phases shows distinct structures on all scales. The
    LNM never exists as a single phase but contributes on average 41\%
    of the \hi. The CNM is prominent only for 22\% of the sky,
    contributes there on average 34\% but locally up to 60\% of the
    \hi\ and is associated with dust at temperatures $T_{\mathrm{dust}}
    \sim 18.6$ K. Embedded cold filaments show a clear anti-correlation
    between CNM and LNM. Also the smoothly distributed WNM is
      anti-correlated with the CNM. It contributes for the rest of the
      sky  39\% with dust associated at temperatures $T_{\mathrm{dust}}
    \sim 19.4$ K.  }
  {The CNM in filaments exists on small scales. Here the observed
    anti-correlation between LNM and CNM implies that both, filaments
    and the surrounding more extended LNM, must have a common
    origin. }

  % -------------------------------
  \keywords{ISM: general -- ISM:  structure --  ISM: dust --  ISM: clouds }
% -------------------------------
  \maketitle
%
%________________________________________________________________

\section{Introduction}
\label{Intro}

The interstellar medium, notably the \hi, exists as a multiphase
medium. \citet{Field1969} pointed out that most of the \hi\ should exist
in two stable phases in pressure eqilibrium, the CNM at excitation
temperatures $T < 300 $ K, and the WNM at $T \la 10^4$ K. But soon
\citet{Salpeter1976}, reviewing the formation and destruction of
interstellar dust grains, objected and argued that there must be a third
unstable phase at intermediate temperatures. This phase was dubbed by
him the luke-warm neutral medium (LNM).

\citet{McKee1977} extended the two-phase model of \citet{Field1969} and
considered supernova (SN) explosions in an in-homogeneous
environment. Their model of the interstellar medium (ISM) has three
phases, including the ionized medium, and is dominated by individual SN
explosions. The interior of SN remnants is filled by the hot ionized
medium (HIM).  Blast waves sweep up the gas inside the bubbles and pile
it up into the shells. As soon as this gas is shocked, it starts to
cool. It recombines rapidly and forms the CNM.

Subsequently the thermal equilibrium gas temperatures of the diffuse
interstellar medium were calculated by \citet{Wolfire1995} an updated by
\citet{Wolfire2003}. These investigations have shown that there can be
conditions where the \hi\ gas is thermally unstable. However, the
conclusion was that the local ISM is probably only marginally in the
regime in which there must be more than two bistable \hi\ phases.

From observations there were early indications for substantial amounts
of unstable \hi\ gas. Among others, significant contributions from
unstable \hi\ gas were observed by
\citet{Dickey1977,Mebold1982,Kalberla1985,Spitzer1995,Fitzpatrick1997}.
However it was \citet{Heiles2001} who initiated a vivid debate with the
statement ``Large amounts of thermally unstable gas are not allowed in
theoretical models of the global interstellar medium.''  From the
Arecibo absorption line survey toward 79 sources there was evidence
that 60\% of the gas is WNM. At least 48\% of this WNM turned out to be
unstable \citep{Heiles2003a}. In other words, 30\% of the gas belongs to
the LNM.  More recently, \citet{Roy2013} obtained for 33 compact
extragalactic radio sources deep high-velocity resolution absorption
line data. They found that at least 28\% of the gas must have
temperatures in the thermally unstable range. 
A more recent study of \hi\ absorption in the direction of 57 background
radio sources with the Very Large Array \citep{Murray2015} has found
an even smaller mass fraction of thermally unstable HI, about 20\%.

Subsequent theoretical investigations have shown that considerable
amounts of unstable gas can be induced by turbulence. \citet{Gazol2001}
conclude that about 50\% of the turbulent gas mass has temperatures that
are characteristic for the LNM. \citet{Audit2005} find also large
fractions of thermally unstable gas. This fraction increases with the
amplitude of the turbulent forcing. The thermally unstable gas tends to
be organized in filamentary structures. Similar, \citet{Avillez2005} find
that up to 49\% of the mass belongs to the thermally unstable regime.
In general, these results show all that turbulence is playing a key role
for phase transition in the \hi\ gas \citep{Saury2014}.

The debate about the multiphase composition of the \hi\ gas was
summarized by \citet{Vazquez-Semadeni2012} and amounts to the question
whether classical discrete-phase models need to be replaced by a “phase
continuum”. For details we refer to this excellent review. 

In this paper we want to study phase dependencies in the local \hi\ gas.
We used high resolution observations of the all-sky \hi\ brightness
temperature distribution with large single dish radio telescopes. The
data are decomposed into Gaussian components. We find three distinct
linewidth-regimes that can be considered to represent the CNM, LNM, and
WNM. Section \ref{Obs} describes the observations, the data reduction,
the Gaussian decomposition and selection criteria applied by us.  Using
Gaussian parameters we model separate \hi\ distributions for the CNM,
LNM, and WNM. These distributions are presented in Sect. \ref{Phases}
and we discuss the column density distributions in detail. The Sky
appears to contain regions that are dominated by the CNM. In Sect.
\ref{CNMdominated} we separate these regions and discuss phase dependent
distributions for column densities and associated dust
temperatures. Cold filamentary gas stands out and is discussed in more
detail in Sect. \ref{Filaments}. The interrelations between \hi\ phases
are shown in Sect. \ref{PhaseRelations}. Our results are discussed in
Sect. \ref {Discussion}, the summary is in Sect. \ref{Summary}.

\section{Observations and data reduction} 
\label{Obs}

\subsection{The HI4PI 21 cm line survey }
\label{HI4PI}

For the northern hemisphere we use the Effelsbeg-Bonn \hi\ Survey
(EBHIS) from observations with the 100-m Effelsberg radio telescope
\citep{Winkel2016a,Winkel2016b} and for the southern hemisphere the
Galactic All Sky Survey (GASS), observed with the 64-m Parkes telescope
\citep{Naomi2009,Kalberla2010,Kalberla2015}.
Both surveys, EBHIS with the first data release \citep{Winkel2016b} and
GASS with its final data release \citep{Kalberla2015}, were gridded in
position to a common nside = 1024 HEALPix database
\citep{Gorski2005}. After calibration, this was first step in merging
both surveys and initially the original velocity vectors, different for
both surveys, were kept. We used this intermediate data product, both
surveys with their genuine spatial resolution and original velocity
vectors, for our Gaussian decomposition as described in
Sect. \ref{Gaussians}.  When merging both surveys by
\citet{Winkel2016c}, the EBHIS beam shape was adapted to the GASS
resolution. In addition the GASS spectra were smoothed and adapted to
the EBHIS velocity grid. In our analysis, to decompose the spectra as
accurate as possible into Gaussian components, we stayed as close as
possible to the original database but used positions on the common
HEALPix grid. To calculate maps for this publication we use Gaussian
components to generate profiles on the HEALPix grid. Next the data are
smoothed to a common effective resolution of 30\arcmin, afterward maps
are generated.

Using an nside = 1024 HEALPix database implies an effective angular
resolution of $\Theta_\mathrm{pix} = 3\farcm44 $ for that grid
\citep{Gorski2005}. The EBHIS data were gridded to an effective beam
sized with FWHM = 10\farcm8, and the GASS to FWHM = 16\farcm2. This
means that our database is over-sampled, neighboring positions are not
independent from each other. Our choice of the nside = 1024 HEALPix
database is motivated by the aim to enable an easy comparison between
all-sky \hi\ data and published {\it Planck} maps.

The Gaussian analysis enables us to separate different phases according
to the linewidth-regimes derived in Sect. \ref{DefPhases}. We generate
separate maps for the CNM, LNM, and WNM.  The EBHIS and GASS overlap for
declinations $ -5\degr < \delta < 1\degr$. When selecting Gaussian
components we use a border line between both surveys at $ \delta =
-2\degr$. To avoid discontinuities in maps, caused by different beam
sizes, we used a linear interpolation between both surveys for $ -4\degr
< \delta < 0\degr$.

%=========================================================================
%=========================================================================
%=========================================================================

\subsection{Gaussian analysis}
\label{Gaussians}

After generating the HEALPix database, we decomposed all brightness
temperature spectra $T_\mathrm{b}(v_i)$ into Gaussian components
\begin{equation}
   T_\mathrm{b}(v_i) = \sum_{j=1}^N T_\mathrm{bc,j}
             \exp \left[-\frac{(v_i - v_{\mathrm{c},j})^2}
             {2\sigma_j^2}\right], \label{Eq1}
\end{equation}
where $T_\mathrm{bc,j}$, $v_{\mathrm{c},j}$ and $\sigma_j$ are the
adjustable parameters of the Gaussian component $j$, the sum is taken
over N components, describing the given profile and $v_i$ is the central
velocity of the spectrometer channel $i$. For the decomposition we used
mostly the same approach, which was described by \citet{Haud2000} and
applied earlier to the Leiden/Argentine/Bonn (LAB) data
\citep{Kalberla2005}. In general, this is a rather classical Gaussian
decomposition, but with two important additions, introduced for reducing
the ambiguities inherent to the decomposition procedure. First of all, our
decomposition algorithm does not treat each \hi\ profile independently,
but assumes that every observed profile shares some similarities with
those in the neighboring sky positions, as expected for an over-sampled
database. In addition, besides adding components into the decomposition,
our algorithm also analyzes the results to find the possibilities for
removing or merging some Gaussians without reducing too much the
accuracy of the representation of the original profile with
decomposition. For the acceptable decomposition we required that the RMS
of the weighted deviations of the Gaussian model from the observed
profile had to be no more than 1.006 times the weighted noise level of
the emission-free baseline regions of this profile. This condition
corresponds to $|\lg(\chi^2/N_{\mathrm{dof}})| < 0.015$, but we must
keep in mind that here both, the estimates of the standard deviations of
the data points and the fitted models are based on the same data set. The
value of the multiplier has been chosen so that the averages of the
noise levels and the RMS deviations of the models over all HEALPix
profiles of the survey are equal (in practice, for the final
decomposition the difference of these averages was less than 0.09\% of
their mean).

For the decomposition of the HI4PI, we also introduced some
modifications to our old algorithm. First of all, the accuracy of the
measured brightness temperatures is not the same for all profile
channels. According to the radiometer equation the noise level depends
on the signal strength and different corrections applied during the
processing of the observed spectral dumps introduce additional
uncertainties. During the decomposition, all this was considered through
the weights assigned to all values of $T_\mathrm{b}(v_i)$ of the profile
in each HEALPix pixel. For the GASS part of the HI4PI, these weights
were calculated from the RMS deviations of the spectral dumps from the
corresponding average profile in each pixel, as described in
\citet{Kalberla2015}. For EBHIS, the $T_\mathrm{sys}(v_i)$ together with
the RFI flags was used as a proxy for the RMS in the individual spectral
dumps contributing to the HEALPix pixel.

As the velocity resolution of the EBHIS data is the lowest among the
decomposed surveys, it revealed a problem with the original
decomposition algorithm: near the steepest gradients in the observed
profiles the decomposition results often contained groups of very high
($|T_\mathrm{bc}| \gg |(T_\mathrm{b}(v_{i}) +
T_\mathrm{b}(v_{i+1})/2|$), but narrow ($\sigma < \Delta v$,
where $\Delta v = v_{i+1} - v_{i}$ is the channel separation of the
survey) Gaussians, which centers were located between the consecutive
spectrometer channels $v_i$ and $v_{i+1}$. The obtained decompositions
fitted well the observed profiles at $v_i$ and $v_{i+1}$, but with
strong fluctuations between the measured data points. To suppress such a
behavior, we used the penalty function approach. During the
decomposition, we calculated at $v_{\mathrm{c},j}$ of all Gaussians the
deviations of the model from the cubic spline representation of the
observed profile and added the weighted squares of these deviations to
the RMS of the fit. The weights of these penalty points were calculated
as $w_j = \frac{\Delta v \delta v_{\mathrm{min}}}{2 \sigma_j} W$, where
$\delta v_{\mathrm{min}}$ is the positive velocity difference between
the $v_{\mathrm{c},j}$ and the nearest $v_i$. $W$ was obtained from the
cubic spline fit of the weights of the observed profile at velocities
$v_{\mathrm{c},j}$.

After finding this problem in the decomposition of the EBHIS, we
performed a special search in the results of the GASS decomposition and
found similar narrow Gaussians also there. The only difference was that
in the GASS such components appeared in general one at a time and
therefore did not cause so obvious oscillations of the resulting model
profiles as in EBHIS. Nevertheless, in the final decomposition, the
modification for suppressing such Gaussians was applied to both the
EBHIS and GASS data.

Considerable change in the decomposition algorithm was made available by
the increased power of the computers. In the LAB survey, we used the
decomposition results of one of the neighboring profiles of any given
profile as an initial estimate of the Gaussian parameters, and we only
started the decomposition of the first profile with one roughly
estimated component at the highest maximum of the profile. In the HI4PI
study, the decomposition was divided into two stages. In the first run,
we decomposed all profiles in the HEALPix database independently of
their neighbors, starting with one Gaussian at the brightest tip of the
profile. In the second stage, we compared the results for neighboring
profiles. To accomplish this, we used the decomposition obtained so far
for each profile as an initial approximation for all eight nearest
neighbors of this profile, and checked whether this led to better
decomposition of these neighboring profiles. If the decomposition of a
neighbor was improved, the new result was used as an initial solution
for the eight neighbors of this profile, and so on. The process was
repeated until no more improvements were found (on the order of 500 runs
through the full database).

During this process, we estimated the goodness of the obtained fits
using two criteria. First of all, as described above, we did not accept
the decompositions for which the RMS of the weighted deviations of the
Gaussian model from the observed profile exceeded more than 1.006 times
the weighted noise level of the emission-free baseline regions of this
profile. If for some profiles the RMS criterion was satisfied for more
than one trial decomposition with different numbers of Gaussians, we
accepted the solution with the smallest number of the components. In the
case of acceptable decompositions with equal numbers of the Gaussians,
we chose the decomposition with the smallest RMS as the best. The
decomposition process is described in more detail in Secs. 3.1. and 3.2.
of \citet{Haud2000}.

Finally, in the case of the LAB, we only used positive Gaussians
($T_\mathrm{bc} > 0~\mathrm{K}$) for decomposition and the parts of the
profiles with $ \overline{T_\mathrm{b}(v_i)} < 0~\mathrm{K}$ were
not considered at all. With the HI4PI, we also fitted negative Gaussians
to the regions of the profiles where the brightness temperature was on
average below zero. For fitting the regions of the profiles, where
$\overline{T_\mathrm{b}(v_i)} \ge 0~\mathrm{K}$, we still used only
positive Gaussians and not a combination of positive and negative
components. Therefore, only strong absorption or the baseline problems,
where the $\overline{T_\mathrm{b}(v_i)}$ of the profile drops below
the continuum level, induced negative Gaussians. In general, absorption
was modeled as a gap between positive Gaussians.

\subsection{Defining different \hi\ phases}
\label{DefPhases}

The hypothesis that the \hi\ component line shapes are Gaussian rests on
the assumption that motions within any \hi\ cloud have a random velocity
distribution that can be described by a Gaussian function. This
assumption is supported by the finding that the \hi\ absorption features
are often well represented by Gaussians in optical depth $\tau(\nu)$
\citep{Mohan2004}. The decomposition of the emission profiles is
physically meaningful only when $\tau(\nu) \ll 1$. Nevertheless, both
the opacity and brightness profiles of most sources are easily
decomposed into Gaussians, so whether or not this model is physically
correct, it works empirically and is convenient.

As a result, the Gaussian decomposition offers an opportunity to
differentiate between components with different line widths
\citep{Haud2007}. In combination with absorption-line studies
\citep{Dickey2003, Heiles2003a}, components with broad or narrow line
widths have been found to arise from the WNM and CNM, respectively
\citep{Dickey1990, Wolfire2003, Hennebelle2012} and there could be
substantial amounts of thermally unstable LNM as well \citep{Heiles2003a,
Haud2007, Hennebelle2012, Roy2013, Saury2014}.

Our decomposition of the full HEALPix database of the HI4PI profiles
gave on average 8.472 Gaussians per profile, but the complexity of the
profiles greatly varies between different sky regions and often the
Gaussians in the decomposition are heavily blended with each other. This
considerably complicates the physical interpretation of the results, as
for such profiles the decomposition may be not unique. This means that
several quite different solutions may approximate the observed profile
almost equally well, and the decomposition provides no satisfactory
means for choosing between these solutions, while others, equally good
or even better ones, may not be found at all. As described in Sec.
\ref{Gaussians}, we have tried to reduce this problem by keeping the
number of the used Gaussians as low as possible and by considering
during the decomposition of each profile the information about the
neighboring profiles.

\begin{figure}
%   \resizebox{\hsize}{!}{\includegraphics{Fig1b.pdf}}
   \resizebox{\hsize}{!}{\includegraphics{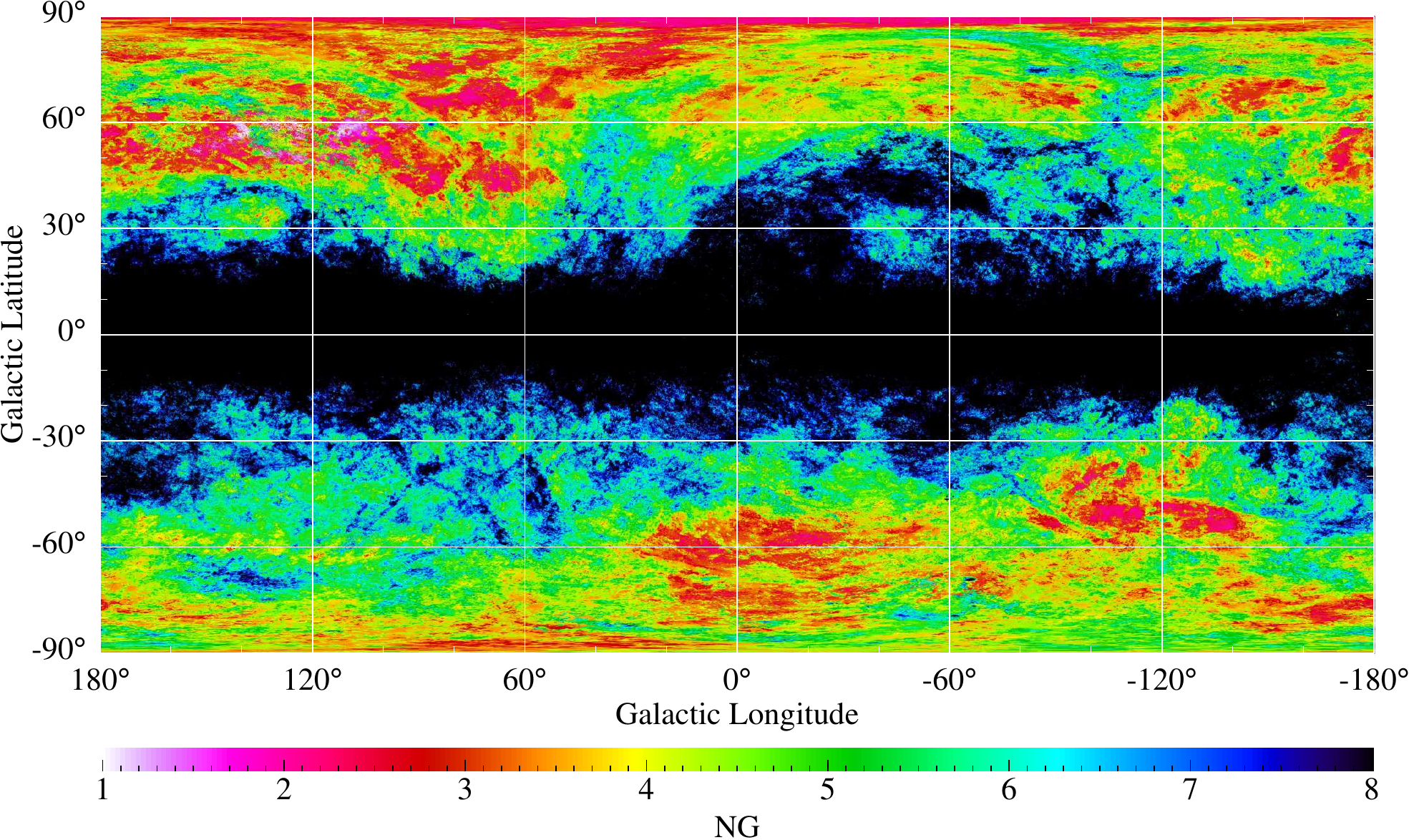}}
   \caption{Sky positions of the 7\,855\,871 profiles with less than
      eight Gaussians in the central emission peak. The positions of the
      simplest profiles are marked by the reddest colors. Among the
      profiles with an equal number of Gaussians those with lower column
      density of the local \hi\ are plotted with redder color.}
   \label{Sky}
\end{figure}

In this way we hope that for similar profiles the algorithm prefers
similar decompositions, which may better represent the general
properties of the \hi\ gas than the set of more independent solutions.
Nevertheless, these precautions cannot guarantee that our decomposition
results mostly describe the properties of the underlying \hi\ gas and
are not dominated by the random factors inherent to the decomposition
process. Therefore, we decided to compare first the results, obtained
from independent observations (the LAB versus HI4PI) with different
decomposition algorithms (as described in Sect.  \ref{Gaussians}). We
expect that the similarity of the results, obtainable from such studies
indicate that the main properties of the results are determined by the
nature of the galactic gas but in cases where uncertainties in the
Gaussian decomposition are dominant the results may yield nearly
incomparable outcomes.

\begin{figure}
%   \resizebox{\hsize}{!}{\includegraphics{Td3.pdf}}
   \resizebox{\hsize}{!}{\includegraphics{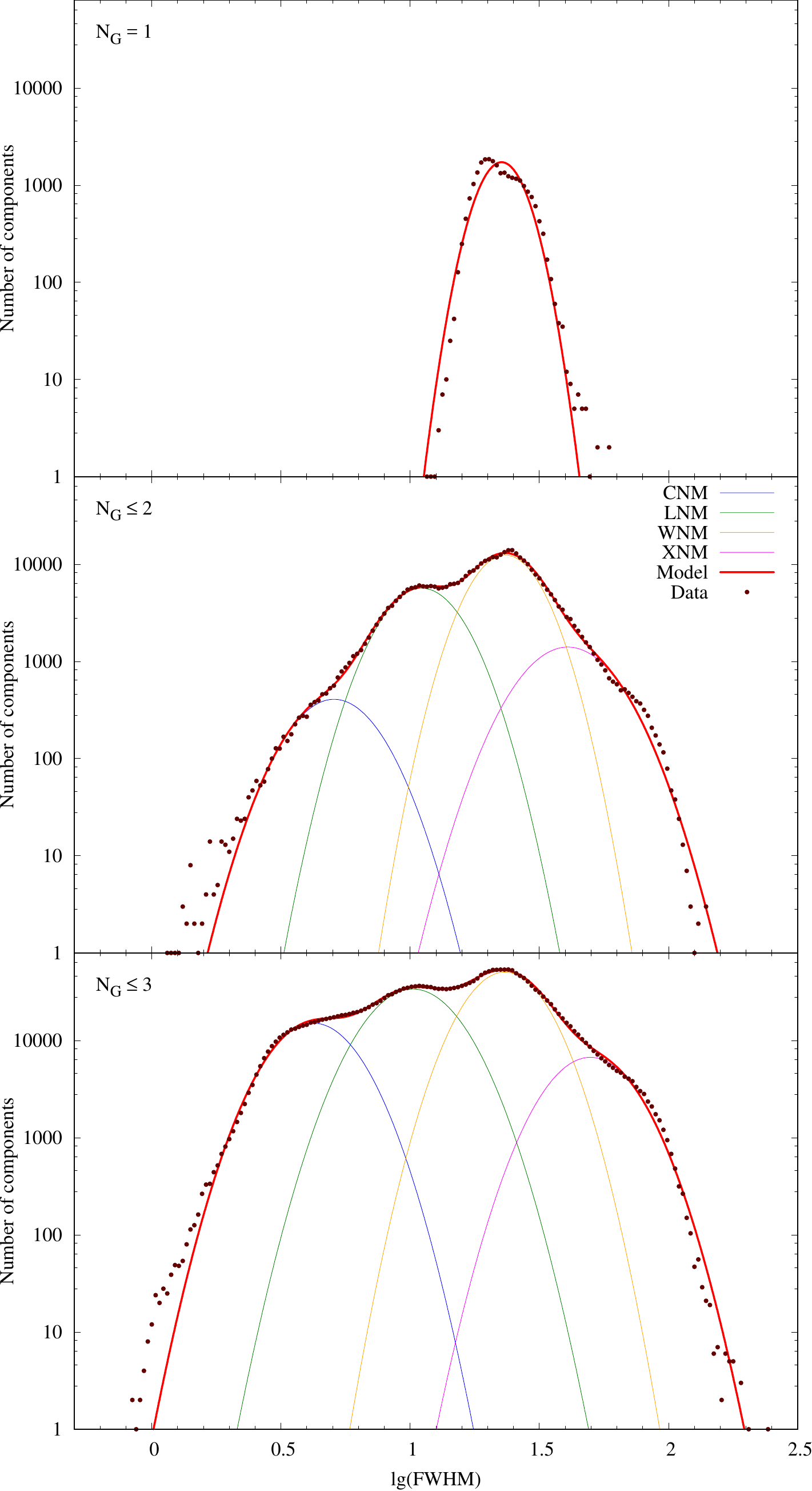}}
   \caption{Distribution of the widths of Gaussians for the
      profiles decomposed to one (upper panel), two (middle panel) or
      three (lower panel) components.}
   \label{Td3}
\end{figure}

It is clear that the blending and non-uniqueness problems are weaker for
the simplest profiles and for the strongest Gaussians in the
decompositions of the more complicated profiles. Therefore, as with
the LAB \citep{Haud2007}, we started the analysis of the HI4PI data with
the simplest profiles in the database, which are located at relatively
high galactic latitudes and followed the changes in the results caused
by gradually adding more complicated profiles at lower latitudes
(Fig.~\ref{Sky}). With the HI4PI data we used for these tests only the
Gaussians which belong to the central peaks of the observed profiles.
We defined the central peak as the maximum of $T_\mathrm{b}(v_i)$
closest to $v_{\mathrm{LSR}} = 0~\mathrm{km\,s}^{-1}$. We considered
this peak to extend to the velocities where the $T_\mathrm{b}(v)$,
calculated from the decomposition  results, drops below the noise level
of the corresponding observed profile.

This approach had an additional advantage that it eliminated from the
analysis nearly all noise Gaussians without applying any formal
selection criteria (as Eq. 3 and 4 in \citet{Haud2007}). Only a rather
small contamination by RFI remained. In some cases such approach
included also intermediate- (IVC) or even high-velocity clouds (HVC)
into our sample, but as in the following we will apply also different
velocity limits on the studied components, this is not a considerable
problem. Moreover, we will use the number of the Gaussians in the
central peak of the profiles, $N_{\mathrm{G}}$, as an indicator of the
profile complexity and study only the profiles with moderate complexity.
As the inclusion of IVCs and HVCs as local gas increases
$N_{\mathrm{G}}$ for the corresponding pixels, such cases have higher
probability to be excluded as too complex profiles. Fig.~\ref{Sky} gives
the sky distribution of the described complexity estimates for
$N_{\mathrm{G}} < 8$.

Using the LAB data, we have demonstrated \citep{Haud2007} that by
considering only the simplest \hi\ profiles and relatively small LSR
velocities ($-9 \le v_\mathrm{c} \le 4~\mathrm{km\,s}^{-1}$) it is
possible to distinguish three groups of preferred line-widths, which are
defined by fitting the sum of log-normal functions to the frequency
distribution of the Gaussian widths. It is found that on the basis of
the LAB profiles decomposed into one, two or three low velocity
Gaussians, the estimates for the mean line-widths of the CNM, LNM and
WNM gas are $4.9 \pm 0.1$, $12.0 \pm 0.3$, and $24.4 \pm
0.2~\mathrm{km\,s}^{-1}$ (Fig. 4 in \citet{Haud2007}).  In the same way,
we obtained from the decomposition of the HI4PI survey the values $4.7
\pm 0.4$, $11.2 \pm 0.6$, and $24.2 \pm 0.4~\mathrm{km\,s}^{-1}$
(Fig.~\ref{Td3}). For CNM and WNM the results from LAB and HI4PI are
rather close to each other. The agreement is slightly worse for the LNM,
but this is understandable, as the frequency distribution of the widths
of the LNM Gaussians is blended by both, CNM and WNM. This makes the
separation of this gas phase more uncertain.

Besides the curves for CNM, LNM and WNM Fig.~\ref{Td3} (also Fig. 4 in
\citet{Haud2007}) contains a fourth curve, labeled here as XNM. At first
sight, the nature of this component is not obvious and therefore we made
a special study of the profiles, responsible for this component (see
Appendix \ref{XNM}). It turned out that the corresponding Gaussians are
weak and in many cases they describe the wide non-Gaussian wings of the
profiles, but sometimes (mostly in the EBHIS part of the HI4PI but in
less then 3\% of all positions) they are caused by instrumental
problems. We found from different models that these Gaussians represent
3.2\% to 5.3\% of the total \hi\ and in the following they are considered
as belonging to WNM.

In the two lower panels of the Fig.~\ref{Td3} a similar, but even weaker
enhancement is visible also on the left edge of the distribution. This
is caused by rare and very narrow noise Gaussians, still remaining in
the analyzed data sample. As their contribution is very small, we did
not use an additional log-normal curve for describing this enhancement
and in the following the corresponding Gaussians are considered as part
of CNM.

As can be seen from Fig.~\ref{Td3}, with increasing complexity of the
profiles the relative number of CNM Gaussians increases and up to a
certain complexity limit this permits better determination of the
parameters of the CNM gas phase from the width distribution of the
corresponding Gaussians. However, at considerably higher complexity
levels ($N_{\mathrm{G}} > 10)$ of the profiles the definition of the LNM
phase becomes more and more questionable as the decompositions of the
complex profiles usually contain many weak Gaussians with rather
ill-defined parameters. Such components add noise to the distribution of
the widths of the Gaussians, and the CNM, LNM and WNM components of the
width distribution start to merge into a relatively structure-less curve,
where the maximum corresponding to LNM is completely blended by CNM and
WNM. Therefore, the separation of CNM, LNM and WNM phases is best
determined for the profiles with some moderate complexity.

\begin{figure}
%   \resizebox{\hsize}{!}{\includegraphics{Td7.pdf}}
   \resizebox{\hsize}{!}{\includegraphics{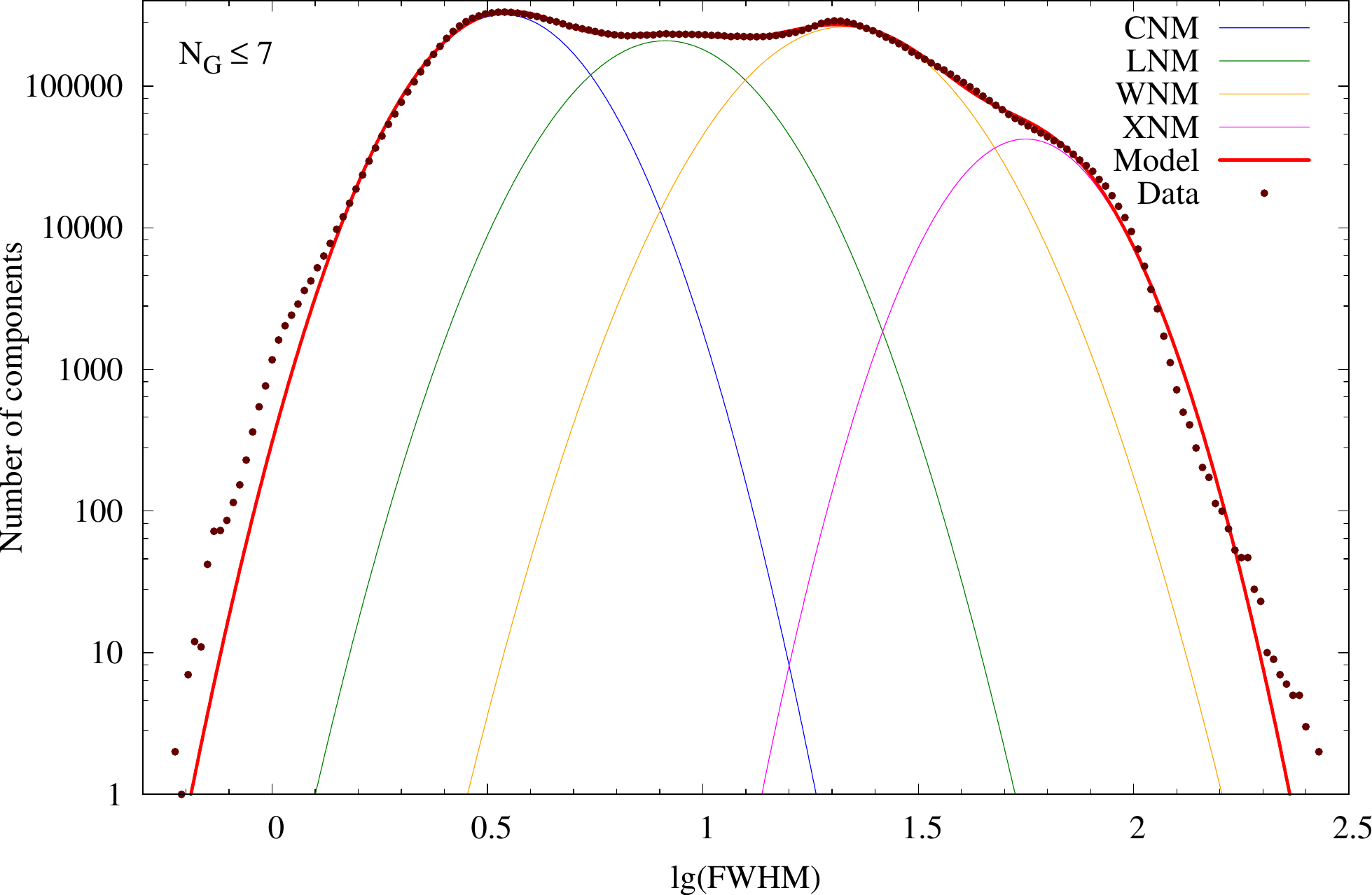}}
   \caption{Distribution of the widths of the Gaussians for the
      profiles with $N_{\mathrm{G}} \le 7$.}
   \label{Td7}
\end{figure}

To estimate the reasonable complexity limit, we analyzed the width
distributions of the Gaussians from the profiles with $1 \le
N_{\mathrm{G}} \le 15$ and found that the uncertainties of the
parameters of the LNM phase started to grow more rapidly for
$N_{\mathrm{G}} \ge 8$. As for $N_{\mathrm{G}} \le 3$ we obtained for
$N_{\mathrm{G}} \le 7$ the following mean line-widths of the CNM, LNM
and WNM gas: $3.6 \pm 0.1$, $9.6 \pm 1.3$, and $23.3 \pm
0.6~\mathrm{km\,s}^{-1}$. The distribution of the Gaussian widths for
$N_{\mathrm{G}} \le 7$ together with the corresponding model curves are
given in Fig.~\ref{Td7}. The sky distribution of all profiles used for
these estimates is in Fig.~\ref{Sky}. Despite the fact that here we used
a considerably simplified approach to the determination of these
averages (we dropped the analysis of the reliability of the
decompositions of the profiles with $N_{\mathrm{G}} \ge 4$, as it was
done by \citet{Haud2007}), these estimates are comparable with the final
results of \citet{Haud2007} ($\mathrm{FWHM} = 3.9 \pm 0.6$, $11.8 \pm
0.5$ and $24.1 \pm 0.6~\mathrm{km\,s}^{-1}$).

As a result, we may recognize that Gaussian fits to observed
\hi\ emission or absorption profiles have been applied by a number of
authors \citep[e.g.,][]{Takakubo1966, Mebold1972, Mebold1982,
  Heiles2003b, Nidever2008, Murray2015}. We have used it with the LAB,
GASS and EBHIS data \citep{Haud2007, Kalberla2015, Kalberla2016,
  Kalberla2017}. We think that it is out of question that such fits are
correct in a mathematical sense. Whether the derived physical parameters
are meaningful (here the $\mathrm{FWHM}$) is a different question and by
selecting example profiles one can always find good or bad
examples. Even for perfect observations it may happen that the derived
parameters do not represent a correct model for the gas distribution. We
cannot exclude such cases but we believe that at least on average the
Gaussian fits deliver a reasonable model for the \hi\
distribution. Similarities in decompositions of the LAB, GASS and EBHIS
profiles support this and regardless of possible ambiguities we believe
that the results give some information on the properties of the CNM, LNM
and WNM gas phases. This agrees with the conclusion, pointed out by
\citet{Murray2017} that Gaussian fits to synthetic \hi\ lines are able
to recover gas structures with excellent completeness at high Galactic
latitude, and this completeness declines with decreasing latitude due to
strong velocity-blending of spectral lines. Moreover, as mentioned by
\citet{Roy2013}, while there are certainly possible drawbacks to using
Gaussian components to model \hi\ 21 cm spectra, there is no obvious
alternative route to determining physical conditions in the ISM.

\subsection{Separating \hi\ phases}
\label{SepPhases}

\begin{figure}
%   \resizebox{\hsize}{!}{\includegraphics{CwO.pdf}}
   \resizebox{\hsize}{!}{\includegraphics{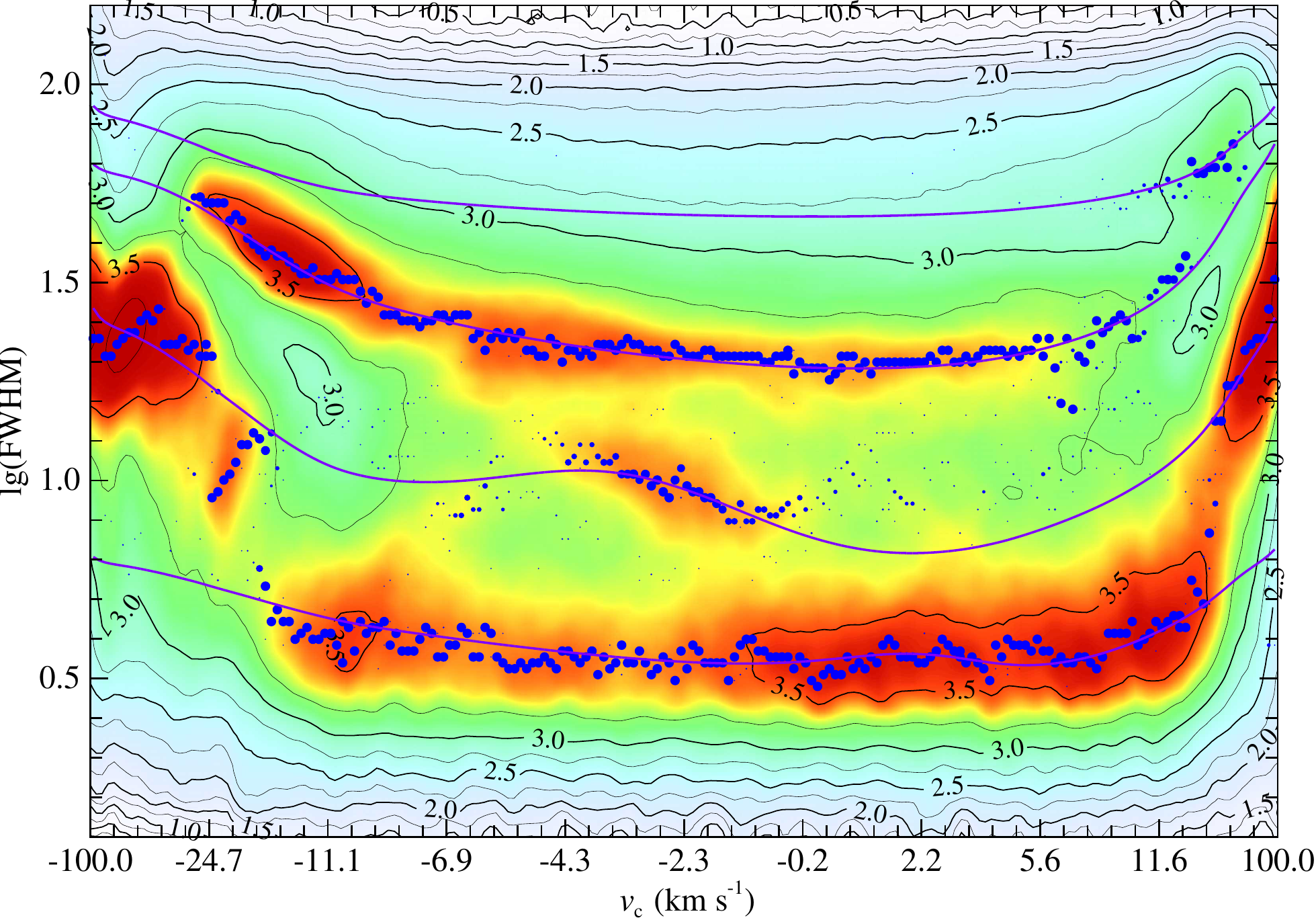}}
   \caption{Frequency distribution of the 42\,488\,690 Gaussians,
      used for the modeling of the line-width distributions of the gas
      phases, in the component central velocity -- line-width plane.
      Blue points mark the local maxima in the frequency distribution of
      the Gaussian line widths at different velocities and the magenta
      lines represent initial estimates of the centers of four
      log-normal model distributions of the Gaussian widths in different
      gas phases. As abscissa we have used the quantiles of the velocity
      distribution of the used Gaussians. The numbers mark the values of
      the deciles.}
   \label{CwO}
\end{figure}

So far we have modeled the average frequency distribution of the widths
of the Gaussians from profiles with $N_{\mathrm{G}} \le 7$ in the
velocity range, preselected in \citet{Haud2007}, but when plotting the
frequency distribution of these Gaussians in the component central
velocity -- line-width plane (Fig.~\ref{CwO}), we can see that at
different velocities the distribution of the widths is different. In our
trial to separate three \hi\ phases, we decided to take into account
also the velocity dependence of the Gaussian widths in different gas
phases. For this we assumed that in each narrow velocity range the
distribution of the widths of the Gaussians can be modeled with the sum
of four log-normal functions, as in Figs.~\ref{Td3} and \ref{Td7}, but
the parameters of the log-normal model distributions may change with the
velocity.

To estimate the reliability of the obtained models, we decided to
construct many different models and to analyze their average results
together with corresponding deviations. Altogether six families of the
models were constructed, each with many sub-models. All these
models were based on the same Gaussian decomposition of the HI4PI \hi\
profiles but differed from each other by the approaches, used for
the modeling of the distribution of the Gaussian parameters in their
central velocity -- line-width plane. For the following we define the
model family as a group of models, based on some more or less general
assumptions or fitting strategies, related to all models of this group. The
sub-models in the family differ then from each other by the usage of
different velocity resolutions, profile complexities and initial
approximations for the non-linear model fitting. For example, in all
families separate models were made for the complexity limits
$N_{\mathrm{G}} \le 3$, 4, 5, 6, or 7 local Gaussians per observed \hi\
profile. A brief overview of our model families is given in Table
\ref{table:0}. The differences between the model families are defined in
the second and the third columns of this table and the last column
describes the velocity ranges, used in the sub-models of the
corresponding family.

\begin{table*}
\caption{Families of the models for the width distributions of the
   Gaussians at different velocities}
\label{table:0}
\centering
\begin{tabular}{c c c c }
\hline\hline
Family & Minimized & Model restrictions  & $|v_{\mathrm{c}}| \le$ \\
\hline
1 & $\sum(N_{\mathrm{O}} - N_{\mathrm{M}})^2$ & $g_{ij} = \sum_{k=0}^6 p_k v_{\mathrm{c}}^k$ & 12.5, 15, 25, 30, 50 \\
2 & $\sum(N_{\mathrm{O}} - N_{\mathrm{M}})^2$ & $g_{ij} = \sum_{k=0}^4 p_k \arctan^k (p_5+p_6 v_{\mathrm{c}})$  & 12.5, 15, 25, 30, 50, 99 \\
3 & $\sum(N_{\mathrm{O}} - N_{\mathrm{M}})^2$ & -- & 50 \\
4 & $\sum(N_{\mathrm{O}} - N_{\mathrm{M}})^2$ & $\mathrm{FWHM}_{CNM} < \mathrm{FWHM}_{LNM} < \mathrm{FWHM}_{WNM} < \mathrm{FWHM}_{XNM}$ & 50 \\
5 & $\sum(\lg(N_{\mathrm{O}}+1) - \lg(N_{\mathrm{M}}+1))^2$ & -- & 50 \\
6 & $\sum(\lg(N_{\mathrm{O}}+1) - \lg(N_{\mathrm{M}}+1))^2$ & $\mathrm{FWHM}_{CNM} < \mathrm{FWHM}_{LNM} < \mathrm{FWHM}_{WNM} < \mathrm{FWHM}_{XNM}$ & 50 \\
\hline\hline
\end{tabular}
   \tablefoot{$g_{ij}$ is the $i$-th parameter ($1 \le i \le 3$) of the
   log-normal model distribution of the gas phase $j$ ($1 \le j \le 4$
   for CNM to XNM). Parameters $p_k$ are adjusted for obtaining the
   two-dimensional model of the frequency distribution of the Gaussians
   in the component central velocity -- line-width plane.
   $N_{\mathrm{O}}$ and $N_{\mathrm{M}}$ are the actual and model
   numbers of the Gaussians in different counting bins. Each family has
   sub-models for $N_{\mathrm{G}} \le 3$, 4, 5, 6, or 7.}
\end{table*}

\begin{figure}
%   \resizebox{\hsize}{!}{\includegraphics{CwM.pdf}}
   \resizebox{\hsize}{!}{\includegraphics{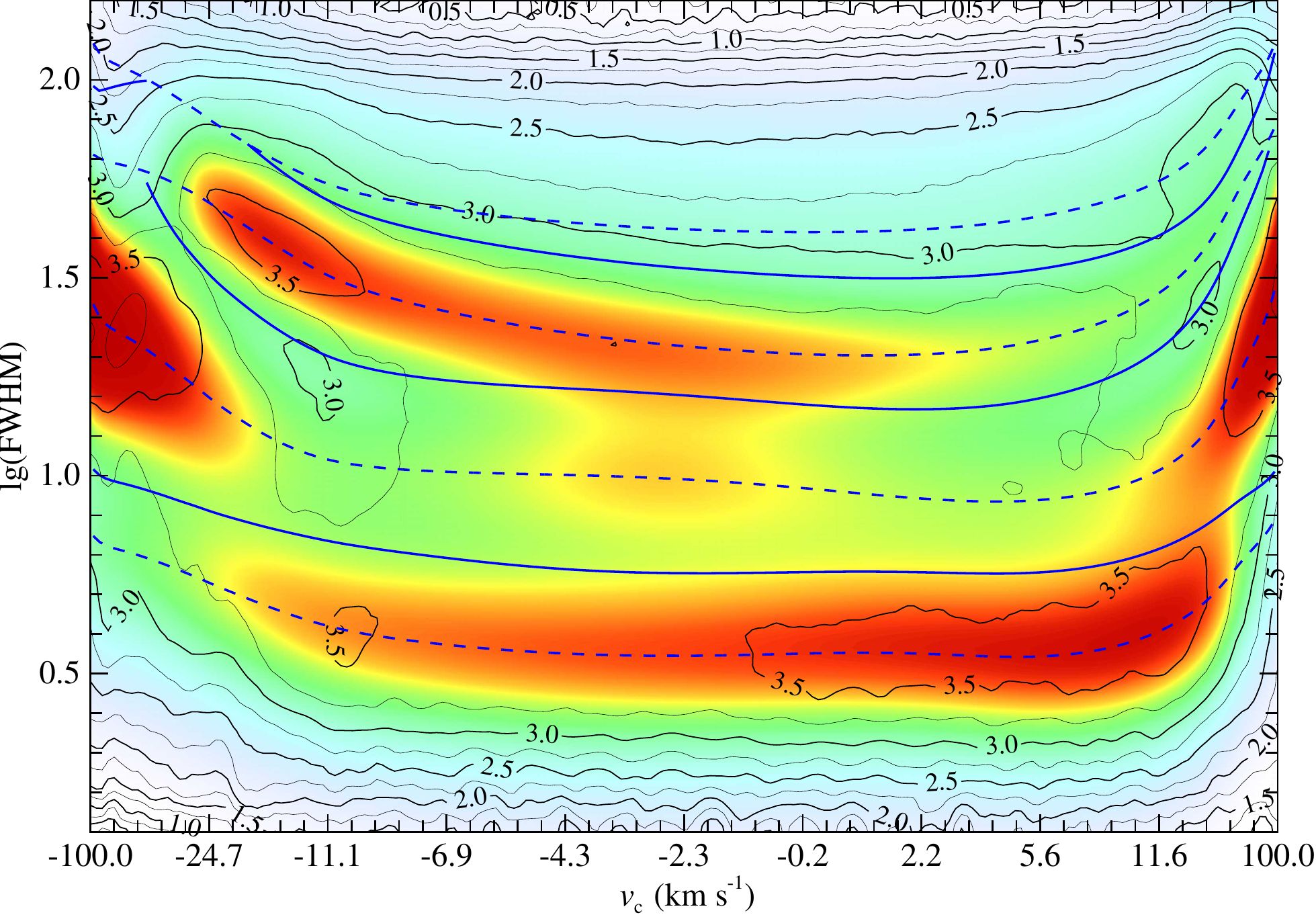}}
   \caption{As Fig.~\ref{CwO}, but the background colors now
      correspond to the sub-model for $|v_{\mathrm{c}}| \le
      99~\mathrm{km\,s}^{-1}$ and $N_{\mathrm{G}} \le 7$ from the second
      model family. The blue dashed lines give the positions of the
      peaks of the fitted log-normal distributions and solid lines
      indicate the formal division between the modeled gas phases.}
   \label{CwM}
\end{figure}

In the first two model families, we assumed that the changes in the
parameters of the four log-normal distributions with the velocity can be
described with the polynomial functions, which contain adjustable
parameters. In the first family we used the sixth order polynomial of
velocity and in the second family the fourth order polynomial of the
function $\arctan (p_5+p_6 v_{\mathrm{c}})$, where $p_5$ and $p_6$ were
also adjustable model parameters. For sub-models we analyzed profiles
with different complexity $N_G$ and their Gaussians from different
velocity ranges (as specified in Table \ref{table:0}). We divided all
Gaussians in the analyzed velocity ranges into 200 groups, each
containing the same number of components with similar velocities. In
this way the usage of different velocity ranges gave us also different
velocity resolutions of the corresponding models. By adjusting the
parameters of the polynomials, describing the velocity dependencies of
three parameters in each of four log-normal distribution, we represented
the complete distribution of the Gaussians (as in Fig.~\ref{CwO}) with
some model distribution (one possible example in Fig.~\ref{CwM}) by
minimizing the RMS deviations between observed and model distributions
in all velocity and line-width bins.

To define the initial approximation for all models in these two
families, we searched in each velocity bin of Fig.~\ref{CwO} for the
line widths, corresponding to the local frequency maxima and recorded
their velocities, Gaussian widths and the breadths, $\Delta$, of the
$\lg(\mathrm{FWHM})$ regions, in which the corresponding point was a
global maximum. The results are given as blue points in Fig.~\ref{CwO}.
The sizes of the points are proportional to $\Delta$. The found maxima
were classified by hand into four groups, corresponding to four
log-normal functions, fitting the frequency distribution of the
Gaussians. The initial dependencies of the centers of the log-normal
functions on the velocity were then defined as fits of the polynomials
through the corresponding points of the local maxima in the observed
frequency distribution. For this fitting, the values of $\Delta$ were
used as weights. The results are given in Fig.~\ref{CwO} by magenta
lines. Initial approximations of the heights of the log-normal
distributions were determined by fitting the family specific polynomials
through the Gaussian frequencies at the points calculated from the
functions found for the centers of the log-normal distributions.

From the preceding experience on the modeling of the Gaussian frequency
distributions in the fixed velocity range (Figs.~\ref{Td3} and
\ref{Td7}), it was clear that the most uncertain property of the
log-normal distributions is their width. Therefore, the initial
approximations for the widths of the log-normal distributions were
determined as constants from the velocity range $-3.39 < v_{\mathrm{c}}
< -1.26~\mathrm{km\,s}^{-1}$, where the LNM is most clearly separated
from CNM and WNM (Fig.~\ref{CwO}). When starting the final model
fitting, these constants were used for all velocities (the coefficients
of the velocity dependent terms of the corresponding functions were all
taken to be equal to zero), but during the fitting of the models the
coefficients of all terms were allowed to vary freely.

For the fitting of the models, we used the multidimensional downhill
simplex method and for the first two families of the models we generated
up to five subfamilies with different algorithms for the generation of
the initial simplex. When the model converged to some solution, we
generated a new large simplex around the obtained result and repeated
the fitting until the repetitions did not yield any considerable
improvements in the results. Due to a large number of the free fitting
parameters, such repetitions generally found deeper minima than those
obtained in the first run of the fitting.

\begin{figure}
%   \resizebox{\hsize}{!}{\includegraphics{Dis.pdf}}
   \resizebox{\hsize}{!}{\includegraphics{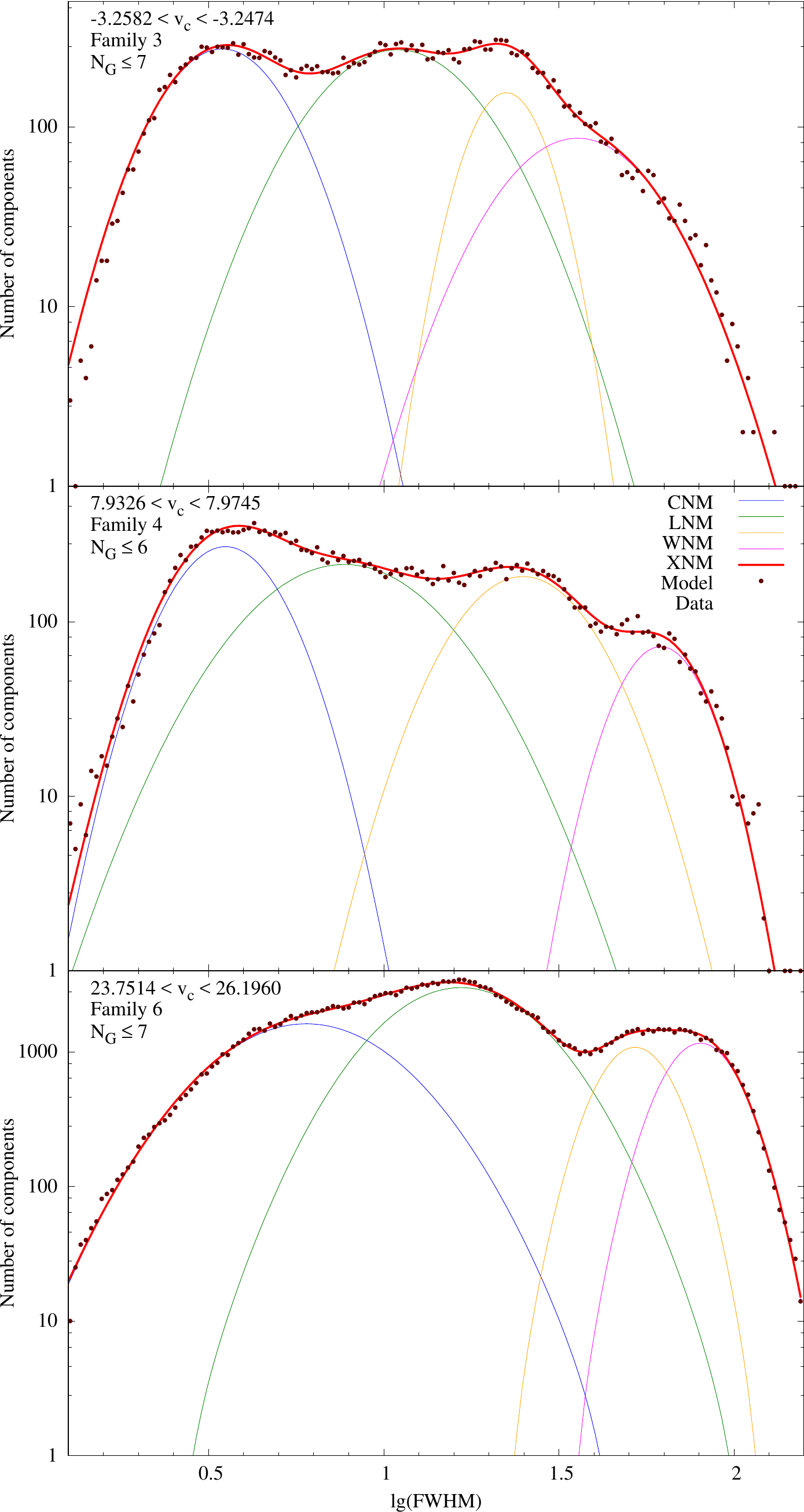}}
   \caption{Three examples of the models from the last four families. In
      the upper panel is the model from the velocity range, where the
      LNM is most clearly visible in Fig.~\ref{CwO}. The middle panel
      corresponds to the velocities, close to the limit
      $|v_{\mathrm{c}}| \le 8~\mathrm{km\,s}^{-1}$, used for the
      discussions in the present paper and the lower panel illustrates
      the situation at extreme velocities $|v_{\mathrm{c}}| =
      25~\mathrm{km\,s}^{-1}$, used for Fig.~\ref{Ave}.}
   \label{Dis}
\end{figure}
All together we obtained in both model families up to 150 different
models. In each obtained model we searched the velocity dependent values
of the line widths at which the modeled frequency of the Gaussians in
two neighboring (on the order of increasing line widths) gas phases
were equal. The obtained values of
$\mathrm{FWHM_{CL}}(v_{\mathrm{c}})$,
$\mathrm{FWHM_{LW}}(v_{\mathrm{c}})$ and
$\mathrm{FWHM_{WX}}(v_{\mathrm{c}})$ for the borders between CNM -- LNM,
LNM -- WNM and WNM -- XNM components were converted into corresponding
Doppler temperatures $T_{\mathrm{D,CL}}(v_{\mathrm{c}})$,
$T_{\mathrm{D,LW}}(v_{\mathrm{c}})$ and
$T_{\mathrm{D,WX}}(v_{\mathrm{c}})$, where
\begin{equation}
   T_\mathrm{D}(v_{\mathrm{c}}) =
      21.8618\,\mathrm{FWHM}^2(v_{\mathrm{c}}) \label{Eq2}
\end{equation}
and averaged over all models of one family. The standard deviations of
the results from different models were considered as error estimates of
the results.

An example of the model from the second family for $|v_{\mathrm{c}}| <
100~\mathrm{km\,s}^{-1}$ and $N_{\mathrm{G}} \le 7$ is given in
Fig.~\ref{CwM}. Here the isolines represent the observed distribution
from Fig.~\ref{CwO} and the background colors give the model
approximation of these lines. Blue dashed curves give the locations of
the peaks of the log-normal model curves (the magenta lines from
Fig.~\ref{CwO} after the model fitting) and the solid blue lines are the
line widths, which separate different gas phases.

The upper left corner of Fig.~\ref{CwM} demonstrates that at higher
velocities the behavior of the models was sometimes so wild that it was
impossible to define the separating line widths. At the same time, at
low velocities the polynomial models were even more stable than we
expected in advance. Therefore, we decided to take a step further
and to construct additional model families, in which we dropped the
requirement of the polynomial dependence of the parameters of the
log-normal distributions on the velocity, but analyzed the width
distributions only for $|v_{\mathrm{c}}| < 50~\mathrm{km\,s}^{-1}$. In
the last four families of the models we turned back to the fitting of
the width distributions in the narrow velocity ranges, as it was
described in Sec. \ref{DefPhases}, but for each fit we defined an
independent velocity region for the fitting. The velocity ranges were
constructed randomly so that they contained at least 19\,000 and at most
210\,000 Gaussians and all Gaussians had equal probabilities to be
included in some velocity range. For each model the upper limit of the
profile complexity was also randomly chosen so that $N_{\mathrm{G}} \le
3$, 4, 5, 6, or 7.

In each of these four additional families we used a different approach
to the fitting of the models and generated 60\,000 independent
sub-models for different velocity ranges and profile complexities. For
all these models the initial approximations were taken from the final
models of the second polynomial family. For families three and four we
minimized the parameter $\sum(N_{\mathrm{O}} - N_{\mathrm{M}})^2$, but
for families five and six we used $\sum(\lg(N_{\mathrm{O}}+1) -
\lg(N_{\mathrm{M}}+1))^2$, where $N_{\mathrm{O}}$ and $N_{\mathrm{M}}$
are the numbers of the Gaussians in different width bins from our
Gaussian decomposition and from the corresponding model, respectively.
In this way the families three and four were mostly sensitive to the ridges of
the frequency distributions, but five and six gave more weight to the wings
of the distributions. In families three and five we did not apply any
additional restrictions to the model fitting, but in families four and six we
also demanded that during the fitting process the $\mathrm{FWHM}$,
corresponding to the peaks of the model distributions, had to follow the
relation $\mathrm{FWHM_{CNM}} < \mathrm{FWHM_{LNM}} <
\mathrm{FWHM_{WNM}} < \mathrm{FWHM_{XNM}}$. It was accomplished through
the penalty function so that each violation of this rule multiplied the
sum of the squares of the residuals by 2.0. Some examples of the
models from the last four families are given in Fig.~\ref{Dis}.

After constructing the models, the interpretation of the results was
similar to that of the polynomial families one and two. We calculated the
line widths at which the modeled frequencies of the Gaussians in the
neighboring gas phases were equal. The obtained values of the
$\mathrm{FWHM}$ were converted into the Doppler temperatures, averaged
in regular velocity bins over all models of each family and the
corresponding standard deviations were calculated as estimates of the
result uncertainties.

\begin{figure}
%   \resizebox{\hsize}{!}{\includegraphics{Ave.pdf}}
   \resizebox{\hsize}{!}{\includegraphics{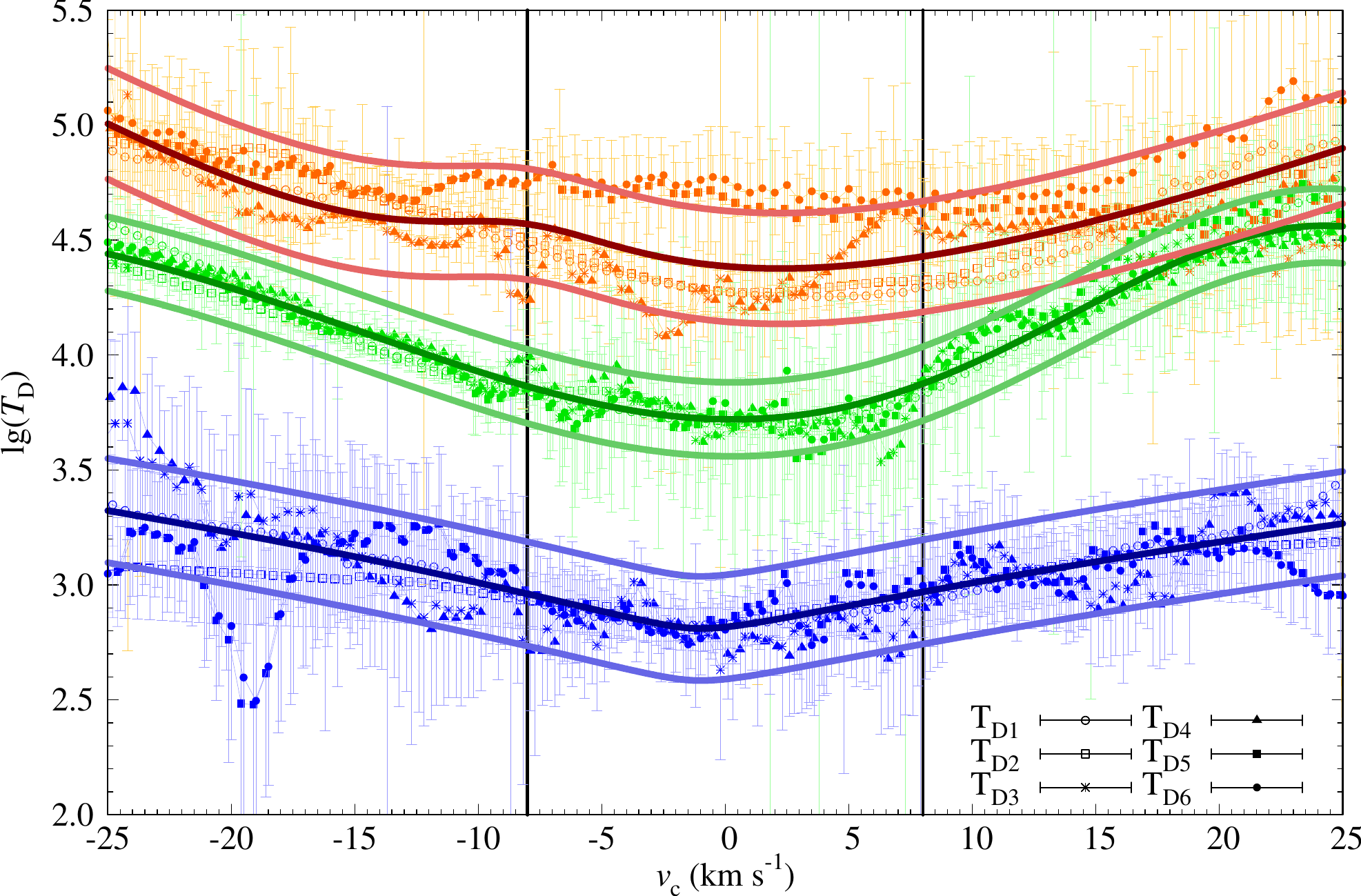}}
   \caption{Results for $T_{\mathrm{D,CL}}$ (blue),
      $T_{\mathrm{D,LW}}$ (green) and $T_{\mathrm{D,WX}}$ (red) from six
      model families $T_{\mathrm{D1}} - T_{\mathrm{D6}}$. Each family is
      plotted with different symbols. The error bars correspond to the
      standard deviations of the sub-models in each family. The
      polynomial fits to the results from all families are plotted with
      the thick dark lines and the thick lighter lines give the
      corresponding median uncertainties in the velocity range
      $|v_{\mathrm{c}}| \le 25~\mathrm{km\,s}^{-1}$. Two vertical black
      lines indicate the velocity range $|v_{\mathrm{c}}| \le
      8~\mathrm{km\,s}^{-1}$, used for the discussions.}
   \label{Ave}
\end{figure}
The comparison of the results indicated that up to about
$|v_{\mathrm{c}}| = 15~\mathrm{km\,s}^{-1}$ the different model families
were in good mutual agreement. At higher velocities the deviations
started to increase (especially at negative velocities for
$T_{\mathrm{D,CL}}$ -- the Doppler temperature, which separates CNM from
the LNM), but remained acceptable up to about $|v_{\mathrm{c}}| =
25~\mathrm{km\,s}^{-1}$. At even higher velocities the deviations grow
rapidly, which makes the results extremely uncertain. This is rather
understandable. We planned to study the local interstellar medium and to
test how far our approach may be viable. In Fig.~\ref{CwO} the bands of
the CNM and WNM are well visible up to $|v_{\mathrm{c}}| \approx
15~\mathrm{km\,s}^{-1}$. The blue dots defining the LNM are considerably
more scattered, but still the presence of this phase is felt. Beyond
$|v_{\mathrm{c}}| \approx 15~\mathrm{km\,s}^{-1}$ the bands of CNM and
WNM turn upward. CNM at negative and WNM at positive velocities becomes
also less obvious and the definition of the LNM becomes rather doubtful.
At $|v_{\mathrm{c}}| > 25~\mathrm{km\,s}^{-1}$ the identification of the
LNM with the intermediate- and high-velocity clouds (the strong
frequency enhancements at about $\lg(\mathrm{FWHM}) = 1.35$ near the
left and right edges of the figure, see \citet{Haud2008}) is already
more than questionable. These results are also understandable from
the comparison of the example models, given in Fig.~\ref{Dis}. At low
velocities the CNM, LNM and WNM phases are well defined in the
distribution of the Gaussian widths, but at higher velocities their
identification becomes more and more doubtful. To avoid possible
misidentifications we eventually decided to restrict our analysis to
$|v_{\mathrm{c}}| < 8~\mathrm{km\,s}^{-1}$, see Sect. \ref{Maps}.

The results for $|v_{\mathrm{c}}| \le 25~\mathrm{km\,s}^{-1}$ from all
model families are summarized in Fig.~\ref{Ave}. The values for
$T_{\mathrm{D,CL}}$ are given in blue, those for $T_{\mathrm{D,LW}}$ are
in green and red corresponds to $T_{\mathrm{D,WX}}$. The error-bars
illustrate the standard deviations of the results in the different model
families. The dark thick curves are the polynomial fits (using
corresponding standard deviations for computing the weights) to all the
presented results and the lighter thick curves illustrate the median
uncertainties (for $|v_{\mathrm{c}}| \le 25~\mathrm{km\,s}^{-1}$) of the
dark curves.

%=========================================================================
%=========================================================================
%=========================================================================

\begin{figure}[th] %%  
   \centering
   \includegraphics[width=9cm]{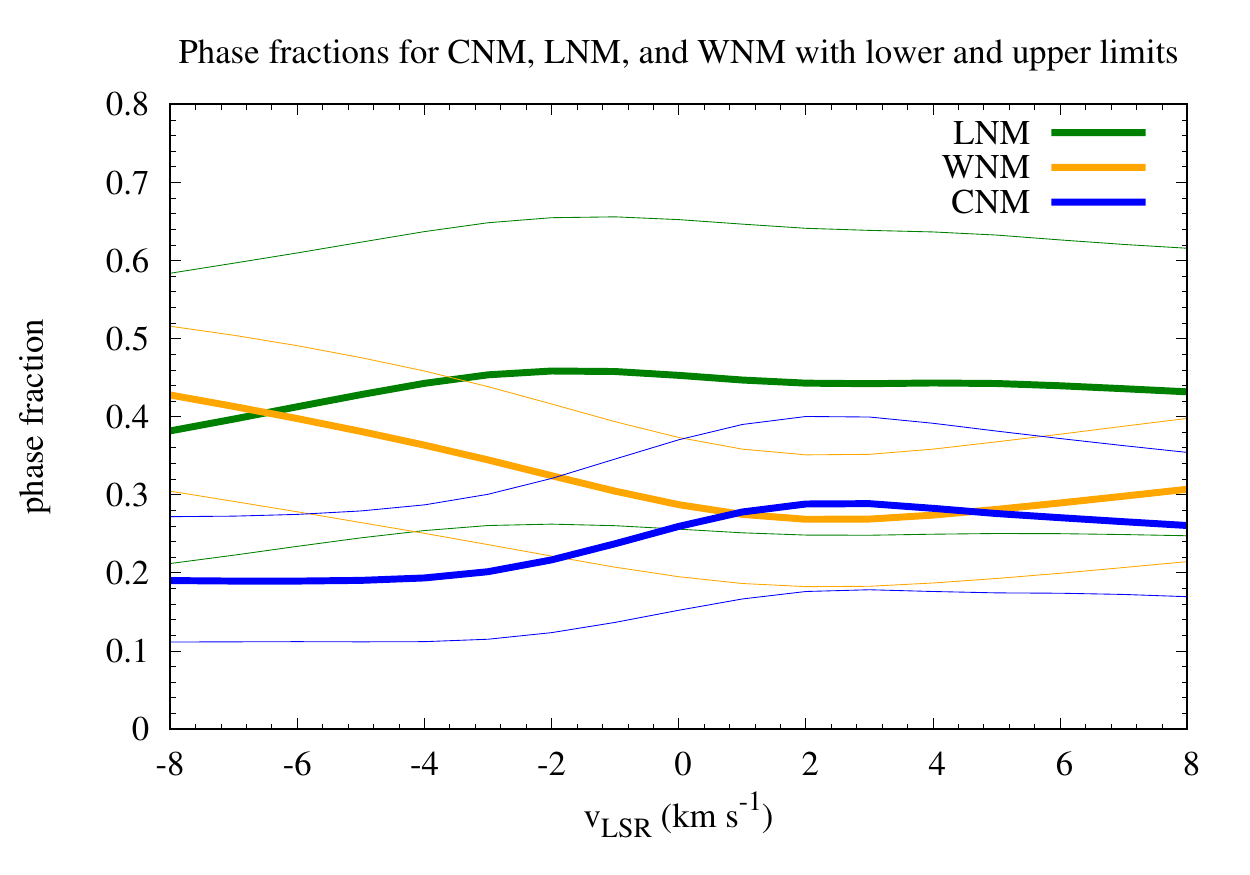}
   \caption{Velocity dependencies for phase fractions $f_{\mathrm{CNM}}$
     (blue), $f_{\mathrm{LNM}}$ (green), and $f_{\mathrm{WNM}}$
     (orange), plotted with thick lines, and their upper and lower limits
     (thin). }
   \label{Fig_VelDist}
\end{figure}

%=========================================================================

\section{Derived properties of discrete \hi\ phases}
\label{Phases}

\begin{figure*}[h] %%  
   \centering
   \includegraphics[width=18cm]{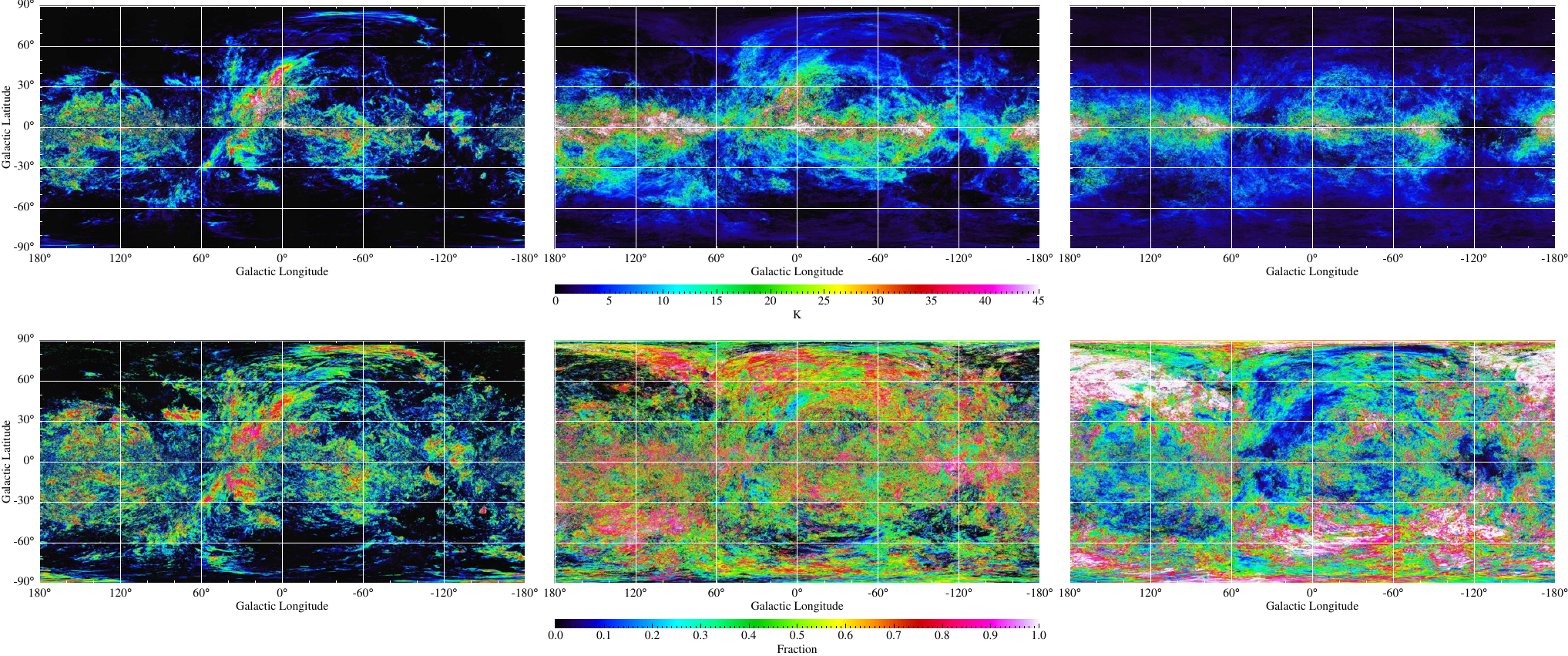}
   \caption{Top: All-sky plate carr{\'e}e (CAR) display of the observed
     \hi\ brightness temperature distributions at a velocity of $
     v_{\mathrm{LSR}} = 0 $ \kms, distinguishing contributions from the
     CNM (left), LNM (middle), and WNM (right). The color scale is
     linear from $T_{\mathrm{B}} = 0 $ K to $T_{\mathrm{B}} = 40$
     K. Bottom: \hi\ phase fractions $f_{\mathrm{CNM}}$ (left),
     $f_{\mathrm{LNM}}$ (middle), and $f_{\mathrm{WNM}}$ (right) for the
     same data. Galactic coordinates are used, the Galactic Center is in
     the middle. }
   \label{Fig_AllSky_1}
\end{figure*}

So far, when deriving criteria to distinguish Gaussians belonging to
different phases, we used only profiles with simple structures,
corresponding to decompositions with $N_{\mathrm{G}} \le 7$
components. For a meaningful analysis of the \hi\ distribution on the
sky we need to generalize our results and assume that the derived
criteria do not depend significantly on the number of clouds along the
line of sight or the complexity of the profiles, represented by
$N_{\mathrm{G}}$.

We sort Gaussian components according to their center velocities and
Doppler temperatures, using the velocity dependent thresholds from
Fig. \ref{Ave}, and group the components for separate phases. These
different groups of Gaussians are then used to generate individual
synthetic profiles $T_\mathrm{bP}$ for the phases P = CNM, LNM, and WNM
with a velocity grid of $\delta v_{\rm LSR} = 1 $ \kms\ on an nside =
1024 HEALPix grid in position. Thus we replace the observed
\hi\ distribution by three separate distributions to model the CNM, LNM,
and WNM.  As mentioned earlier, the WNM includes the contributions from
the XNM. The sum of these phase dependent distributions fits to the
observed brightness distribution $T_\mathrm{b}$; we only omit Gaussians
representing obvious instrumental problems \citep{Kalberla2015}. When
generating maps we smooth these brightness temperatures to an effective
beam-size of 30\arcmin~ FWHM.

%=========================================================================
\begin{figure*}[bht] %%  
   \centering
   \includegraphics[width=18cm]{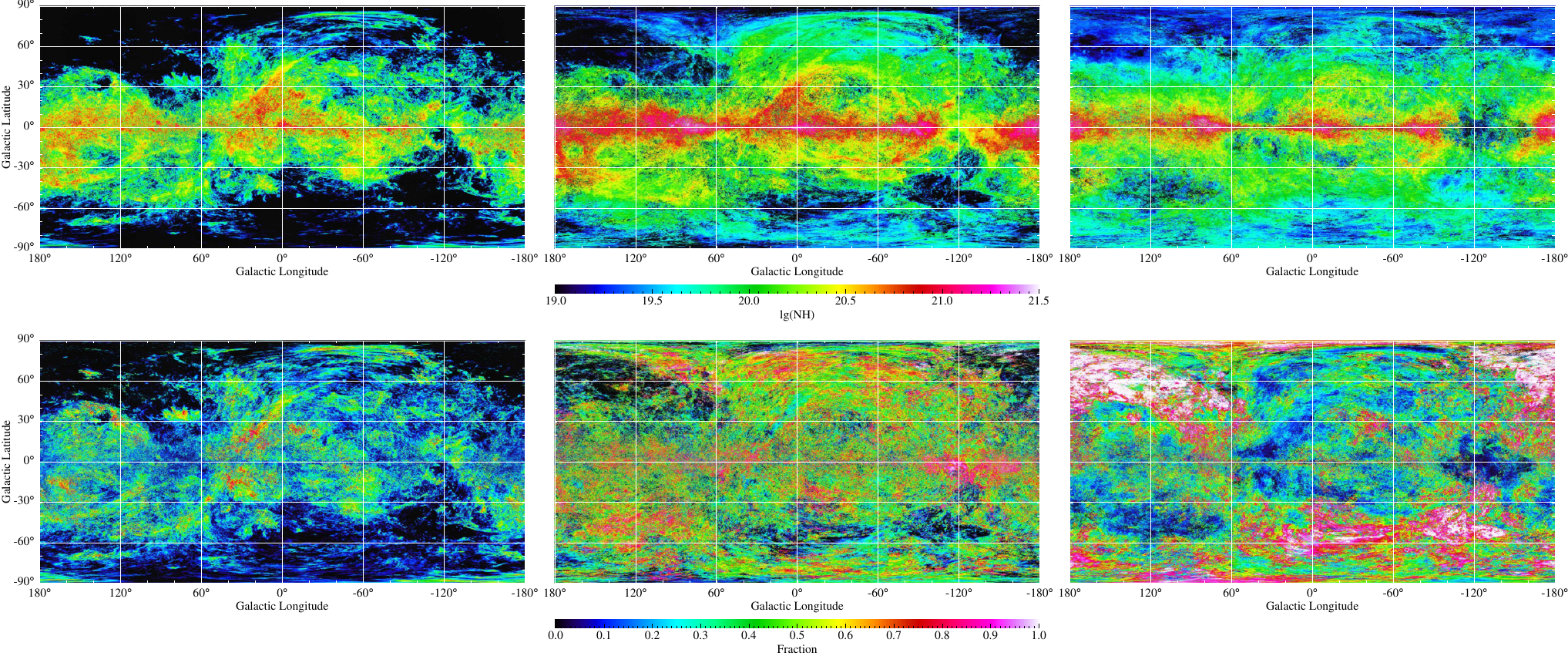}
   \caption{Top: All-sky display of the observed \hi\ column density
     distributions, integrating over a velocity range $ -8 <
     v_{\mathrm{LSR}} < 8 $ \kms\ and distinguishing contributions from
     the CNM (left), LNM (middle), and WNM (right). The color scale is
     logarithmic, ${\mathrm{lg}}(N_{\mathrm{H}}/[{\mathrm{cm}}^{-2}])$.
     Bottom: \hi\ phase fractions $f_{\mathrm{CNM}}$ (left),
     $f_{\mathrm{LNM}}$ (middle), and $f_{\mathrm{WNM}}$ (right) for the
     same data.  Galactic coordinates are used, the Galactic Center is
     in the middle.}
   \label{Fig_AllSky_2}
\end{figure*}

To measure, how much of the \hi\ gas belongs to a particular phase we
define phase fractions $f_{\mathrm{P}}(_{v1}^{v2})$ 

\begin{equation}
f_{\mathrm{P}}(_{v1}^{v2}) = \frac { \int_{v1}^{v2} T_\mathrm{bP}(v_{\rm LSR})
  \delta v_{\rm LSR} } {\int_{v1}^{v2} T_\mathrm{b}(v_{\rm LSR})\delta
  v_{\rm LSR} }
\label{EQ_f}
\end{equation}
for P = CNM, LNM, and WNM.

Phase fractions are in general velocity dependent but for simplicity we
drop this notation in the following. We will see later that phase
fractions depend significantly on environmental effects.

\subsection{Velocity dependencies}
\label{VelDep}

Velocity dependencies of derived single channel phase fractions are
shown in Fig. \ref{Fig_VelDist}. We display a comparison of the average
all-sky phase fractions for $ -8 < v_{\mathrm{LSR}} < 8 $ \kms\ (thick
lines) for the individual phases. Velocity dependencies are obvious, in
particular $f_{\mathrm{CNM}}$ and $f_{\mathrm{WNM}}$ show systematic
fluctuations. 

\begin{figure*}[!thbp] %%  $ |b| > 20\degr$.
   \centering
   \includegraphics[width=6cm]{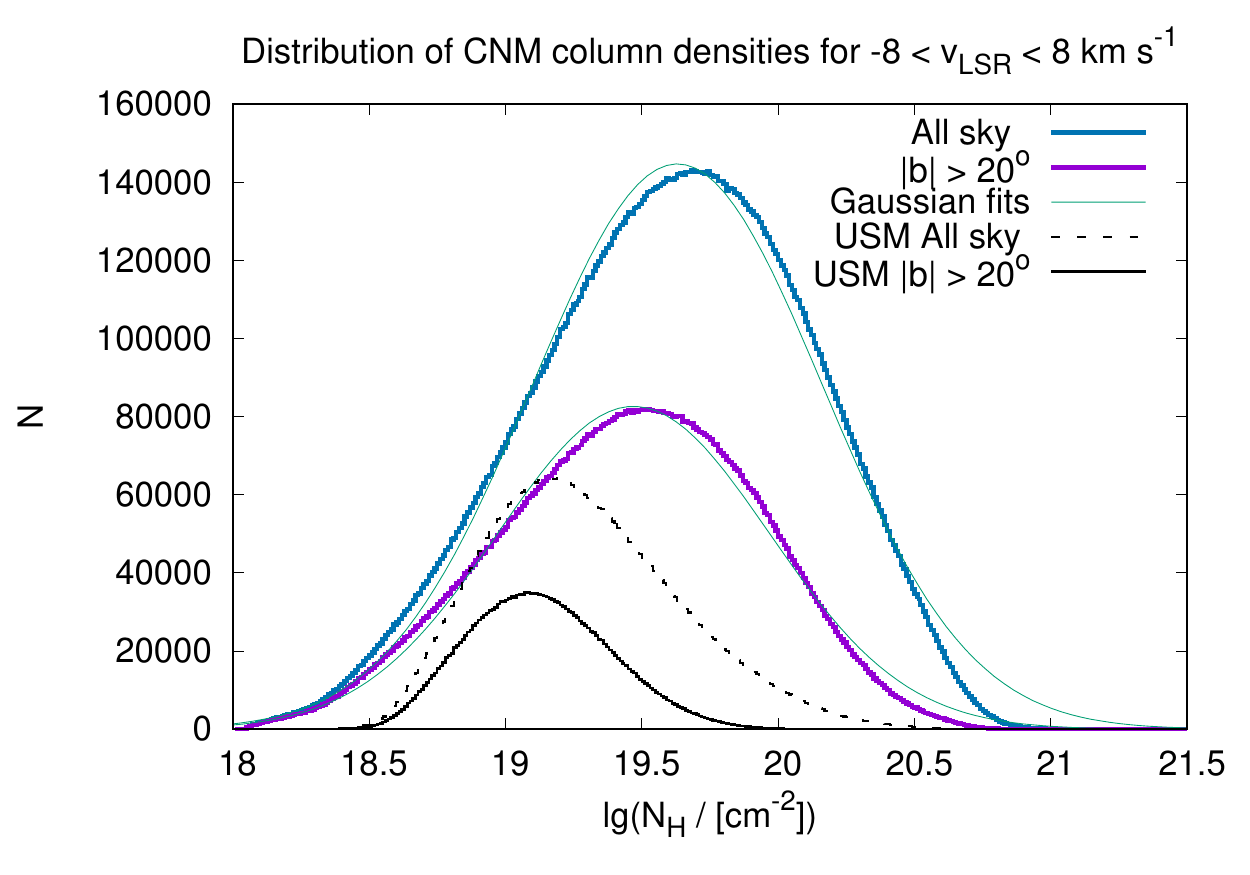}
   \includegraphics[width=6cm]{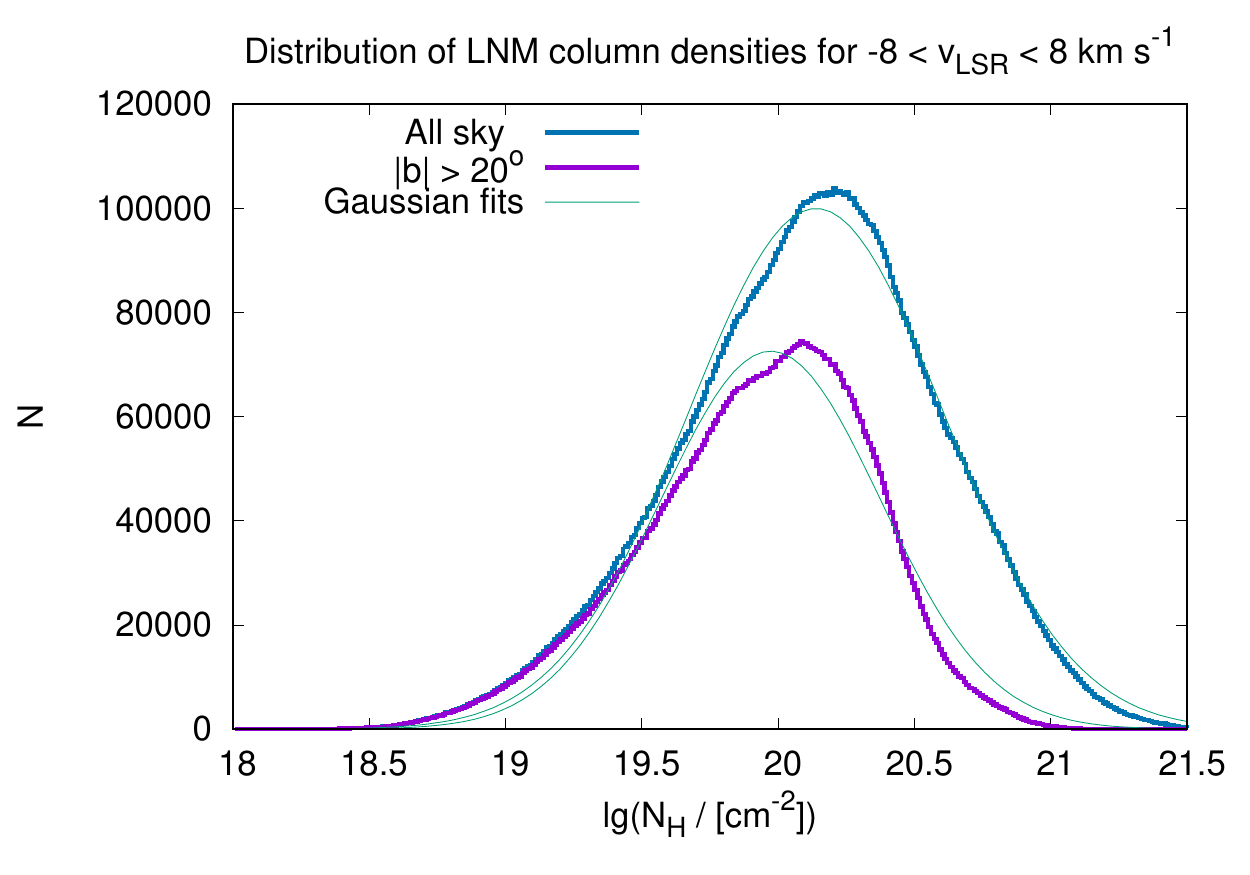}
   \includegraphics[width=6cm]{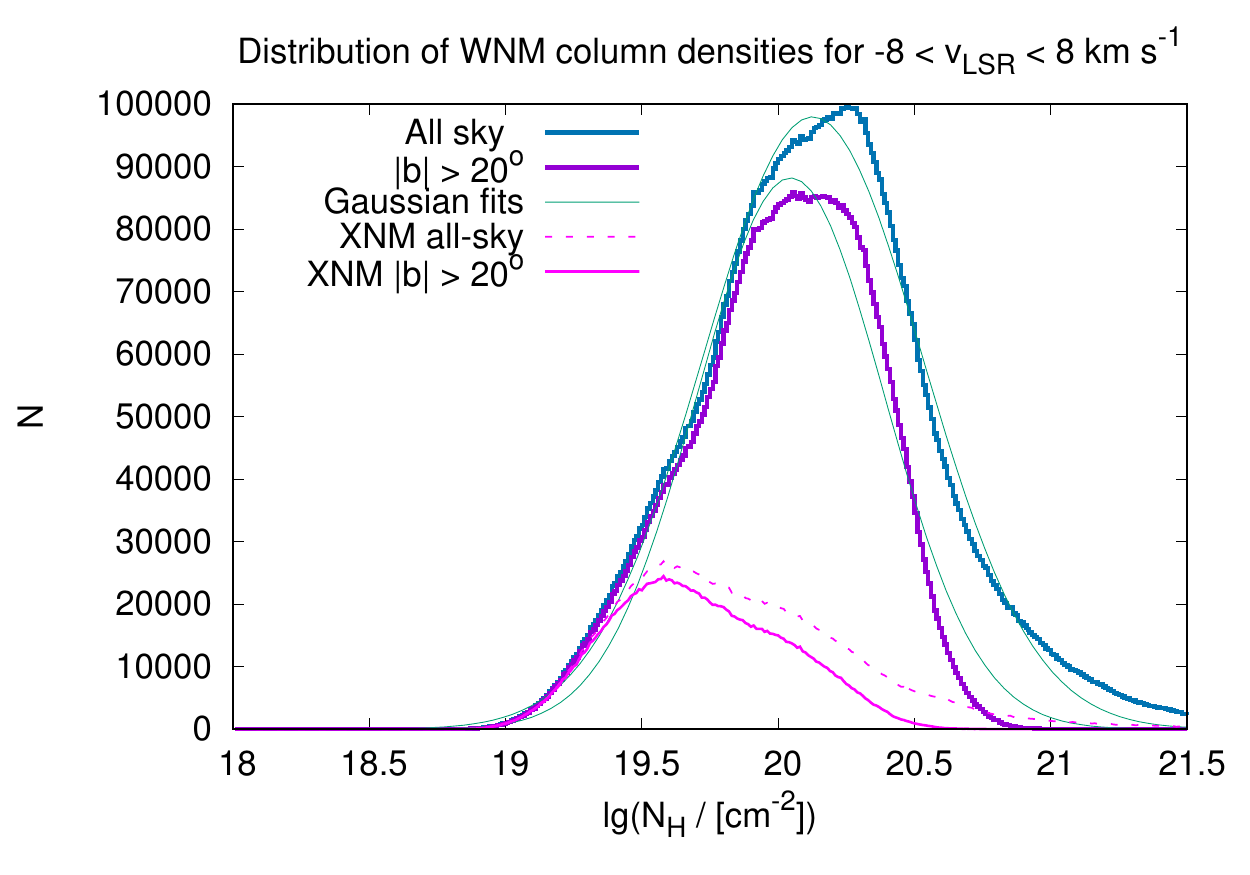}
   \caption{Distribution functions for column densities $
     {\mathrm{lg}}(N_{\mathrm{H}}/[{\mathrm{cm}}^{-2}]) $ of the CNM
     (left), LNM (middle), and WNM (right), derived from Gaussian
     components in the velocity range $ -8 < v_{\mathrm{LSR}} < 8 $
     \kms. The light and dark blue colors distinguishes distributions
     for all Sky and $ |b| > 20\degr$. The green lines show in each case
     least square Gaussian fits, resulting in log-normal
     distributions. For the CNM (left panel) we include the
     distributions of major fliamentary USM structures as observed by
     \citet[][their Fig. 12]{Kalberla2016}. On the right hand side the
     XNM components are also included for comparison.}
   \label{Fig_histo_gauss}
\end{figure*}

Figure \ref{Fig_VelDist} includes also estimates for the uncertainties
of the derived average phase fractions (thin lines). For all of the
phases we used median standard deviations for the thresholds between the
linewidth regimes (see Fig. \ref{Ave} ) to derive upper and lower limits
for the phase fractions. The uncertainties are significant, up to 50\%,
in particular for the LNM since this phase is affected by uncertainties
in the thresholds for both, the CNM and the WNM phase.

\subsection{All-sky maps for CNM, LNM, and WNM}
\label{Maps}

In this section we show all-sky distributions of separate phases to give
a first impression of morphological differences between the different
phases. We use a plate carr{\'e}e projection.\footnote[1]{ On request
  HEALPix nside = 1024 FITS files that can be used to generate other
  projections are available from the first author.}

In Fig. \ref{Fig_AllSky_1} we use a single velocity channel at $
v_{\mathrm{LSR}} = 0 $ \kms\ to show the separate components. At the top
we display the brightness temperature contributions for CNM (left), LNM
(middle), and WNM (right). At the bottom we show with the same sequence the
corresponding phase fractions $f_{\mathrm{CNM}}$, $f_{\mathrm{LNM}}$,
and $f_{\mathrm{WNM}}$ according to Eq. \ref{EQ_f}.

These all-sky maps display a surprising wealth of large and small scale
structures.  For the CNM and the LNM there are dominant filamentary
features but even the $f_{\mathrm{WNM}}$ map shows remarkable structures,
either with high or low $f_{\mathrm{WNM}}$ fractions. The WNM maps are
barely compatible with a diffuse medium, the usual description we
find in textbooks. In fact, in case of ongoing thermal instabilities we
expect some anti-correlations between CNM on one hand and LNM or WNM on
the other side. Such anti-correlations will be discussed in
Sects. \ref{CNMdominated} and \ref{PhaseRelations}.

%=========================================================================

\begin{figure*}[!htbp] %%  
   \centering
   \includegraphics[width=6cm]{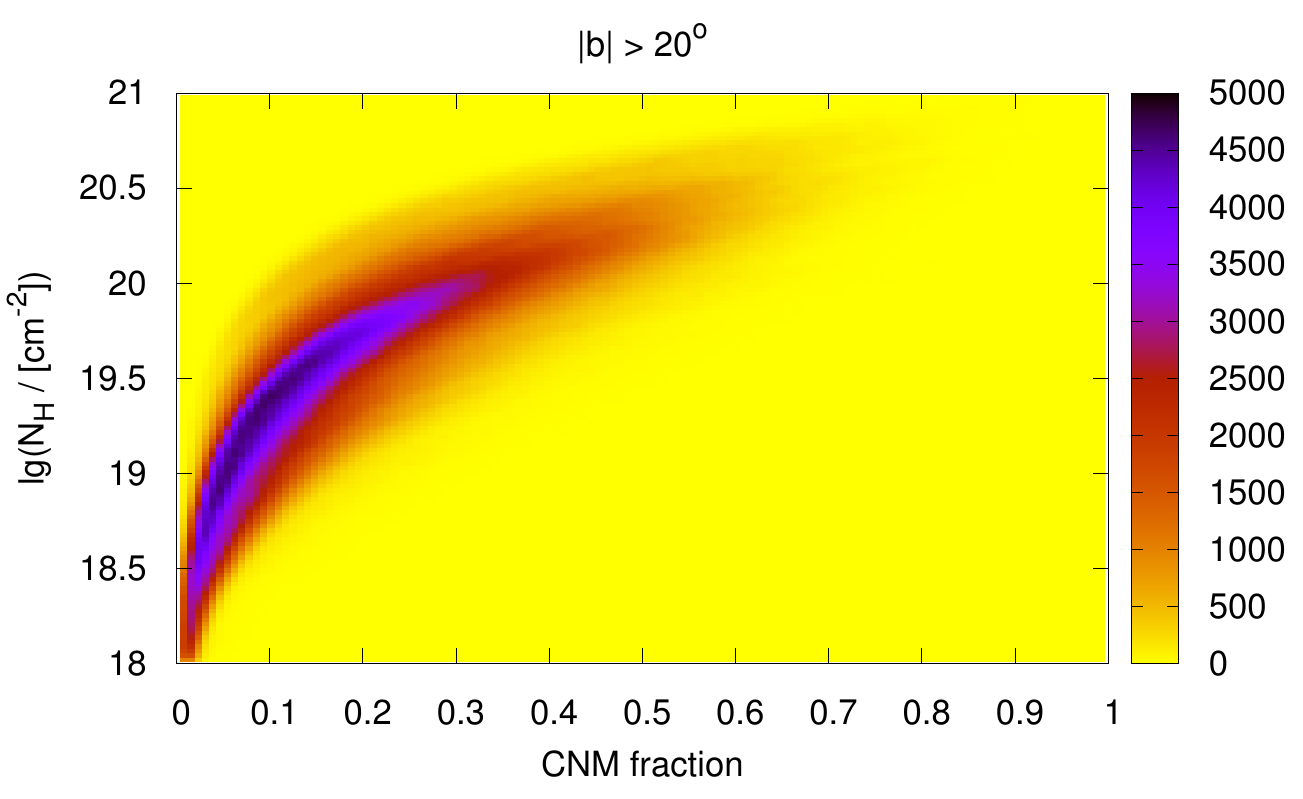}
   \includegraphics[width=6cm]{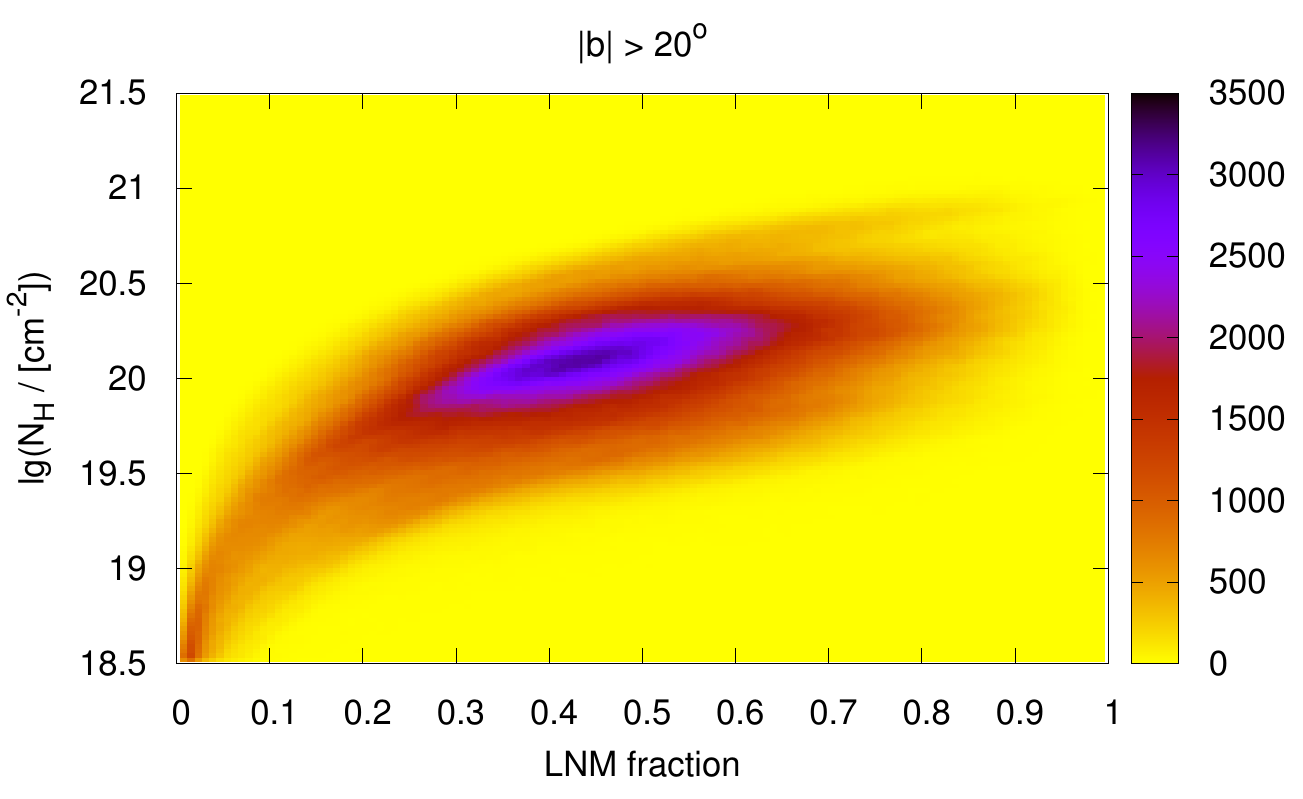}
   \includegraphics[width=6cm]{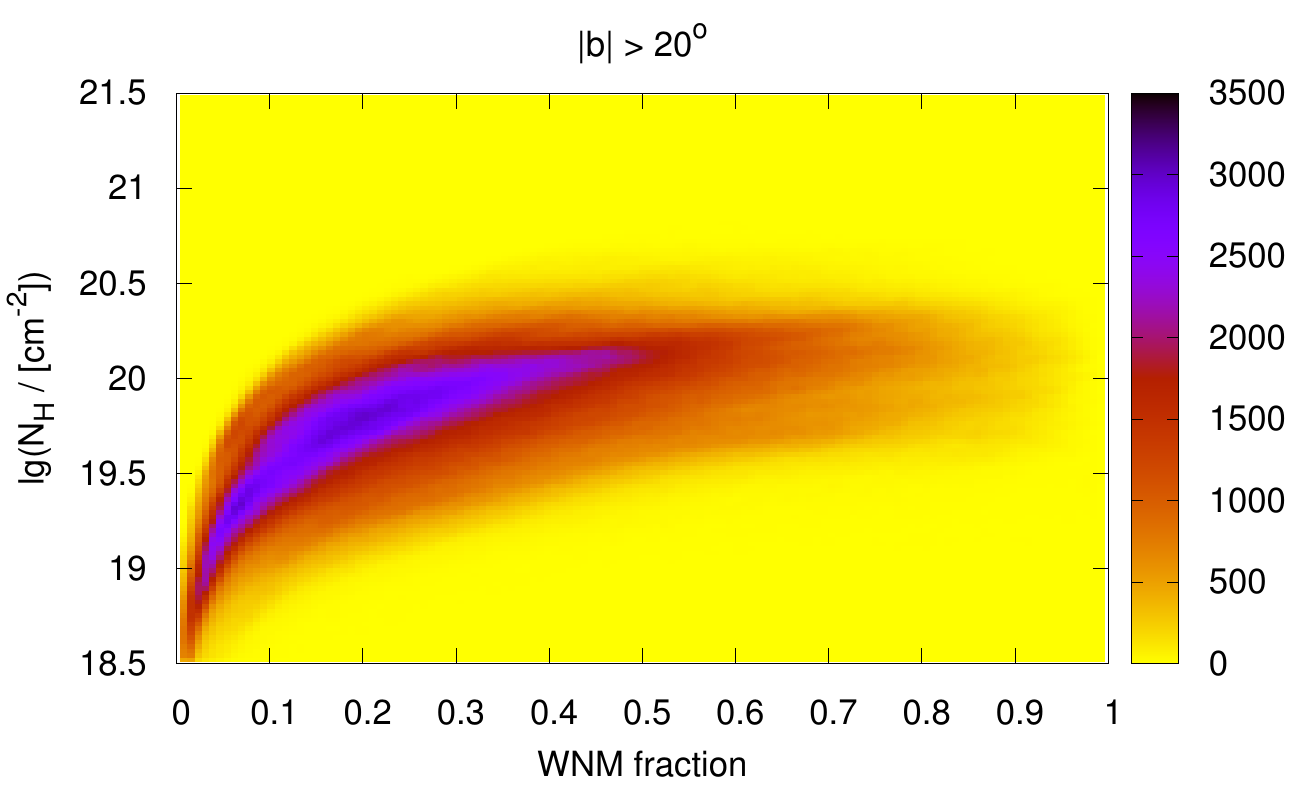}
   \includegraphics[width=6cm]{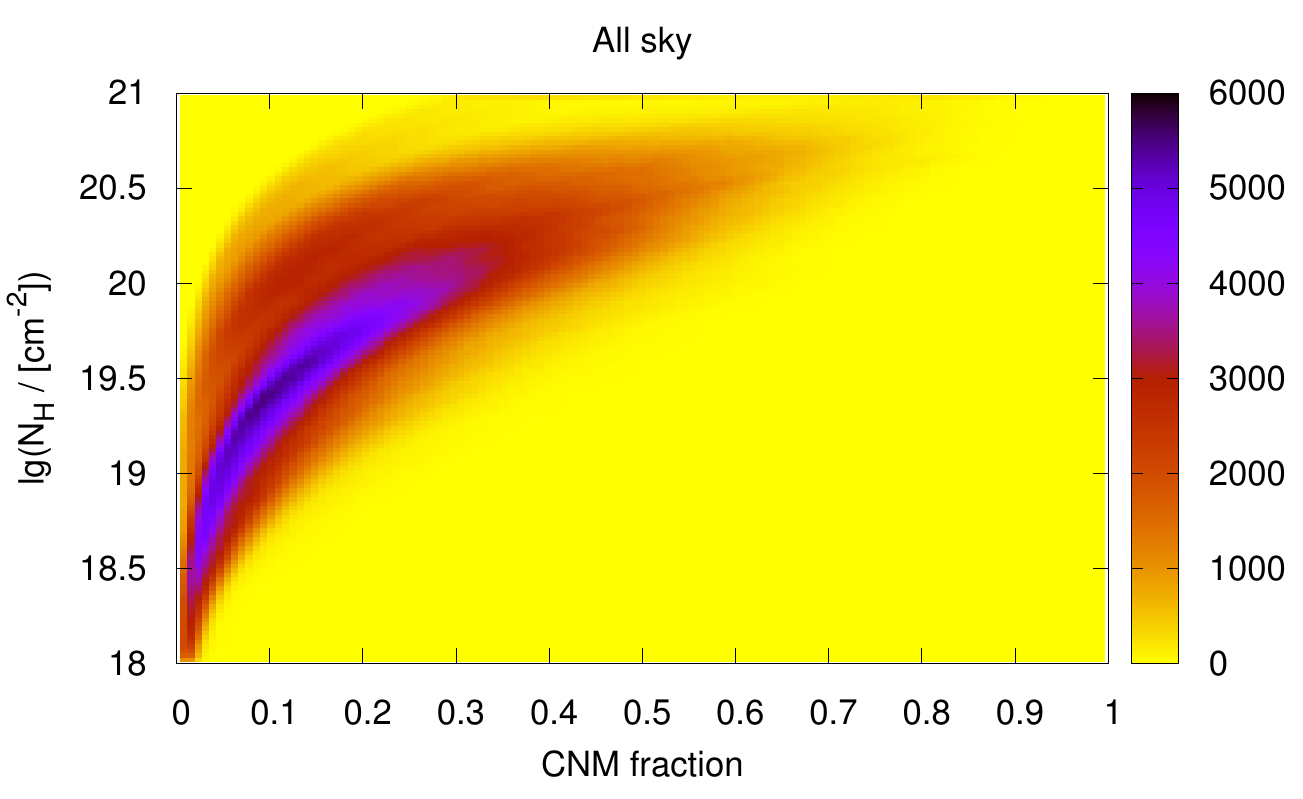}
   \includegraphics[width=6cm]{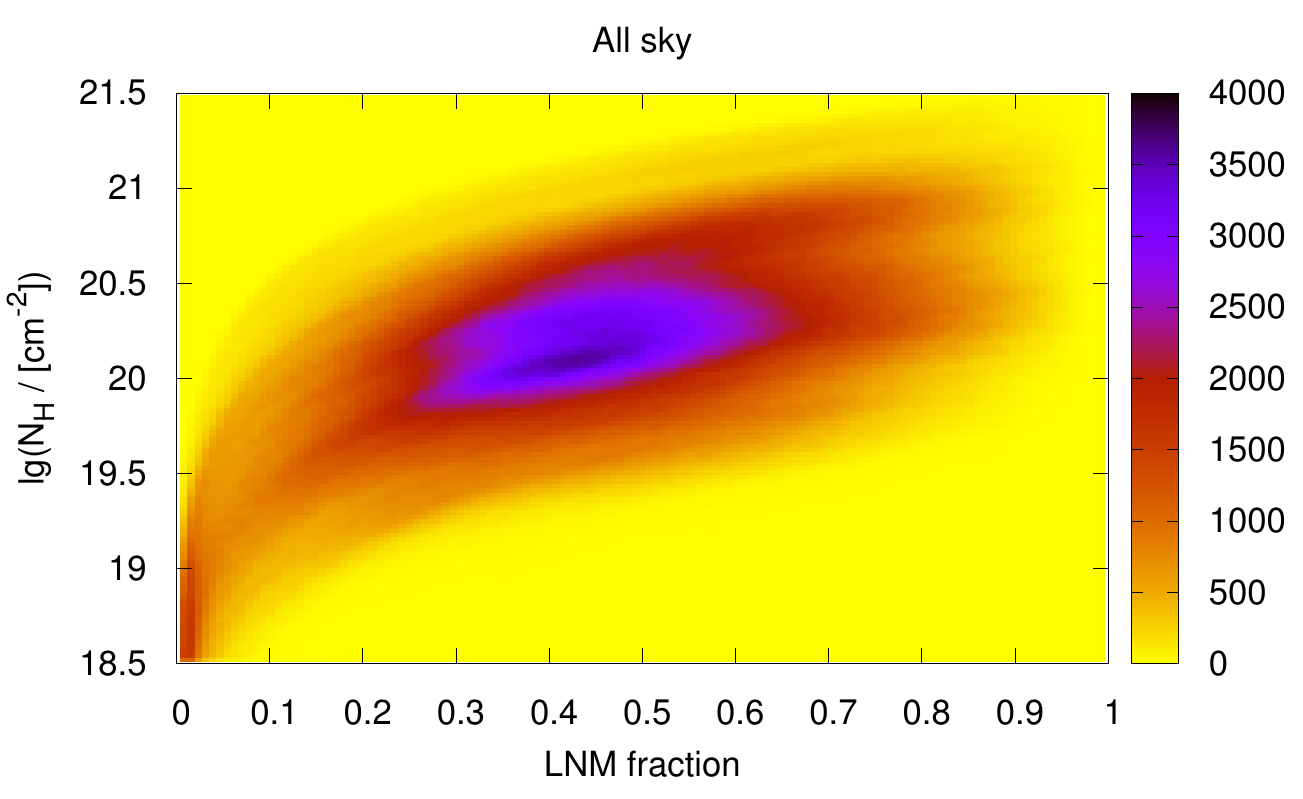}
   \includegraphics[width=6cm]{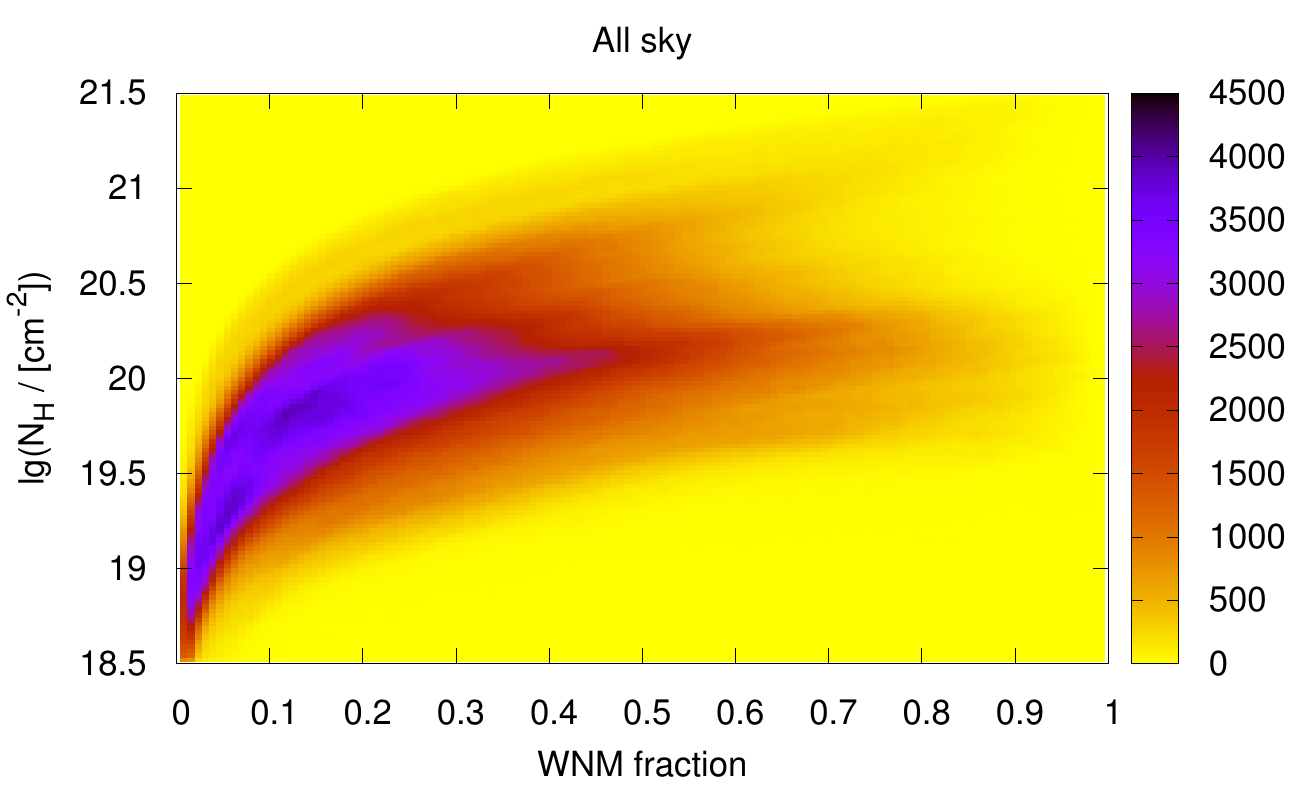}
     \caption{2D density distribution functions in the velocity range $ -8
     < v_{\mathrm{LSR}} < 8 $ \kms\ for CNM (left), LNM (middle), and
     WNM (right), showing the relations between phase fractions and
     column densities $
     {\mathrm{lg}}(N_{\mathrm{H}}/[{\mathrm{cm}}^{-2}])$ for each
     phase. Top: positions with $ |b| > 20\degr$, bottom: all sky data,
   }
   \label{Fig_fcorrelNH}
\end{figure*}

Figure \ref{Fig_AllSky_2} gives an alternative representation of the
all-sky column density distribution for the velocity range $ -8 <
v_{\mathrm{LSR}} < 8 $ \kms. With the same sequence of the phases as in
Fig. \ref{Fig_AllSky_1} we display at the top
${\mathrm{lg}}(N_{\mathrm{H}}/[{\mathrm{cm}}^{-2}])$. At the bottom we
show the corresponding phase fractions. Interestingly, phase fractions
in Figs. \ref{Fig_AllSky_1} and \ref{Fig_AllSky_2} (bottom) show in
general very similar structures. This result can be explained by the
velocity distribution of dominant local \hi\ structures
\citep[][Sect. 5.13]{Kalberla2016} that can be fit well with a Gaussian
of FWHM of 16.8 \kms, centered at $v_{\mathrm{LSR}} = 0 $ \kms. The
local \hi\ gas is dominated by low velocity structures in this range.

The top panels of Figs. \ref{Fig_AllSky_1} and \ref{Fig_AllSky_2} for
the WNM (right hand side) show enhancements of the \hi\ emission in the
Galactic plane with a $\mathrm{sin} (2l)$ modulation. A similar but less
obvious effect is visible for the LNM (middle panels). These
enhancements are caused by differential Galactic rotation
\citep{Mebold1972} and imply for the Galactic plane that some of the
observed gas must originate from large distances. At these positions our
basic assumption, that we consider only local \hi\ gas, is
violated. Surprisingly, when calculating phase fractions (lower panels
of Figs. \ref{Fig_AllSky_1} and \ref{Fig_AllSky_2}) these contamination
are less obvious. Apparently the contaminations cancel mostly when
calculating phase fractions. Anyhow, our caveat is that phase fractions
at low Galactic latitudes are in general less reliable, this includes
also larger uncertainties from the Gaussian analysis at low latitudes.
 
The separation of individual \hi\ phases in Sect. \ref{SepPhases} leads
to significant velocity dependencies of the $T_{\mathrm{D}}$ thresholds,
as shown in Fig. \ref{Ave}. These velocity dependencies are relatively
weak for the velocity range $ -8 < v_{\mathrm{LSR}} < 8 $ \kms~ but
increase significantly at higher velocities. Since this velocity range
covers most of the local \hi\ gas features, as presented in
Figs. \ref{Fig_AllSky_1} and \ref{Fig_AllSky_2}, we decided to restrict
our analysis to this velocity range.

%=========================================================================
%=========================================================================

%\clearpage
%\newpage

%=========================================================================

\subsection{Average column density distributions}
\label{ColDist}

The column density distributions for different phases can be derived
in two ways. Either by selecting Gaussian components according to the
velocity dependent thresholds derived in Sect. \ref{SepPhases} and
generating then maps for a FWHM beam-size of 30\arcmin. This results in
the spatial distributions displayed in Figs. \ref{Fig_AllSky_1} and
\ref{Fig_AllSky_2}.  Alternatively we may derive the column density
distributions directly from the Gaussian database, using appropriate
phases and center velocities. Figure \ref{Fig_histo_gauss} shows the
results for the velocity range $ -8 < v_{\mathrm{LSR}} < 8 $ \kms. We
display data for the CNM, LNM, and WNM (left to right) restricted to
latitudes $ |b| > 20\degr$ but also for all-sky.

Figure \ref{Fig_histo_gauss} shows in all cases column density
distributions that are close to log-normal, the best fit Gaussians are
included for comparison. For the CNM we plot also the distributions for
filamentary structures derived from unsharp masked (USM) maps
\citep[][their Fig. 12]{Kalberla2016}. Such features emphasize the
\hi\ distribution at high spatial frequencies. This gas is cold and
aligned with polarized dust emission \citep{Clark2014,Kalberla2016}. The USM
structures have low column densities and represent only a subset of the
CNM emission. These features have barely direct counterparts in the list
of Gaussian components.
In the right hand panel of Fig. \ref{Fig_histo_gauss} we show also the
\hi\ distribution for the XNM, the fourth phase from Figs. \ref{Td3} to 
\ref{Ave} that we consider as part of the WNM. 

%\clearpage
%\newpage

%=========================================================================

%=========================================================================
\begin{figure*}[thp] %%  
%  \vspace*{10mm}
  \centering
   \includegraphics[width=18cm]{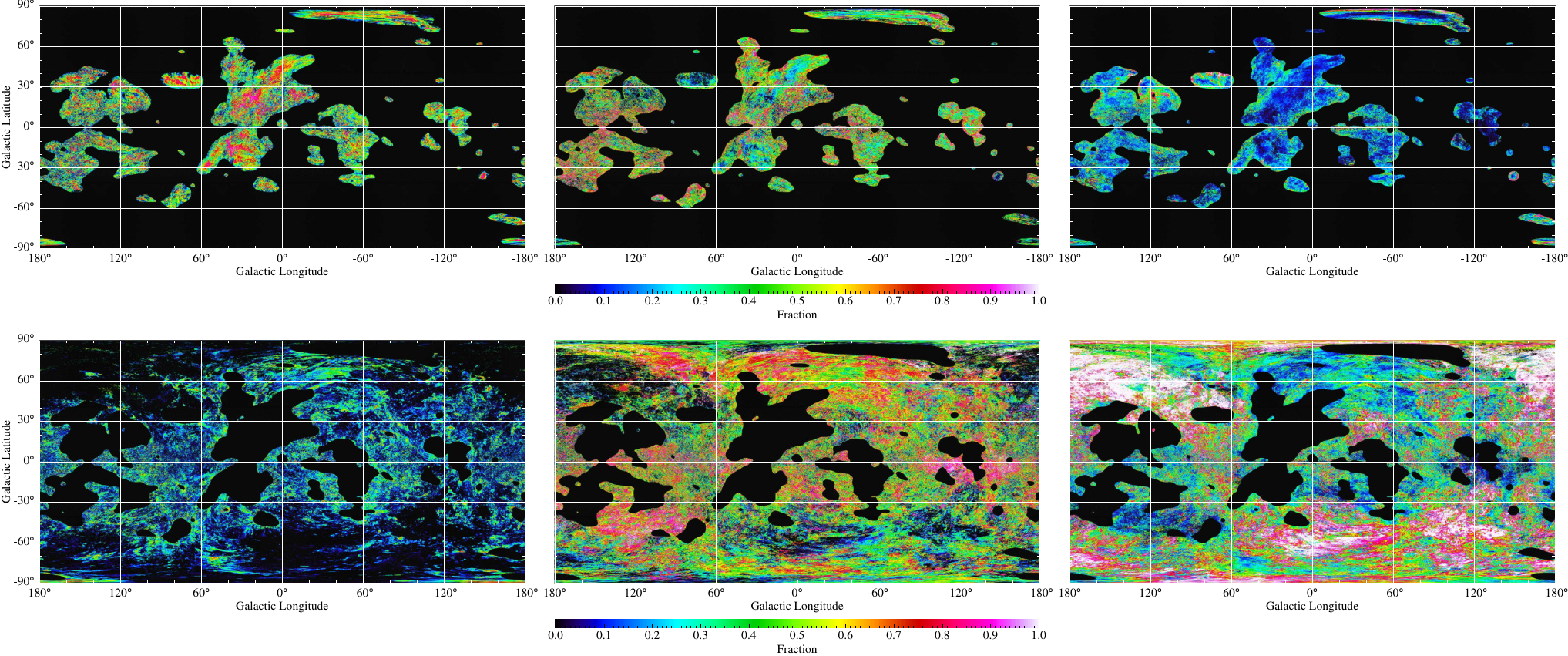}
   \caption{All Sky distribution functions for phase fractions in the
     velocity range $ -1 < v_{\mathrm{LSR}} < 1 $ \kms. Left: CNM,
     middle: LNM, and right: WNM. We distinguish CNM dominated regions,
     shown on top, and WNM dominated regions at bottom.  }
   \label{Fig_LB_CLW}
\end{figure*}

\subsection{Column densities depending on phase fractions}
\label{ColFract}

Here we derive dependencies between column densities and phase fractions
according to Eq. \ref{EQ_f}. We use the velocity range $ -8 <
v_{\mathrm{LSR}} < 8 $ \kms. Figure \ref{Fig_fcorrelNH} displays
separately for each of the gas phases 2D density distributions for
column densities and phase fractions. On the top the data are restricted
to high latitudes $ |b| > 20\degr$, below we plot all-sky data.

High latitude data show the cleanest relations between column densities
and phase fractions. The all-sky distributions are more complex due to
some additional structures originating from regions close to the
Galactic plane. However, in both cases the LNM stands out with a
distribution that differs significantly from the CNM and the WNM.  The
LNM exists mostly with phase fractions $0.3 \la f_{\mathrm{LNM}} \la
0.6$.

%\clearpage
%\newpage

Figure \ref{Fig_fcorrelNH} shows a clear dependency between column
density distributions of individual \hi\ phases and the corresponding
phase fractions. This may imply that phases with distinct different
temperatures are separated from each other. Figures \ref{Fig_AllSky_1}
and \ref{Fig_AllSky_2} show in fact some evidence for regions that are
dominated by one or the other phase. 

%\vspace*{10mm}

\subsection{Average phase fractions}
\label{AvPhase}

For the velocity range $ -8 < v_{\mathrm{LSR}} < 8 $ \kms\ we determine
all-sky average phase fractions $\overline{f_{\mathrm{CNM}}} = 0.24 \pm
.04$, $\overline{f_{\mathrm{LNM}}} = 0.44 \pm .02$, and
$\overline{f_{\mathrm{WNM}}} = 0.32 \pm .05$. Here the listed
uncertainties come only from the velocity dependent fluctuations of
phase fractions for $ -8 < v_{\mathrm{LSR}} < 8 $ \kms. Systematic
errors may be much larger, see Sect. \ref{SepPhases} and
Fig. \ref{Fig_VelDist}. Restricting our sample to high Galactic
latitudes $ |b| > 20\degr$ we obtain $\overline{f_{\mathrm{CNM}}} = 0.25
\pm .02$, $\overline{f_{\mathrm{LNM}}} = 0.41 \pm .02$, and
$\overline{f_{\mathrm{WNM}}} = 0.34 \pm .03$. As detailed in
  Sect. \ref{Maps} we consider the derived parameters close to the
  Galactic plane as uncertain. Still the all-sky results agree
within the errors with average phase fractions from high latitudes.

\begin{figure*}[thbp] %%  
   \centering
   \includegraphics[width=6cm]{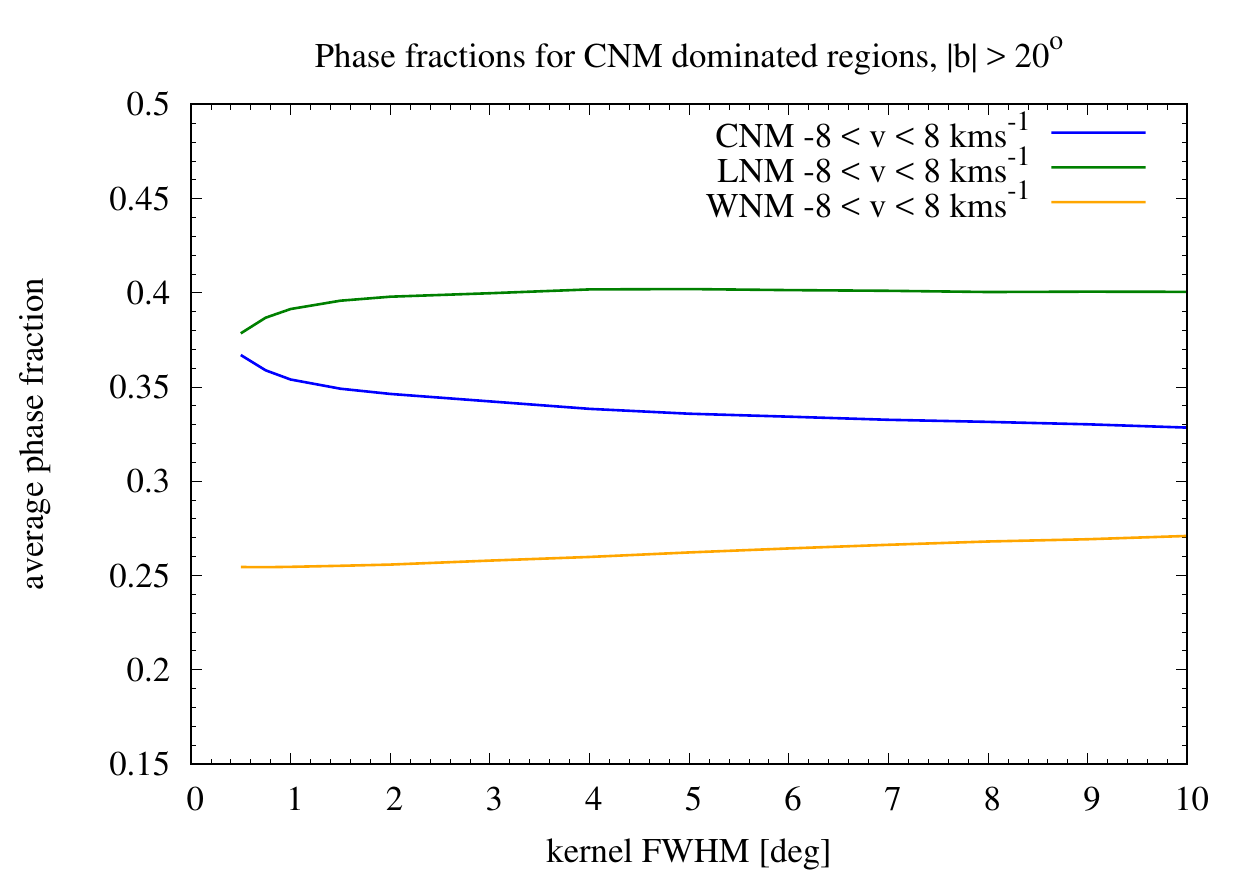}
   \includegraphics[width=6cm]{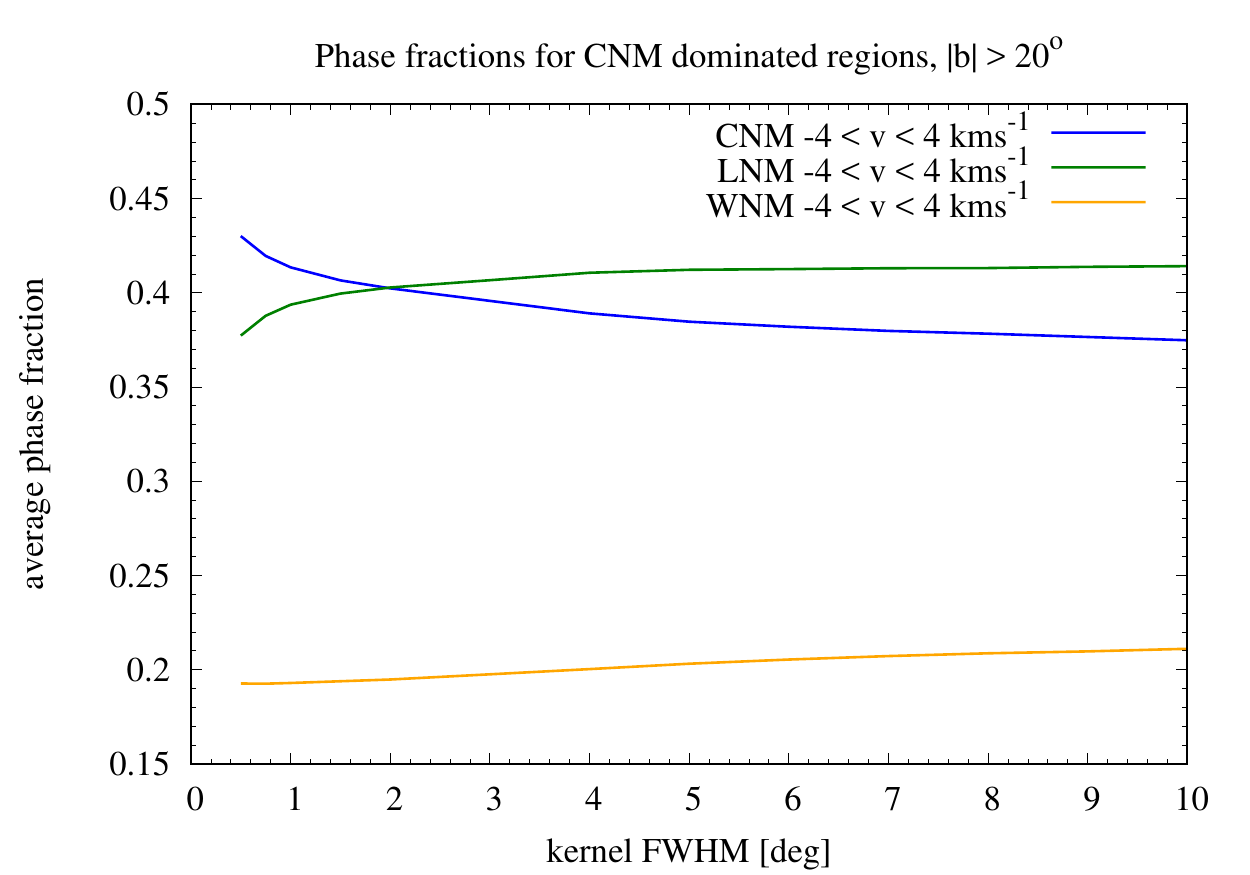}
   \includegraphics[width=6cm]{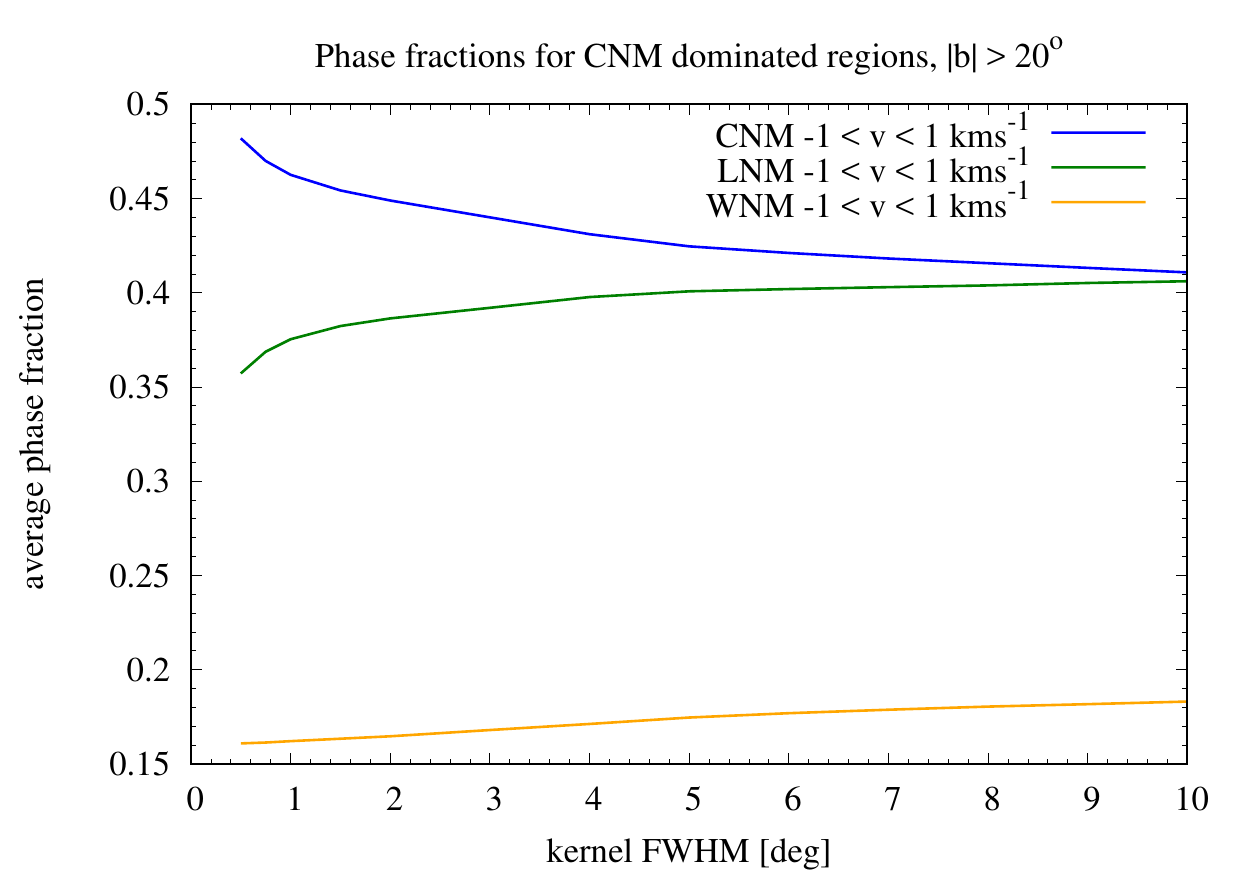}
   \includegraphics[width=6cm]{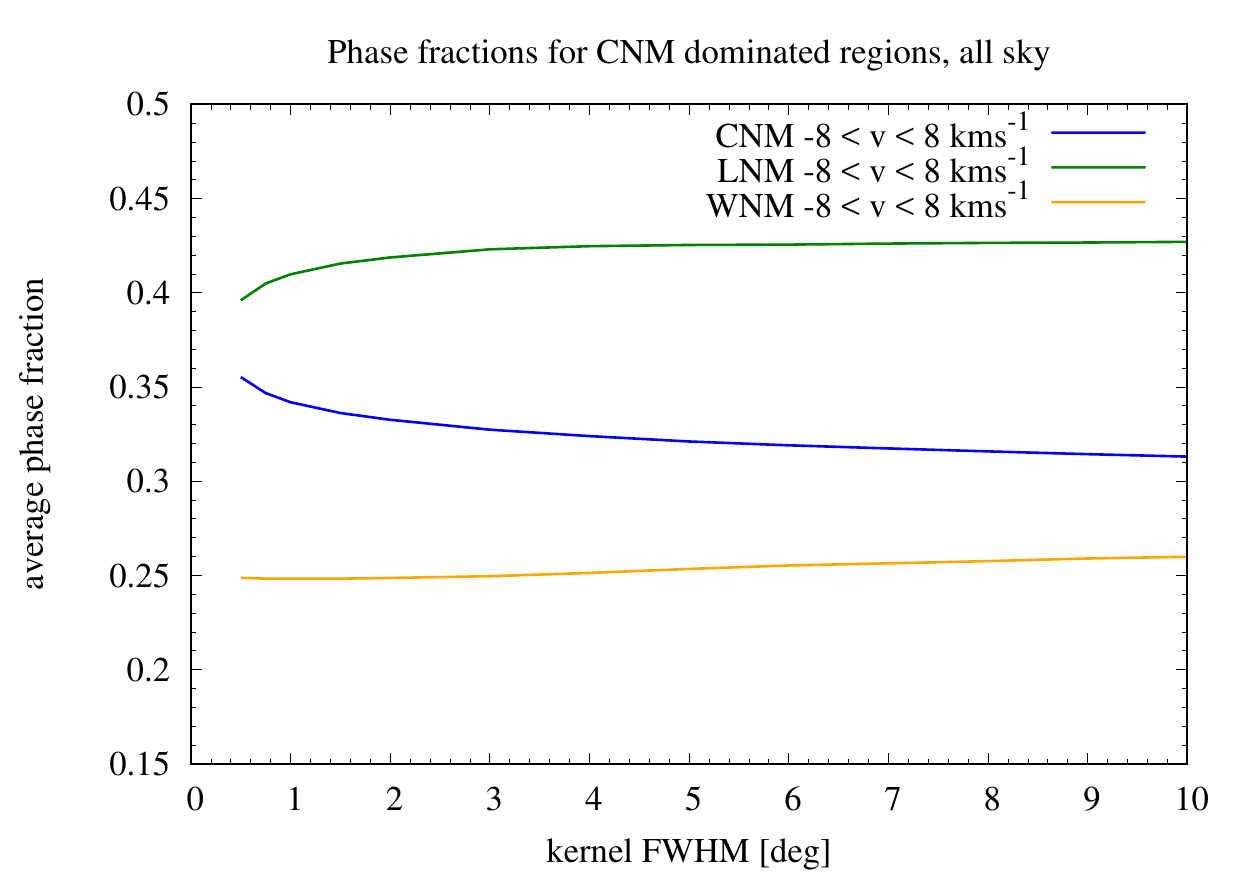}
   \includegraphics[width=6cm]{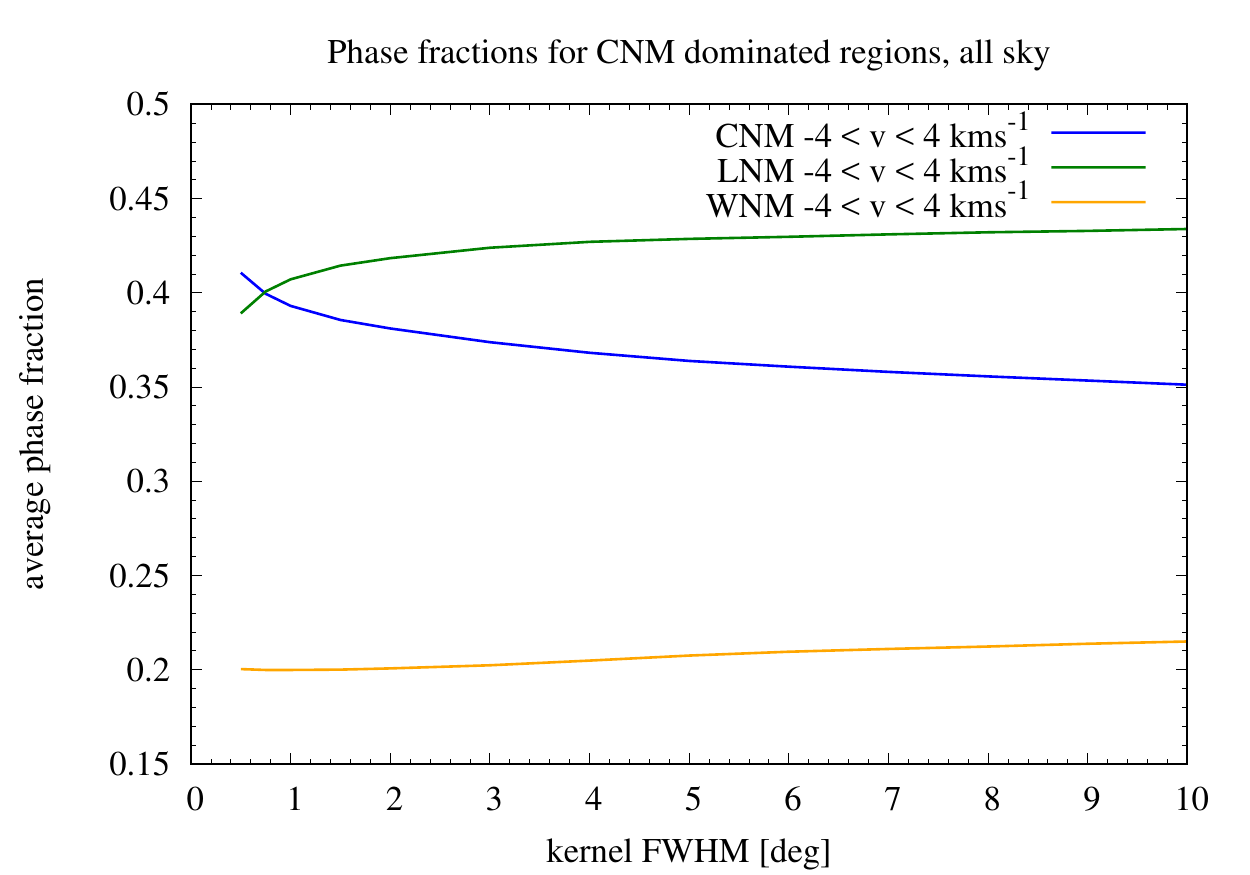}
   \includegraphics[width=6cm]{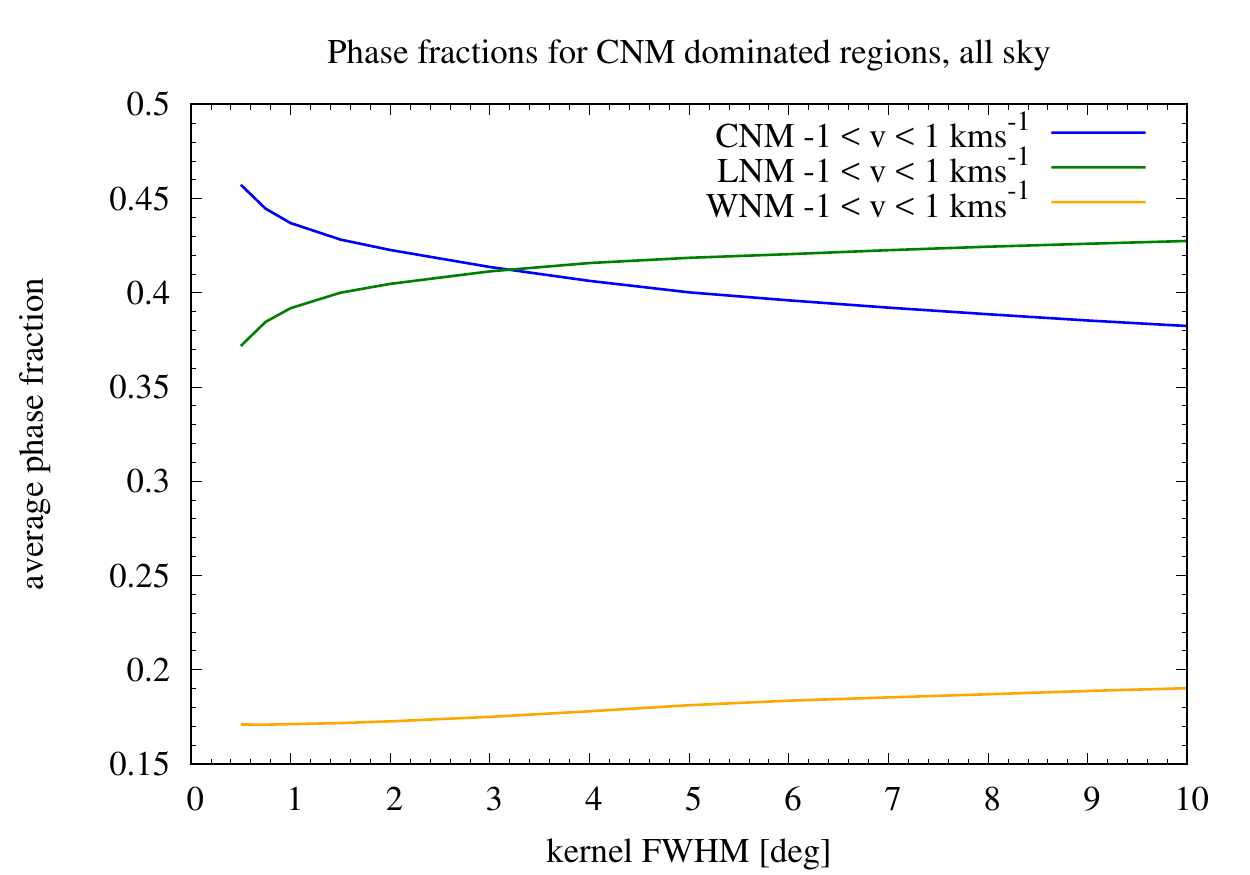}
   \caption{Phase fractions for CNM dominated regions depending on the
     width of the smoothing kernel used for the generation of the
     mask. Top: high latitudes $ |b| > 20\degr$ only, bottom: all sky
     data. From left to right phase fractions for $ | v_{\mathrm{LSR}} |
     < 8 $, $ | v_{\mathrm{LSR}} | < 4 $, and $ | v_{\mathrm{LSR}} | < 1
     $ \kms\ are shown. }
   \label{Fig_islands}
\end{figure*}

\section{CNM and WNM dominated regions}
\label{CNMdominated}

A comparison between Figs. \ref{Fig_AllSky_1} and \ref{Fig_AllSky_2}
shows that the all-sky distribution of the CNM phase is most prominent
at velocities close to zero. We use therefore the range $ -1 <
v_{\mathrm{LSR}} < 1 $ \kms\ to separate CNM dominated regions from the
rest of the \hi\ distribution by applying a mask. To define CNM
dominated regions we first select only positions with $f_{\mathrm{CNM}}
> 0.5$. This map of delta functions, representing cold spots, is
smoothed by a Gaussian function with FWHM = 5\degr; the resulting map is
truncated at 10\% of the peak.  Values above this threshold are set to
one and values below to zero respectively. Approximately 22\% of the
total sky is this way masked as CNM dominated.

The application of this mask allows a separation of CNM dominated
regions including a homogeneous area around the most prominent CNM
structures. Figure \ref{Fig_LB_CLW} shows in the upper panel CNM
dominated areas separated from the rest of the sky. The CNM structures
are surrounded by LNM but there is little WNM associated with the
CNM. Some borders of the CNM dominated regions appear to have enhanced
WNM as a transition to the regions displayed on the lower panels of
Fig. \ref{Fig_LB_CLW} that have been masked out from the CNM dominated
areas. Our masking procedure is subjective, we have tried to work out an
environment around the CNM with a smooth boundary at the transition
between cold (CNM dominated) and warm (WNM dominated)
\hi\ regions.

\subsection{Average phase fractions in CNM and WNM dominated
    regions }

Calculating average phase fractions for $ |b| > 20\degr$ and $ -8 <
v_{\mathrm{LSR}} < 8 $ \kms\ we come for CNM dominated regions to the
striking result that these regions are actually not dominated by the CNM
but by LNM: $\overline{f_{\mathrm{LNM}}} = 0.40 \pm .03$,
$\overline{f_{\mathrm{CNM}}} = 0.34 \pm .07$ and
$\overline{f_{\mathrm{WNM}}} = 0.26 \pm .08$. The situation is similar
for the rest of the sky, the WNM dominated part:
$\overline{f_{\mathrm{LNM}}} = 0.41 \pm .03$,
$\overline{f_{\mathrm{WNM}}} = 0.39 \pm .01$, and
$\overline{f_{\mathrm{CNM}}} = 0.20 \pm .03$. Thus we obtain the
unexpected result that in both cases the LNM is dominant. The average
LNM phase fraction $\overline{f_{\mathrm{LNM}}} = 0.41 \pm .02$,
determined in Sect. \ref{AvPhase}, appears to be characteristic for most
of the sky if we consider the velocity range $ -8 <
  v_{\mathrm{LSR}} < 8 $ \kms.

\subsection{Phase fractions around cold spots}
\label{coldspots}

To define the CNM dominated regions shown in Fig. \ref{Fig_LB_CLW} on
top we selected a smoothing kernel $K$ with a FWHM of 5\degr. For a more
general analysis, proposed by the referee, we extend here our
calculations for the range $ 0.5\degr \le K \le 10\degr$. In addition we
use several velocity ranges for the determination of average phase
fractions.

The top left panel of Fig. \ref{Fig_islands} shows a generalization of
the results from the previous subsection. $\overline{f_{\mathrm{CNM}}}$
decreases for increasing $K$. At the same time we find an increase of
$\overline{f_{\mathrm{LNM}}}$. This implies that the CNM is on average
localized; as we increase the area around the cold spots we observe on
average less CNM but more LNM. $\overline{f_{\mathrm{CNM}}}$ and
$\overline{f_{\mathrm{LNM}}}$ are anti-correlated; the LNM surrounds the
cold spots. In addition we find that also $\overline{f_{\mathrm{WNM}}}$
is gradually increasing with $K$. The response of the WNM is much
smoother than that of the LNM, implying than the WNM is distributed on
average over large scales. This confirms essentially the visual
impression we get from the right hand side of Fig. \ref{Fig_LB_CLW}.

The middle and right hand side panels on top of Fig. \ref{Fig_islands}
show even clearer correlations between CNM, LNM, and WNM for restricted
velocity windows $ -4 \le v_{\mathrm{LSR}} \le 4 $ \kms\ and $ -1 \le
v_{\mathrm{LSR}} \le 1 $ \kms. $\overline{f_{\mathrm{CNM}}}$ increases
with decreasing velocity width but $\overline{f_{\mathrm{LNM}}}$ remains
almost unaffected. The implication is that the CNM in cold spots is not
only spatially localized but also restricted in velocity space. The LNM
is spatially extended, surrounding the CNM, and covers in addition a
larger velocity spread. The WNM has a line-width of typically 24 \kms,
$\overline{f_{\mathrm{WNM}}}$ contributes correspondingly less with
decreasing velocity width. This effect apparently tends to balance the
LNM phase fraction, we observe $\overline{f_{\mathrm{LNM}}} \sim 0.4$
for different choices of the averaging line-width.
In the lower panels of Fig. \ref{Fig_islands} we present our results for
an all-sky analysis, confirming essentially the conclusions from high
latitudes.

\begin{figure*}[thbp] %%  
   \centering
   \includegraphics[width=6cm]{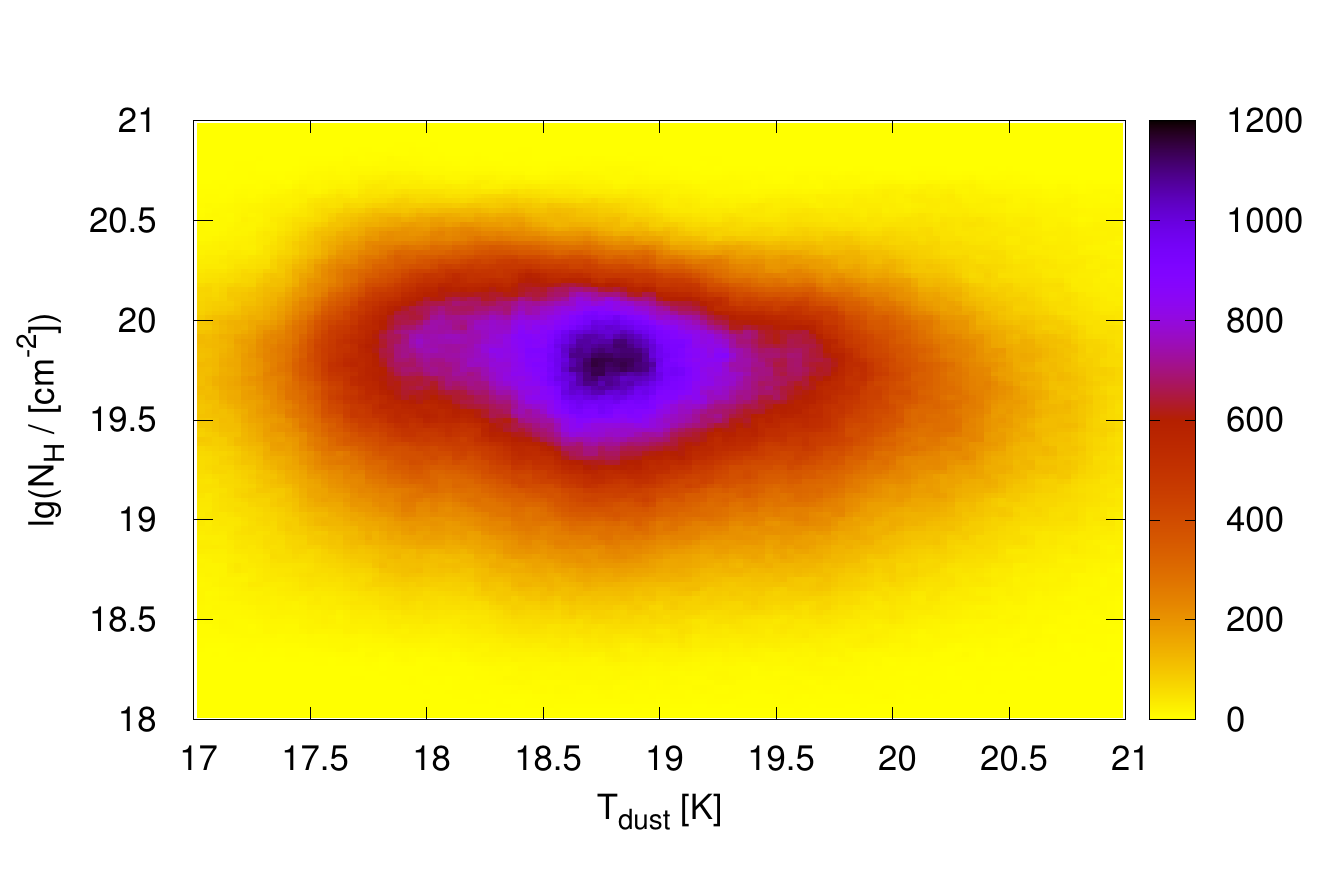}
   \includegraphics[width=6cm]{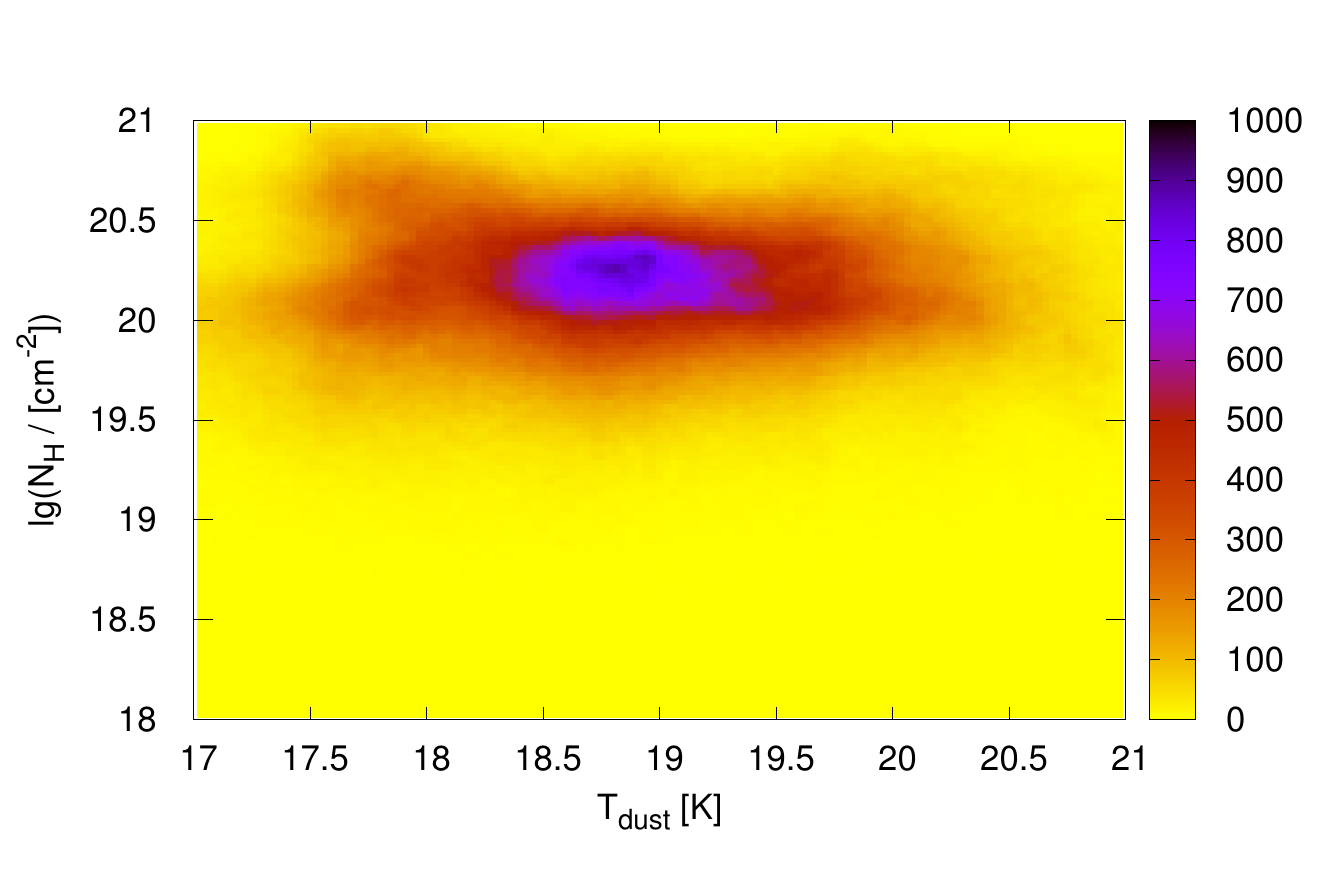}
   \includegraphics[width=6cm]{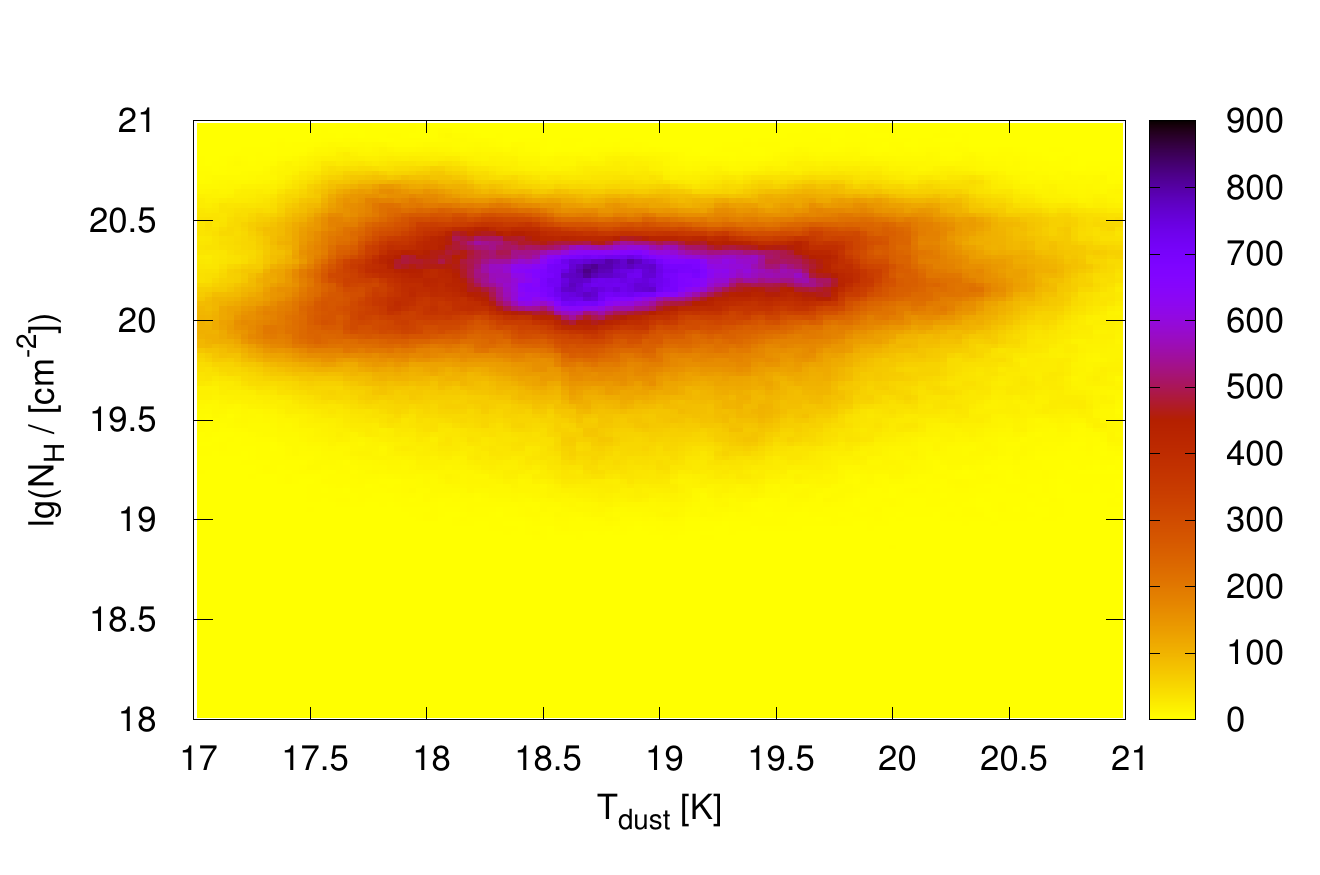}
   \includegraphics[width=6cm]{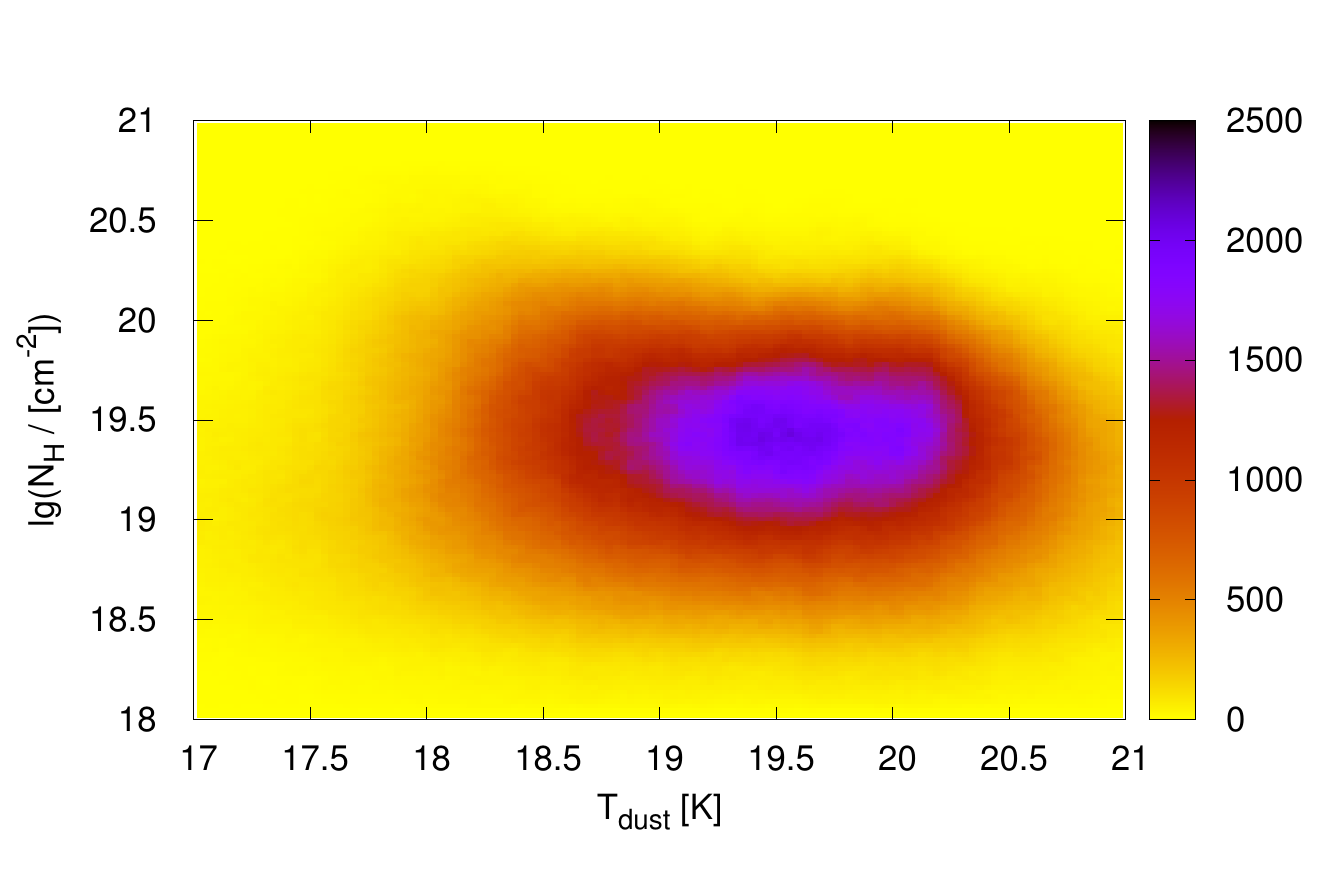}
   \includegraphics[width=6cm]{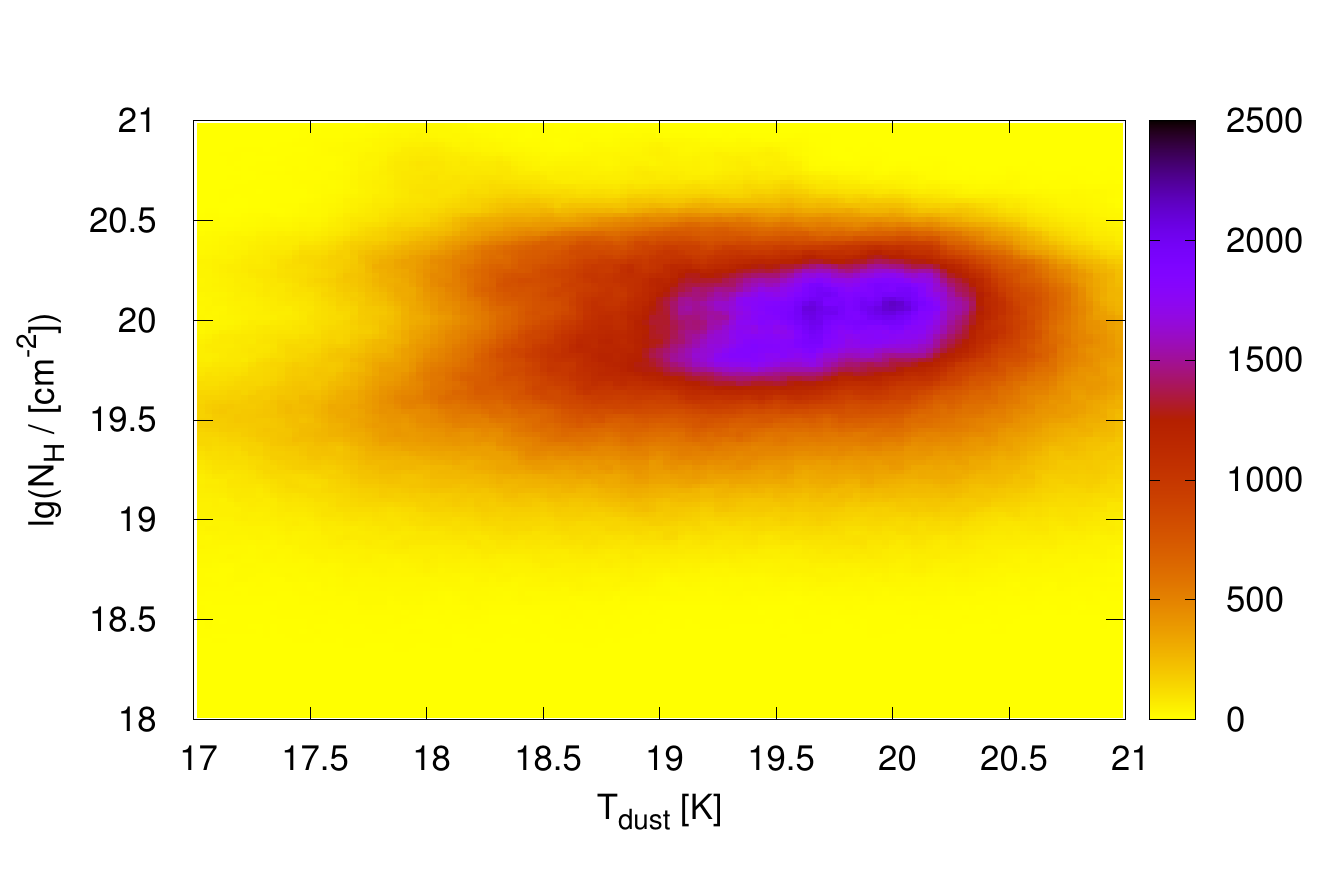}
   \includegraphics[width=6cm]{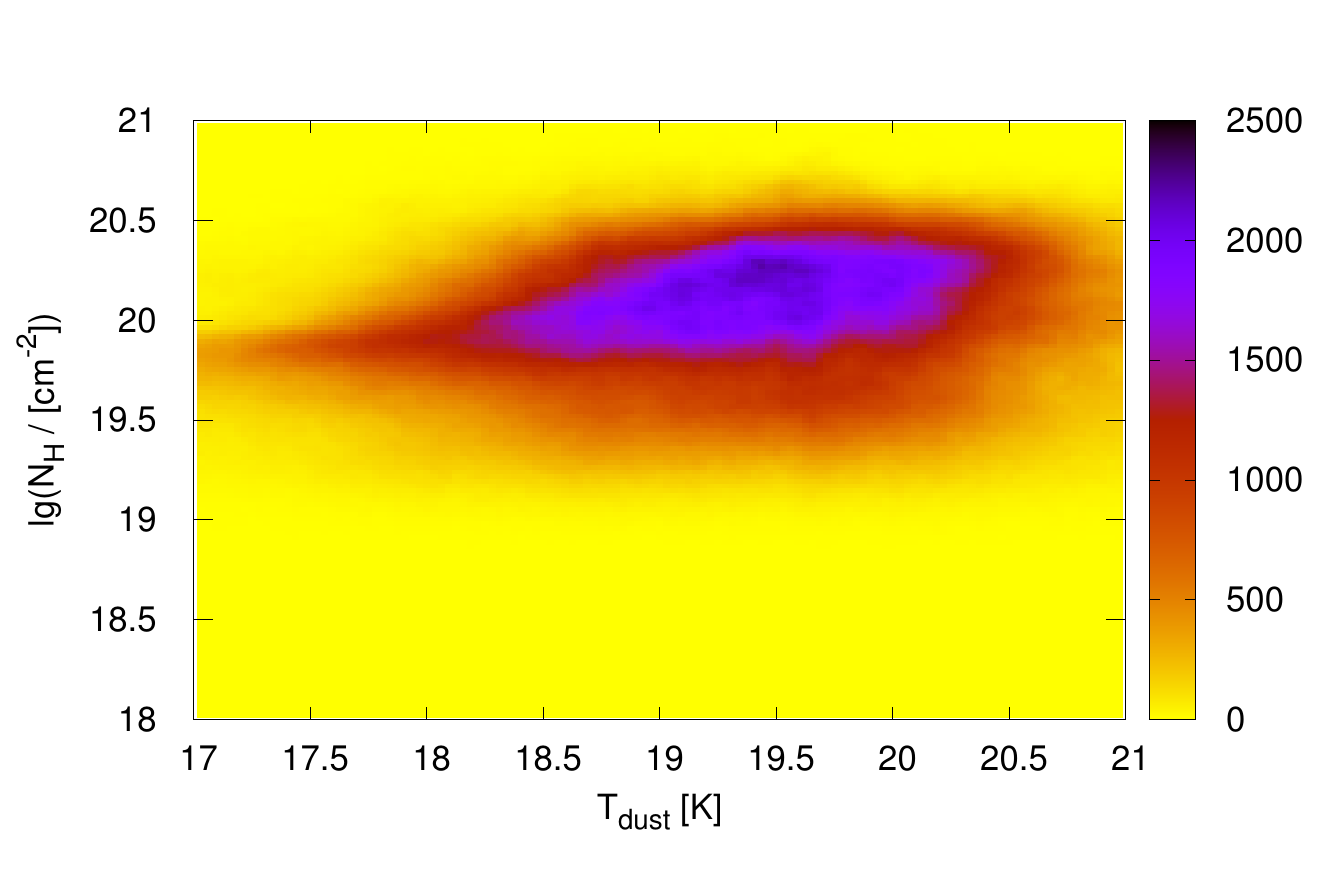}
   \caption{2D density distribution functions showing the relations
     between dust temperatures and column densities for Gaussian
     components in the velocity range $ -8 < v_{\mathrm{LSR}} < 8 $ \kms\
     at latitudes $ |b| > 20\degr$. Components from CNM dominated
     regions are shown on top, the remaining sample at bottom. Left to
     right: CNM, LNM, and WNM components.}
   \label{Fig_dust_CLW2}
\end{figure*}

\subsection{Dust temperatures associated with \hi\ phases}
\label{PhaseDust}

The all-sky maps displayed in Figs. \ref{Fig_AllSky_1},
\ref{Fig_AllSky_2}, and \ref{Fig_LB_CLW} show for the CNM some common
structures with filamentary \hi\ gas and the {\it Planck} HFI Sky Map at
353 GHz \citep[see][their Fig. 2]{Kalberla2016}; the CNM is associated
with polarized dust emission. To find out whether our masking for
cold and warm \hi\ gas results also in regions with physically
distinct different dust temperatures we intend to correlate now the
\hi\ distribution with the dust temperature distribution.

Currently the most reliable dust temperatures are available from
\citet{PlanckXLVIII}.  These authors used a tailored
component-separation method, the so-called generalized needlet internal
linear combination (GNILC) method, which uses spatial information
(angular power spectra) to disentangle the Galactic dust emission and
 anisotropies in the cosmic infrared background.

\begin{figure*}[th] %%  
   \centering
   \includegraphics[width=6cm]{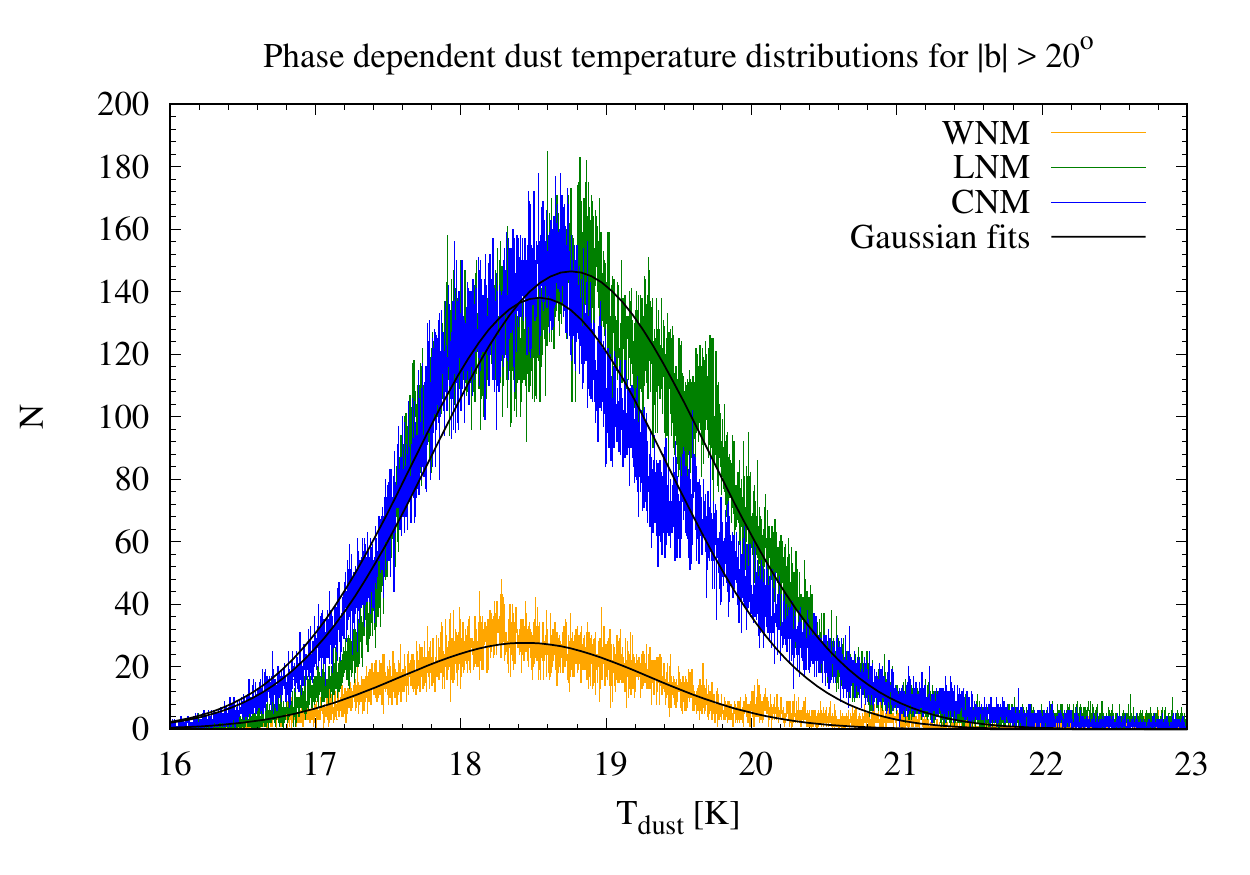}
   \includegraphics[width=6cm]{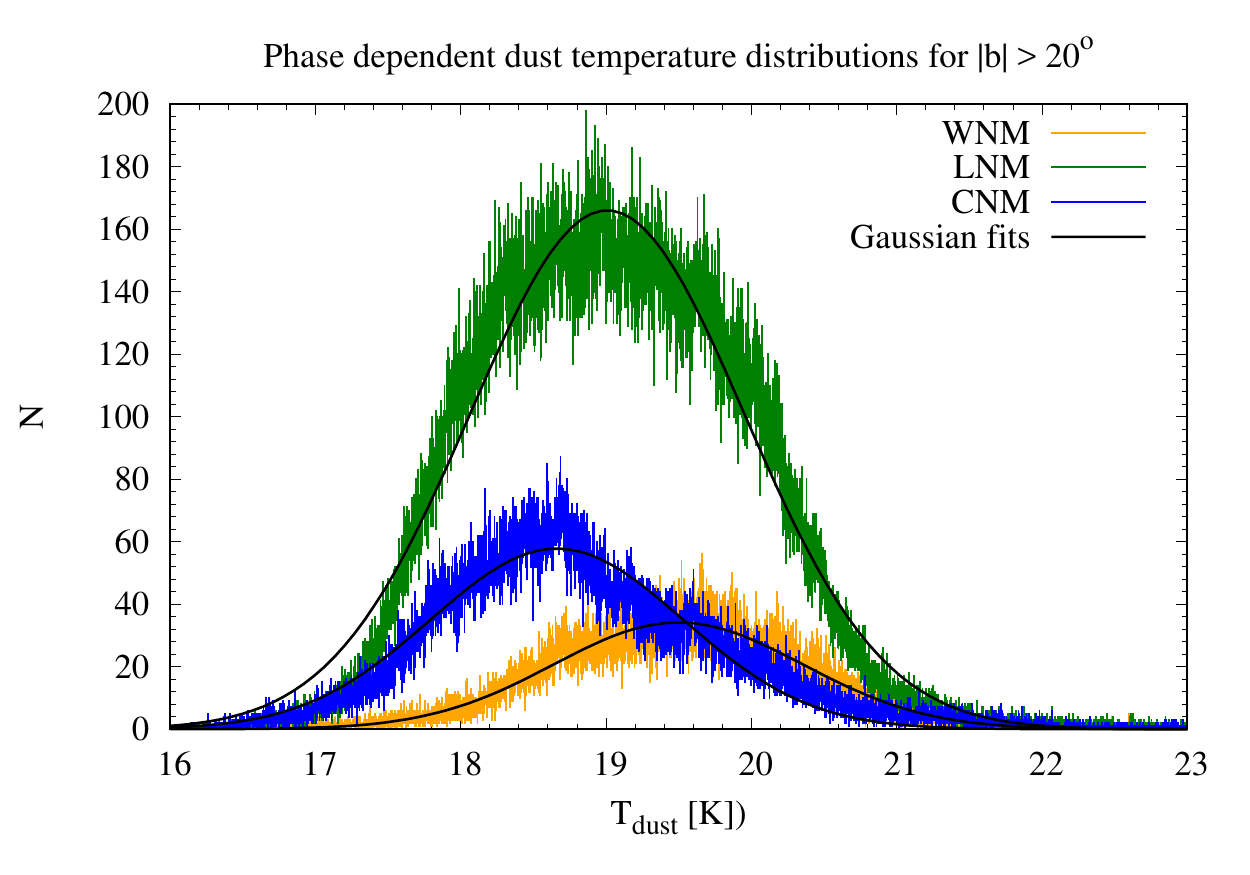}
   \includegraphics[width=6cm]{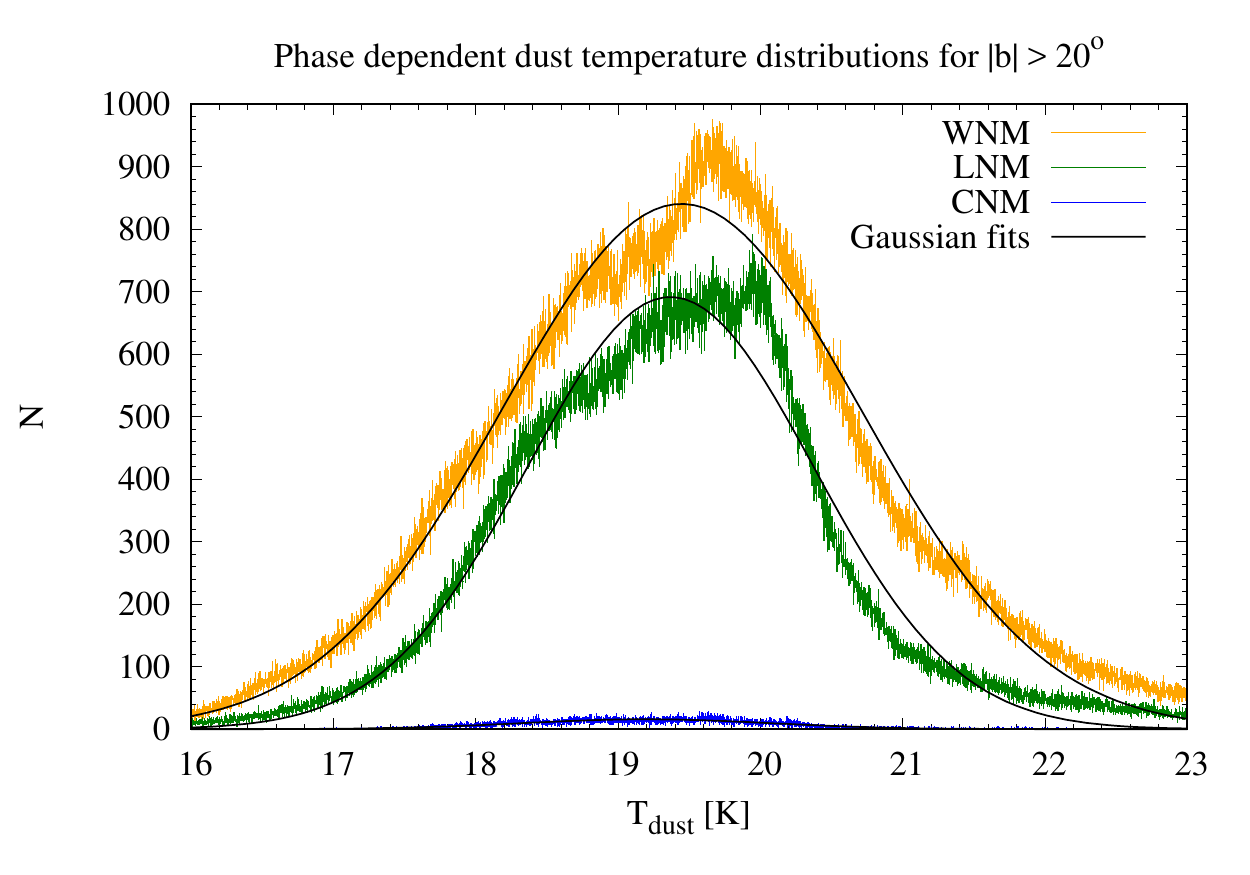}
   \includegraphics[width=6cm]{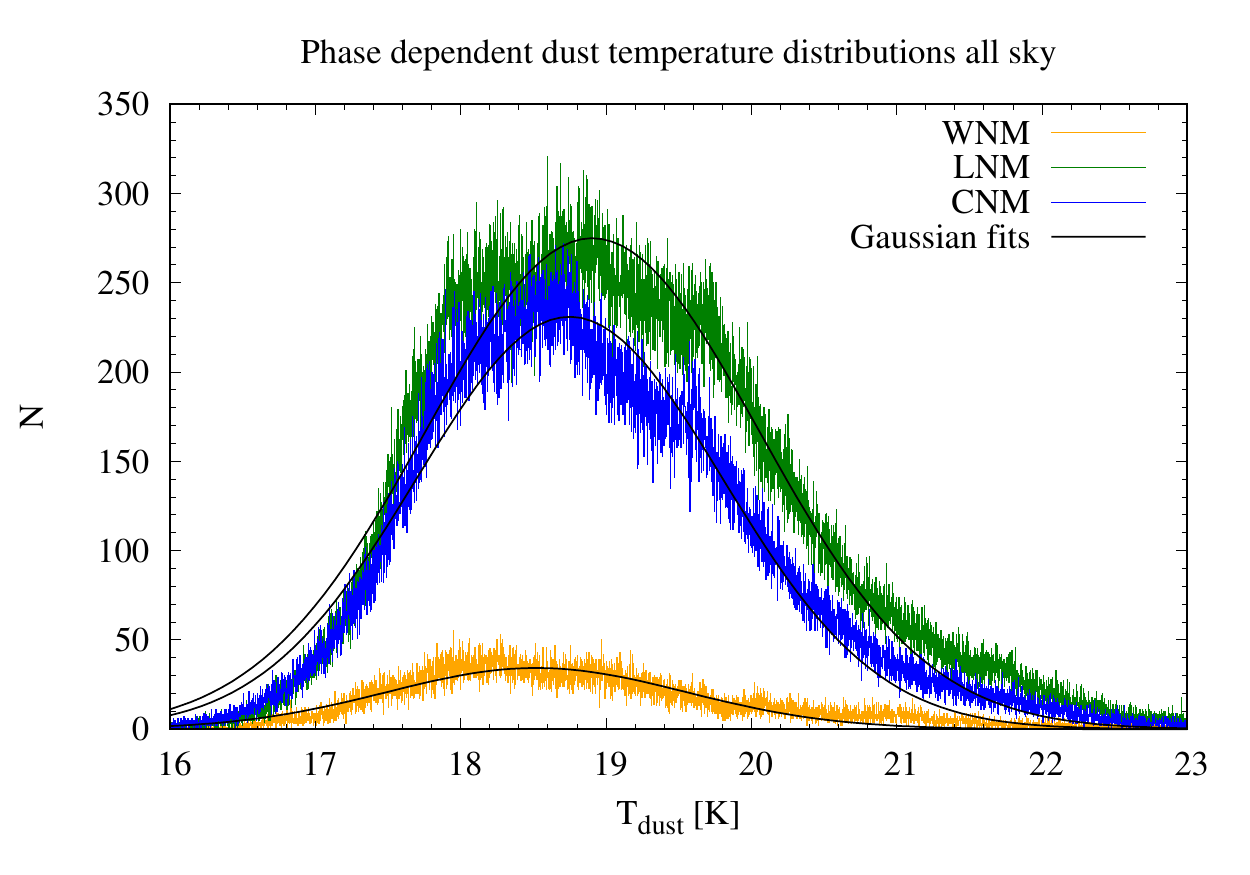}
   \includegraphics[width=6cm]{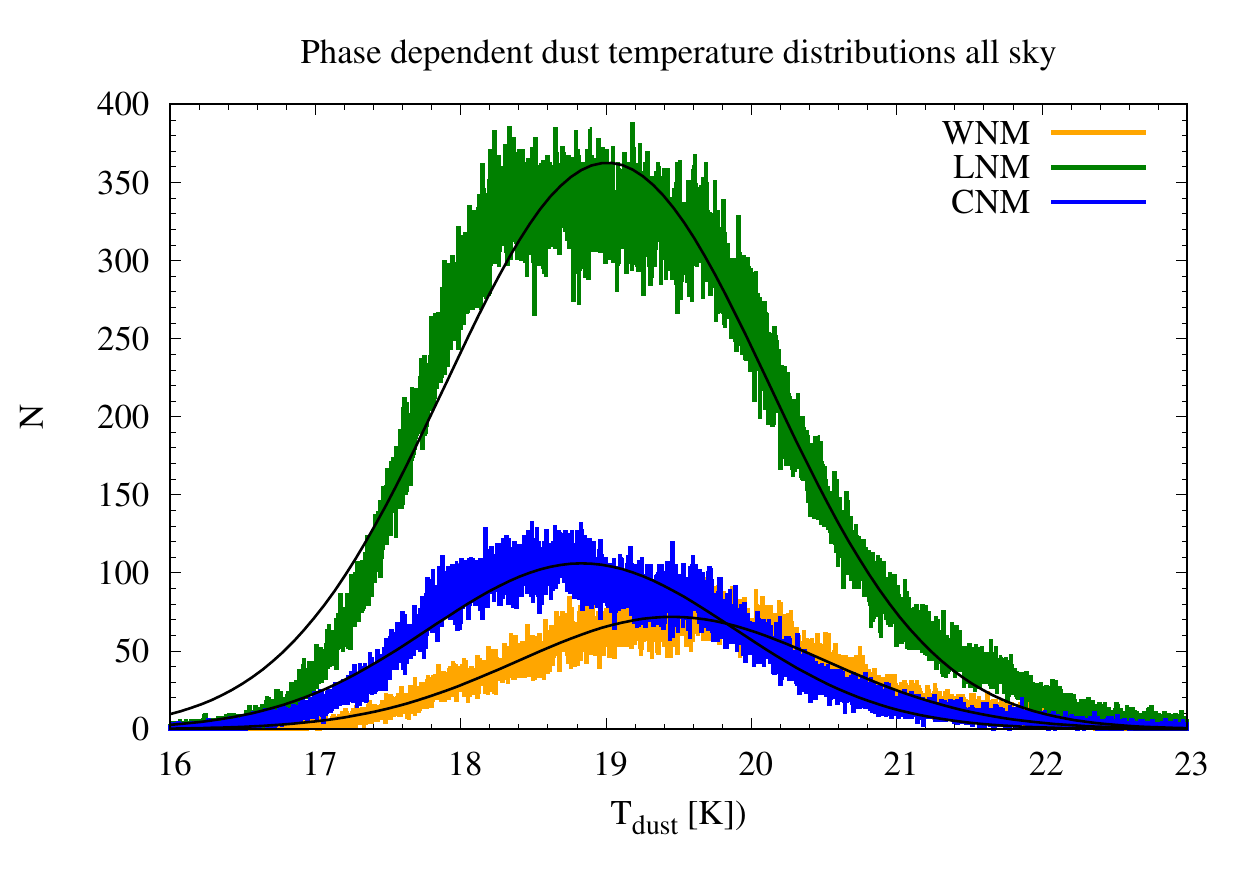}
   \includegraphics[width=6cm]{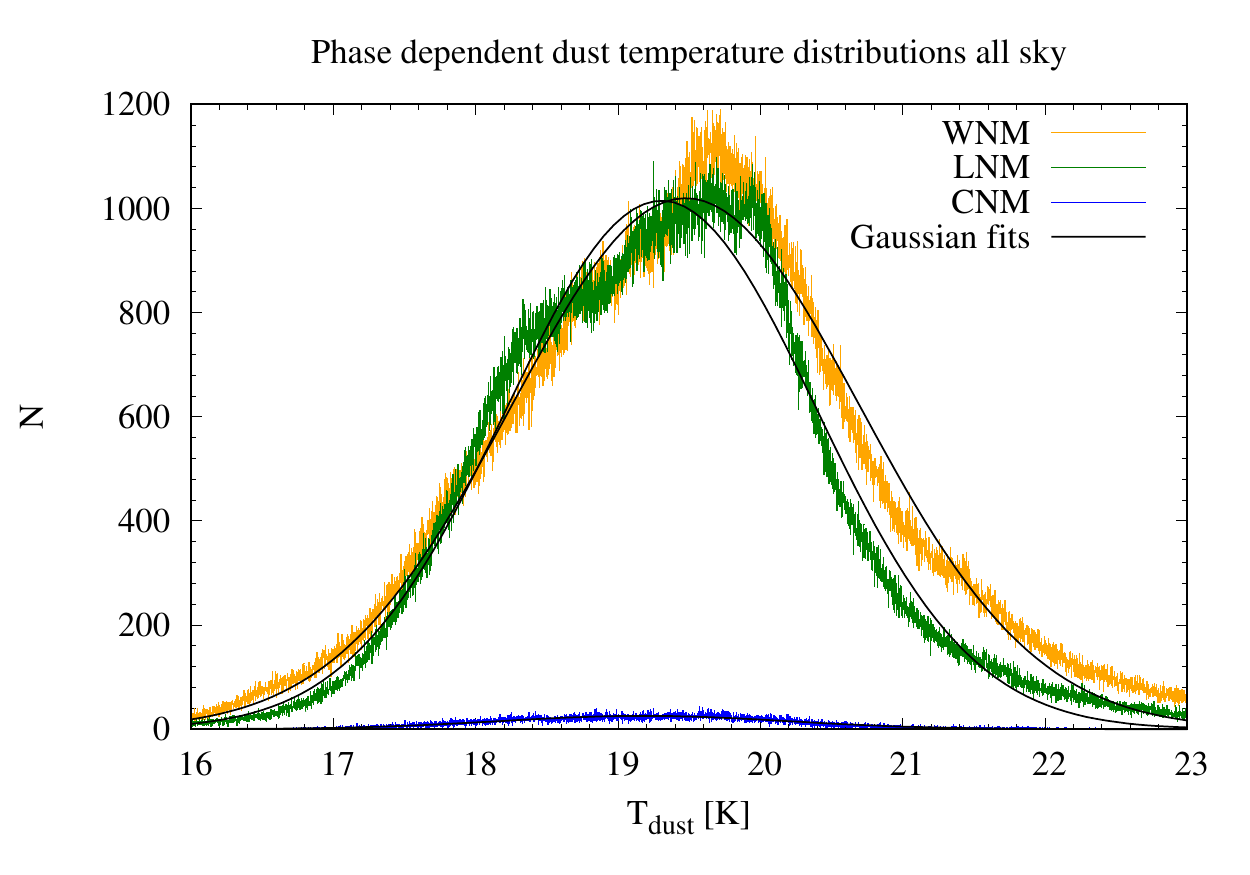}
   \caption{Frequency distribution for dust temperatures in selected
     regions on the sky. From left to right: CNM dominated,
     unconstrained, and WNM dominated according to
     Fig. \ref{Fig_LB_CLW}. Top: data at $ |b| > 20\degr$ in the
     velocity range $ -8 < v_{\mathrm{LSR}} < 8 $ \kms\ are used for
     $f_{\mathrm{CNM}} > 2/3$ (blue), $f_{\mathrm{LNM}} > 2/3$ (green),
     and $f_{\mathrm{WNM}} > 2/3$ (yellow). Bottom: the same
     distribution for all-sky. }
   \label{Fig_dust_CLW1}
\end{figure*}

\begin{table*}
\caption{Dust temperatures associated with CNM, LNM, and WNM at phase
  fractions $f > 2/3$}              % title of Table
\label{table:1}      % is used to refer this table in the text
\centering                                      % used for centering table
\begin{tabular}{l c c c c c c }          % centered columns (4 columns)
\hline\hline                        % inserts double horizontal lines
Selection criteria & CNM  $T_{\mathrm{dust}}$  & dispersion & LNM
$T_{\mathrm{dust}}$  & dispersion & WNM  $T_{\mathrm{dust}}$ & dispersion  \\    % table heading
\hline                                   % inserts single horizontal line
CNM dominated, $|b| > 20\degr$  & 18.54 K  & 0.89 K & 18.76 K & 0.95 K & 18.44 K & 0.85 K \\     
unconstrained, $|b| > 20\degr$  & 18.66 K  & 0.87 K & 19.01 K & 0.95 K & 19.50 K & 0.86 K \\
WNM dominated, $|b| > 20\degr$  & 19.20 K  & 0.87 K & 19.37 K & 1.01 K & 19.45 K & 1.27 K \\
\hline                                             %inserts single line
CNM dominated, all-sky  & 18.75 K  & 1.06 K & 18.91 K & 1.15 K & 18.52 K & 1.03 K \\     
unconstrained, all-sky  & 18.83 K  & 1.05 K & 19.01 K & 1.12 K & 19.45 K & 1.06 K \\
WNM dominated, all-sky  & 19.18 K  & 1.01 K & 19.31 K & 1.09 K & 19.48 K & 1.23 K \\     
\hline                                             %inserts single line
\hline                                   % inserts single horizontal line
\end{tabular}
\tablefoot{Parameters for the Gaussian fits to the dust temperature
  distributions shown in Fig. \ref{Fig_dust_CLW1}; on top for latitudes
  $ |b| > 20\degr$, all-sky below. }
\end{table*}

%\subsubsection{Dust temperatures in CNM and WNM dominated regions}

To determine phase dependent relations between \hi\ column densities and
dust temperatures we use Gaussian components with center velocities in
the range $ -8 < v_{\mathrm{LSR}} < 8 $ \kms\ at latitudes $ |b| >
20\degr$. Corresponding to the cases selected in Fig. \ref{Fig_LB_CLW}
(CNM top and WNM bottom) we plot in Fig. \ref{Fig_dust_CLW2} phase
dependent 2D density distribution functions for dust temperatures and
column densities for different \hi\ phases.

The average dust temperature determined by \citet{PlanckXLVIII} is
$T_{\mathrm{dust}} = 19.41 \pm 1.54$ K. All \hi\ phases in the CNM
dominated part of the sky (Fig. \ref{Fig_dust_CLW2} top) are
characterized by low dust temperatures. \hi\ column densities are high,
in particular for the CNM and LNM. For the rest of the sky (bottom) we
find warmer dust with temperatures around 19.4 K for the CNM and WNM,
but for the LNM even some of the \hi\ gas reaches dust temperatures
close to 20 K.

\subsection{Phase dominated positions for $f > 2/3$ }

Next we consider the question how far dust temperatures depend on
\hi\ phase fractions. As before, we consider CNM and WNM dominated
regions on the sky (Fig. \ref{Fig_LB_CLW}).  We select there positions
where each of the \hi\ phases is dominant. To decide whether a phase is
predominant we use the condition $ f / (1-f) > C $. Thus we demand that
the amount of \hi\ in an individual phase has to be a factor of $C > 1$
larger than the sum of the remaining phase fractions for the other
phases. Figure \ref{Fig_dust_CLW1} shows the distribution of dust
temperatures derived for the case $C = 2$ or $f > 2/3$. This factor is
somewhat arbitrary (though widely used to define a qualified majority)
but our results do not depend significantly on $C$. For all $C > 1 $ we
get well defined dust temperature distributions that can be approximated
by Gaussians. To the left of Fig. \ref{Fig_dust_CLW1} we plotted the
selected cold CNM dominated part, to the right the warm WNM
dominated sky and in the middle the unconstrained sample. Each of the
distributions is fitted by a Gaussian to obtain mean dust temperatures
with dispersions that are listed in Table \ref{table:1}.

These histograms show for all phases that the CNM dominated part of the
local \hi\ distribution is cold with typical dust temperatures
$T_{\mathrm{dust}} \sim 18.7$ K. There is little WNM with
$f_{\mathrm{WNM}} \ga 2/3$. Opposite, the WNM dominated regions have
$T_{\mathrm{dust}} \sim 19.4$ K and there is only little CNM. The
unconstrained \hi\ distribution is dominated by the LNM with
intermediate temperatures, $T_{\mathrm{dust}} \sim 19$ K. The upper plots
in Fig. \ref{Fig_dust_CLW1} are for high Galactic latitudes, $ |b| >
20\degr$, the lower plots display the all-sky distribution. The reliability
of the all-sky histograms is somewhat questionable but we include these
to demonstrate that they share the same trend as the histograms for high
latitudes.

Our results are consistent with the average $T_{\mathrm{dust}} = 19.41
\pm 1.54 $ K as determined by \citet{PlanckXLVIII} at high
latitudes. However Fig. \ref{Fig_dust_CLW1} indicates a general
dependency that dust temperatures decrease in regions that are dominated
by the CNM. Each of the dust components can be fitted by Gaussians (see
Table \ref{table:1}).  For the CNM dominated sample the fits are well
defined with low dispersions around one K. The distributions for the WNM
dominated sky are broader and may contain several components with some
scatter. The lower panel in Fig. \ref{Fig_dust_CLW1} indicates that also
some more scatter may exist at latitudes $|b| \la 20\degr$.

Figure \ref{Fig_dust_CLW1} confirms the trends from
Fig. \ref{Fig_dust_CLW2} but we need to point out that in particular
positions with $f_{\mathrm{CNM}} \ga 2/3$ that are associated with cold
dust are almost exclusively found in the CNM dominated part of the
sky. Opposite, positions with $f_{\mathrm{WNM}} \ga 2/3$ and warm dust
are less frequent in this areas. For the rest of the sky we observe a
dominance of the WNM, but surprisingly also there the LNM is
significant. In all cases, for $f \ga 2/3$, we find low dust
temperatures for CNM dominated regions and opposite high temperatures in
case of the WNM dominated sky, see Table \ref{table:1}. Thus the dust
temperatures are more smoothly distributed compared to the \hi\ gas 
  that may have significant local fluctuations in phase fractions.

%\clearpage
%\newpage

\section{Cold and filamentary \hi}
\label{Filaments}

\begin{figure*}[!thp] %%  
   \centering
   \includegraphics[width=6cm]{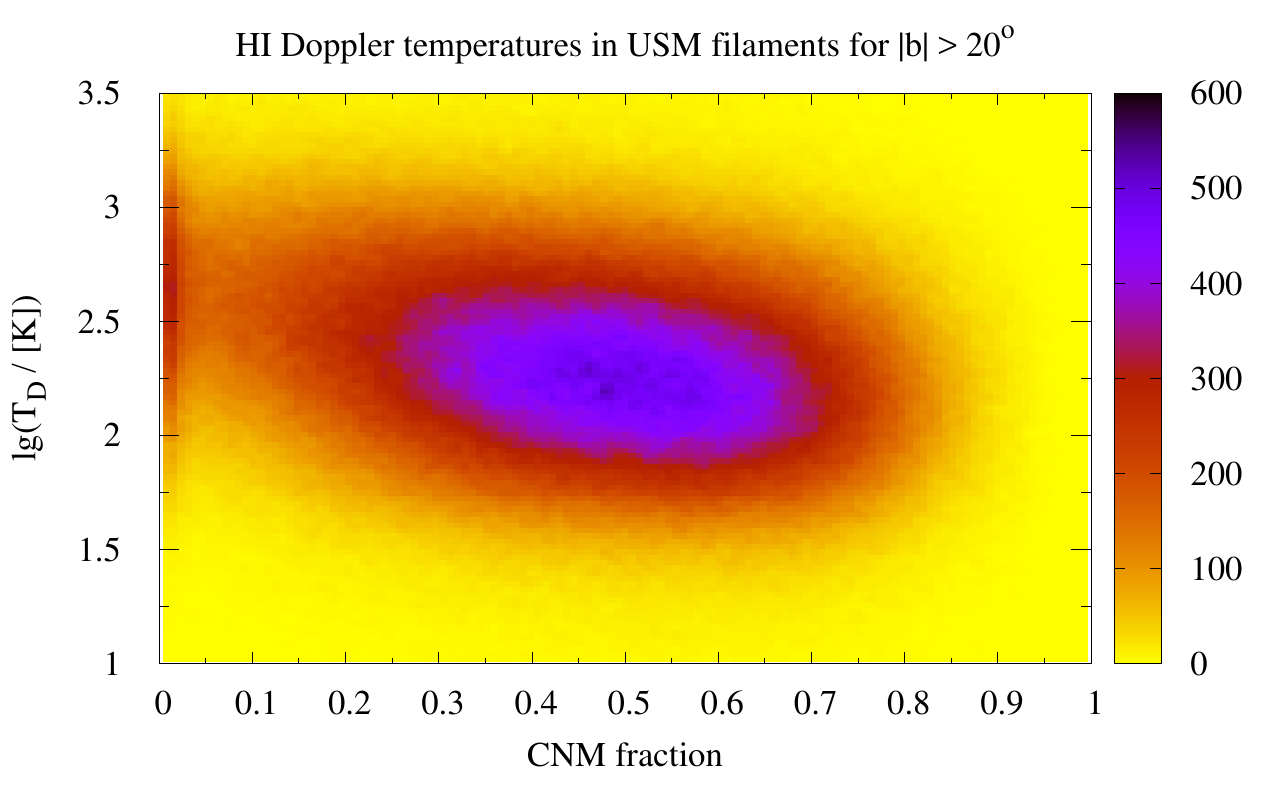}
   \includegraphics[width=6cm]{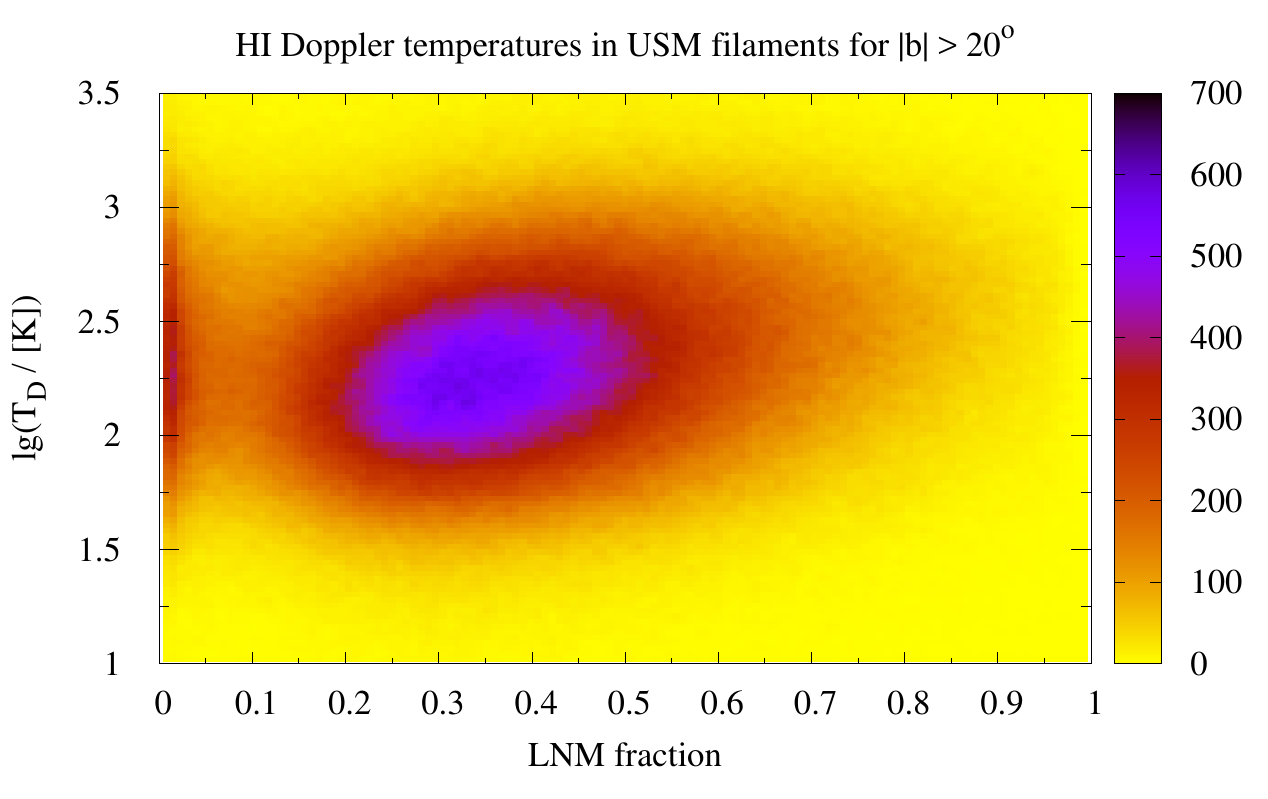}
   \includegraphics[width=6cm]{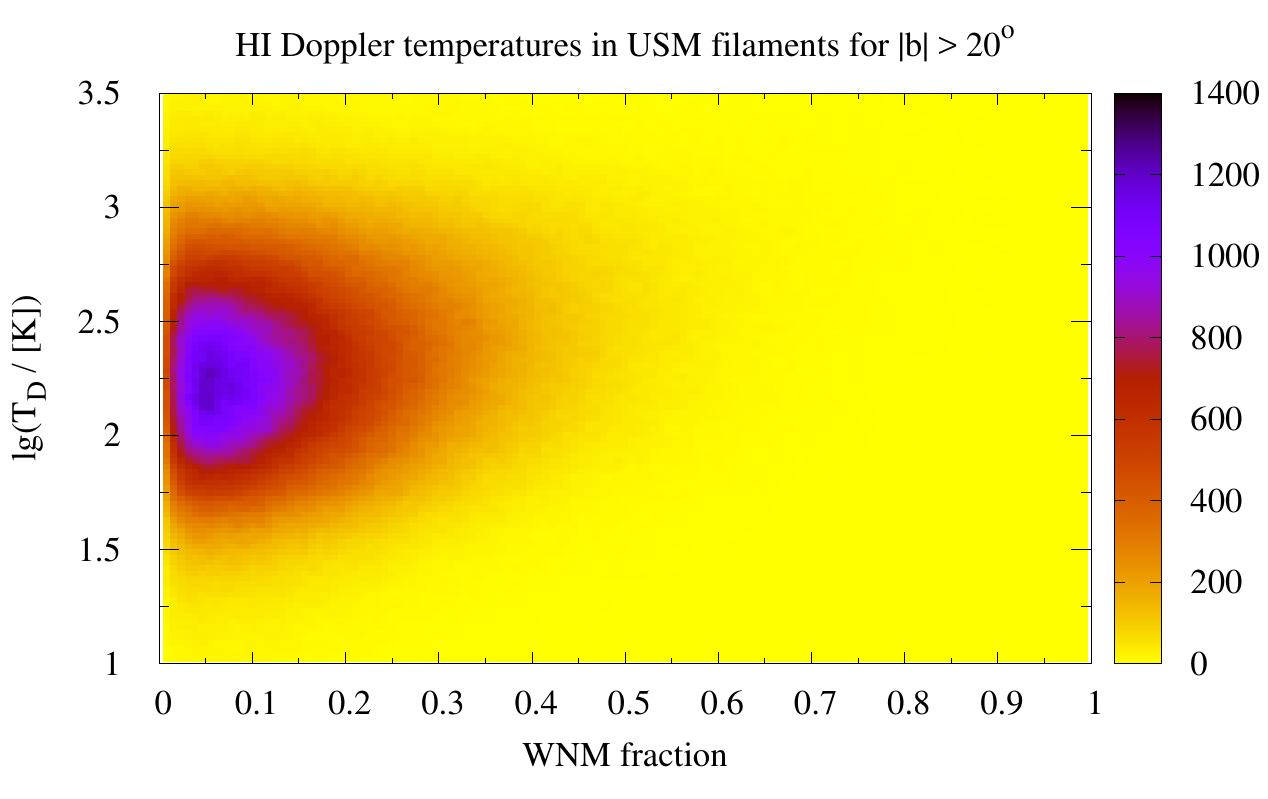}
   \includegraphics[width=6cm]{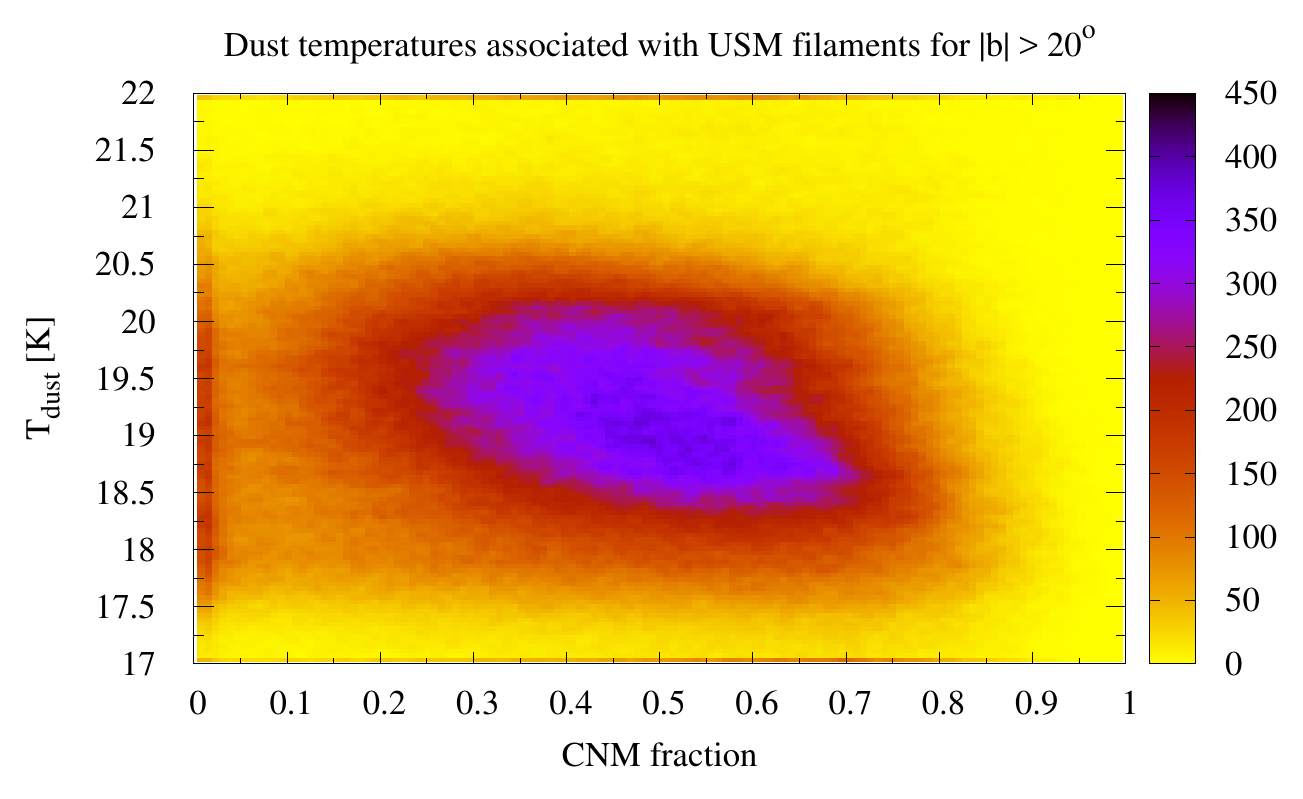}
   \includegraphics[width=6cm]{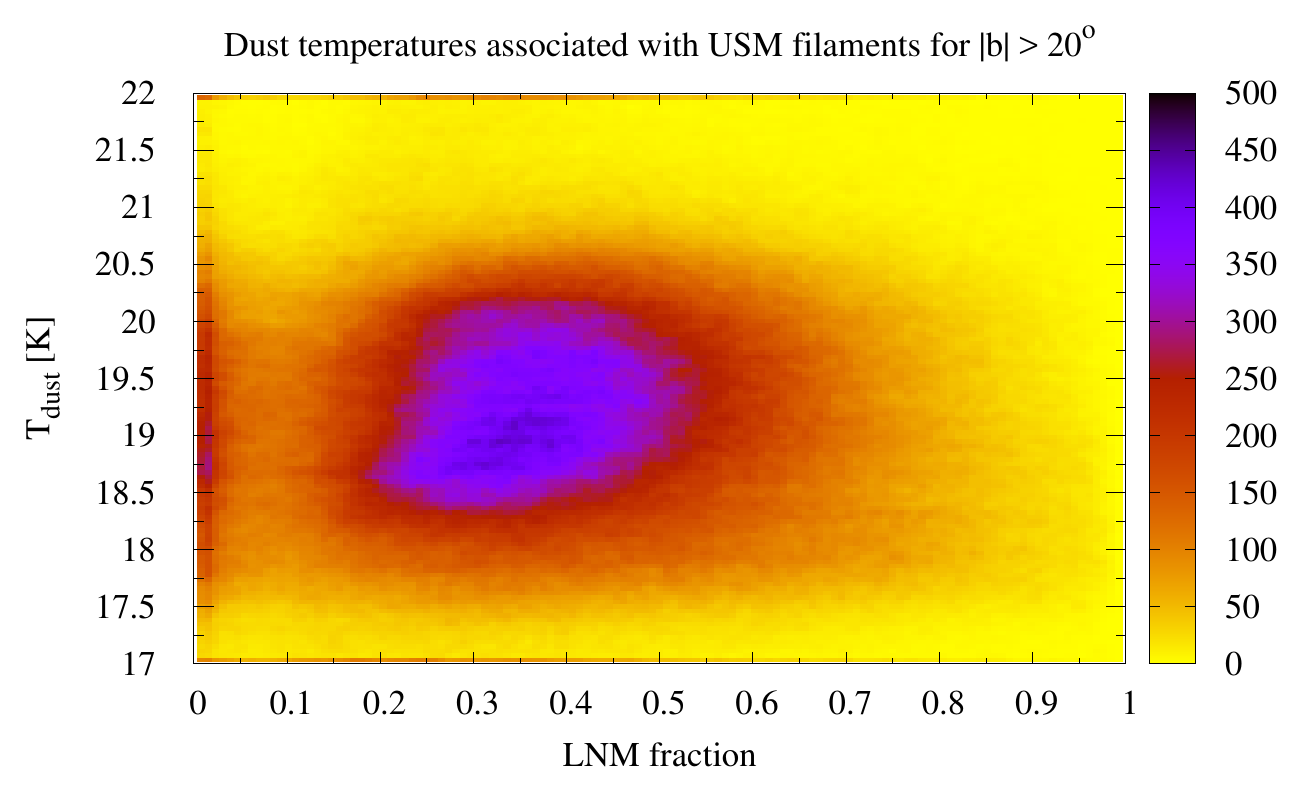}
   \includegraphics[width=6cm]{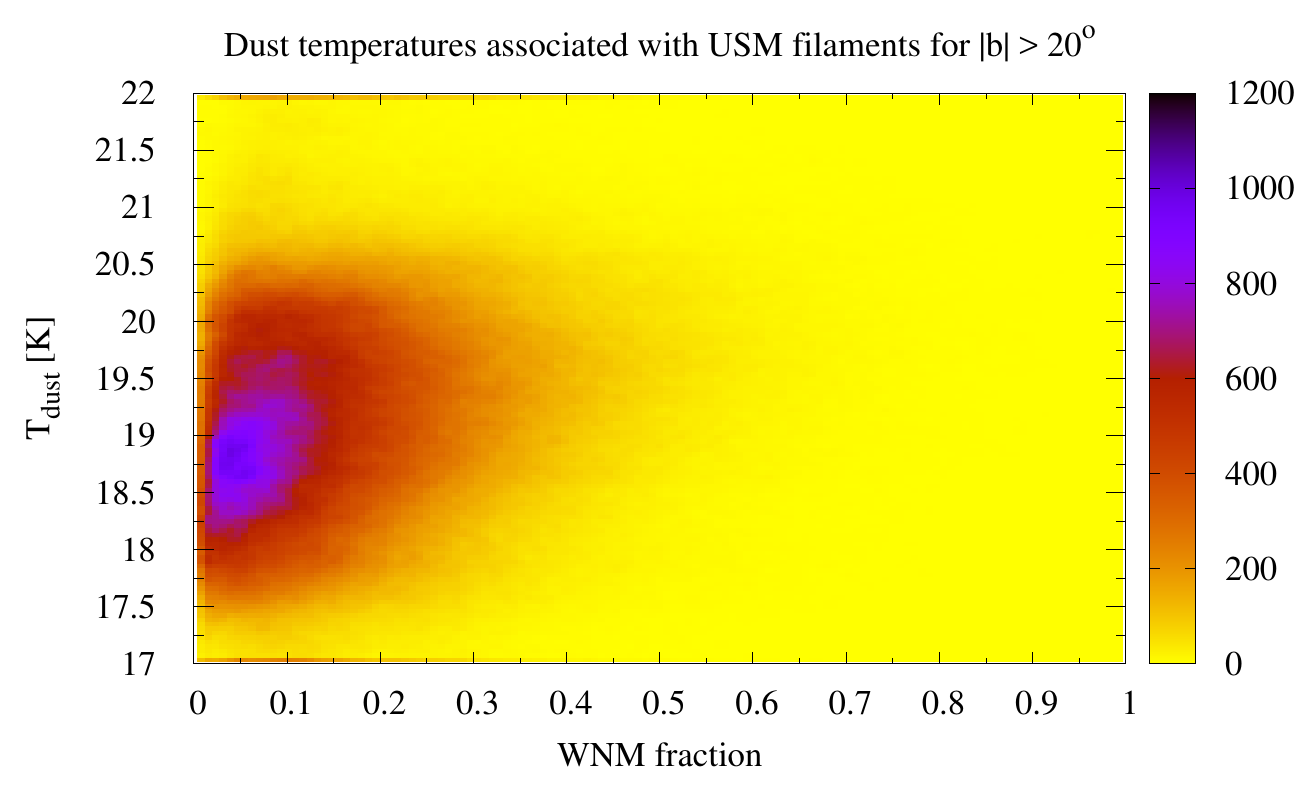}
   \caption{\hi\ in filaments. 2D density distribution functions for
     phase fractions and USM Doppler temperatures $T_{\mathrm{D}}$ (top)
     as well as associated dust temperatures $T_{\mathrm{dust}}$
     (bottom) in filaments at $ |b| > 20\degr$; left: CNM, middle: LNM,
     and right: WNM phase fractions.  }
   \label{Fig_TS_USM}
\end{figure*}

A significant fraction of the cold \hi\ gas exists in filamentary
structures that can be worked out for high resolution \hi\ observations
by unsharp masking \citep{Kalberla2016}. Many of these filaments
are associated with dust and the most prominent filaments are aligned
with the magnetic field. In this section we want to explore
temperature dependencies of the different phases in presence of such
filamentary structures.

Using the USM data from \citet{Kalberla2016}, we search at each position
for the strongest filament in the velocity range $ -8 < v_{\mathrm{LSR}}
< 8 $ \kms\ and calculate its Doppler temperature $ T_{\mathrm{D}}$. We
demand an USM peak temperature of at least one K. If such a feature is
found we determine the velocity channel corresponding to its center
velocity $v_0$. We use one additional channel on both sides and
determine for this velocity window of three \kms\ the phase fractions
$f_{\mathrm{CNM}}$, $f_{\mathrm{LNM}}$, and $f_{\mathrm{WNM}}$.  Our
analysis is restricted to filamentary \hi\ structures at high latitudes
$ |b| > 20\degr$.  Figure \ref{Fig_TS_USM} displays on top the derived
2D density distributions for Doppler temperatures and phase fractions
$f$.

Next we consider the question how the observed dust temperatures might
depend on \hi\ phase fractions for filamentary \hi. For a comparison
between dust temperatures and \hi\ Doppler temperatures we determine at 
each position with a known USM Doppler temperature $T_{\mathrm{D}}$ the
corresponding GNILC dust temperature $T_{\mathrm{dust}}$. In
Fig. \ref{Fig_TS_USM} we present on bottom 2D density distributions for
dust temperatures associated with filamentary \hi\ structures. Comparing
Doppler temperatures at top of Fig. \ref{Fig_TS_USM} with dust
temperatures at bottom we find similar trends for different phase
fractions; $f_{\mathrm{CNM}}$, $f_{\mathrm{LNM}}$, and
$f_{\mathrm{WNM}}$ (left to right).

These plots show that Doppler and dust temperatures in filaments
decrease on average with increasing CNM fraction $f_{\mathrm{CNM}}$.
The vertical stripe at $f_{\mathrm{CNM}} \sim 0 $ in the left panels of
Fig. \ref{Fig_TS_USM} indicates that some of
the USM filaments are not identified with proper Gaussian
components. For an approximate threshold $f_{\mathrm{CNM}} \la 0.05 $ we
find that less than 4\% of the USM filaments are affected. For the other
data we fit $\mathrm{lg}(T_{\mathrm{D}}/\mathrm{[K]}) = 2.5222 \pm 0.0008 - (0.451
\pm 0.002) \cdot f_{\mathrm{CNM}} $ for the \hi\ and $T_{\mathrm{dust}}
= 19.455 \pm 0.002 - (0.643 \pm 0.004) \cdot f_{\mathrm{CNM}}$ K for the
dust. The median Doppler temperature is $T_{\mathrm{D}} = 203$ K,
corresponding to a median dust temperature of $T_{\mathrm{dust}} =
19.08$ K.

\begin{figure}[th] %%  
   \centering
   \includegraphics[width=4.4cm]{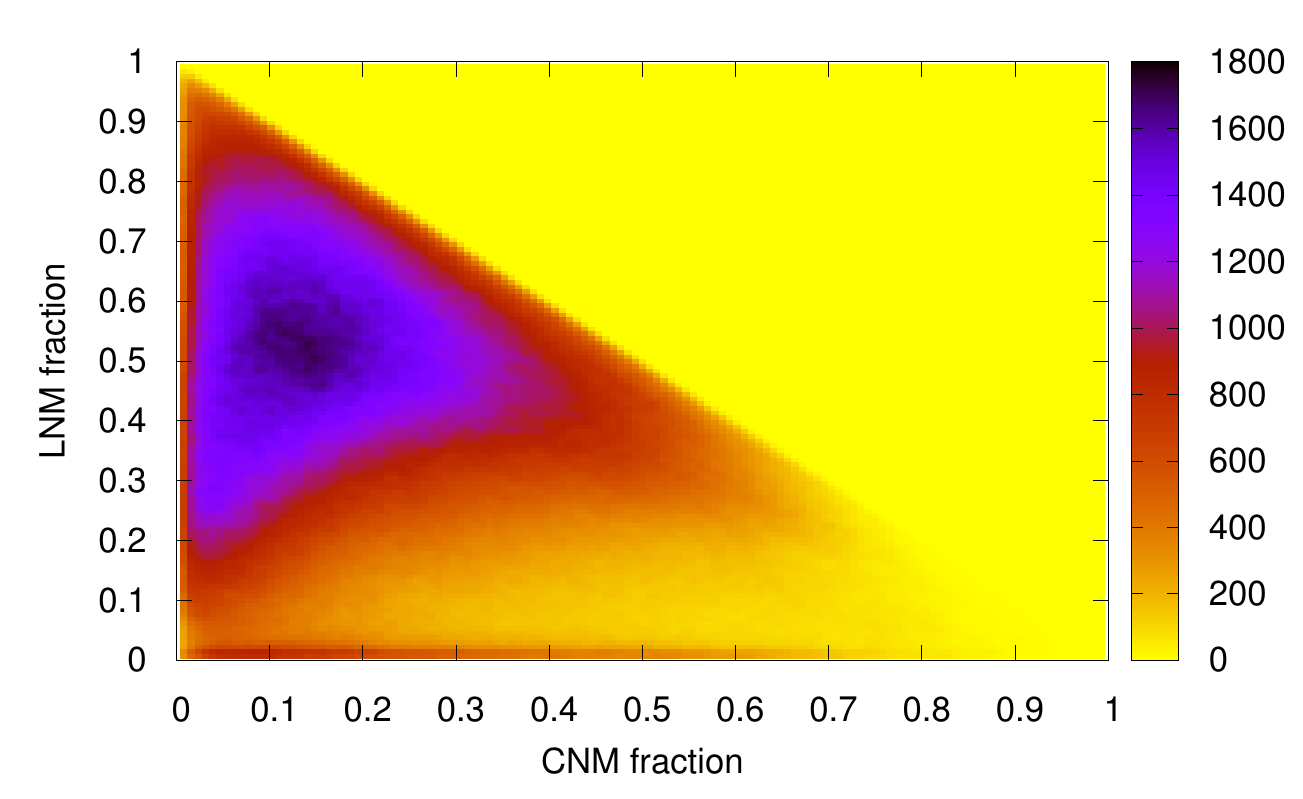}
   \includegraphics[width=4.4cm]{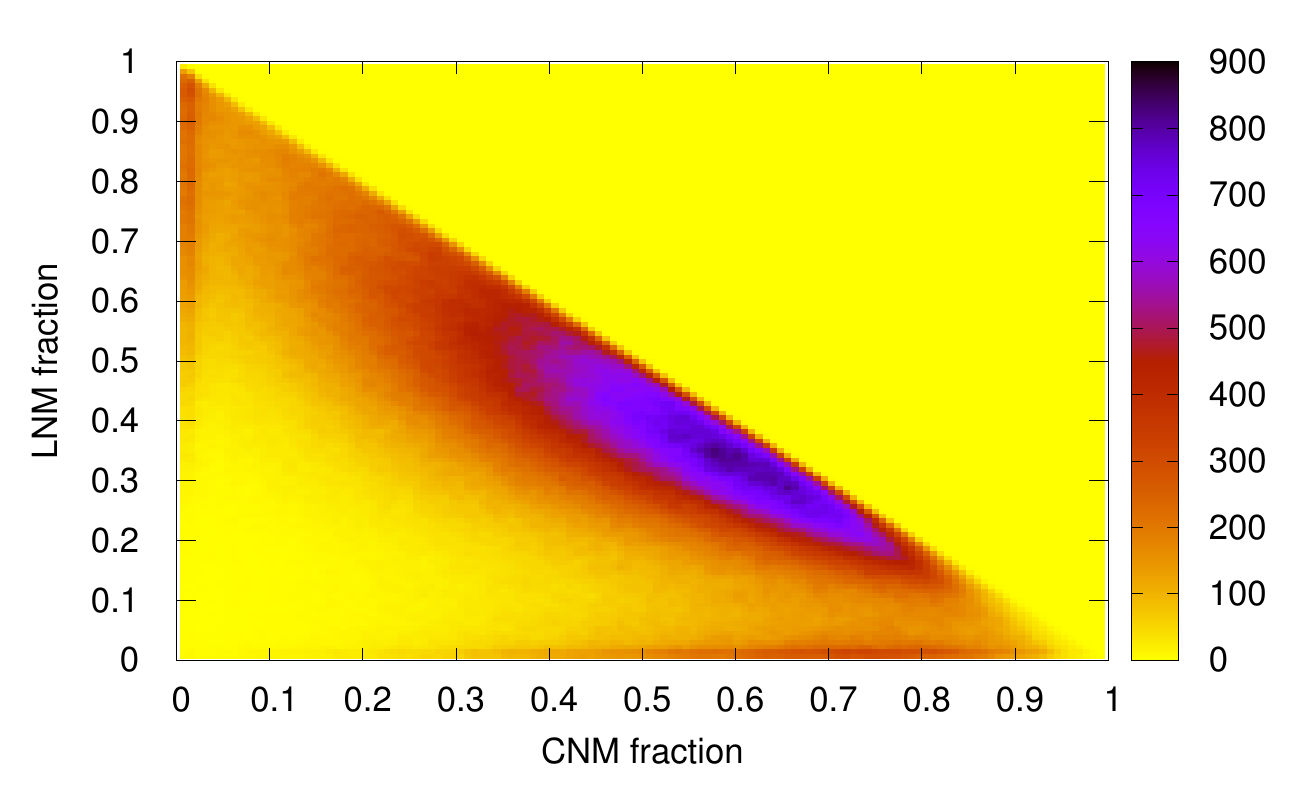}
   \includegraphics[width=4.4cm]{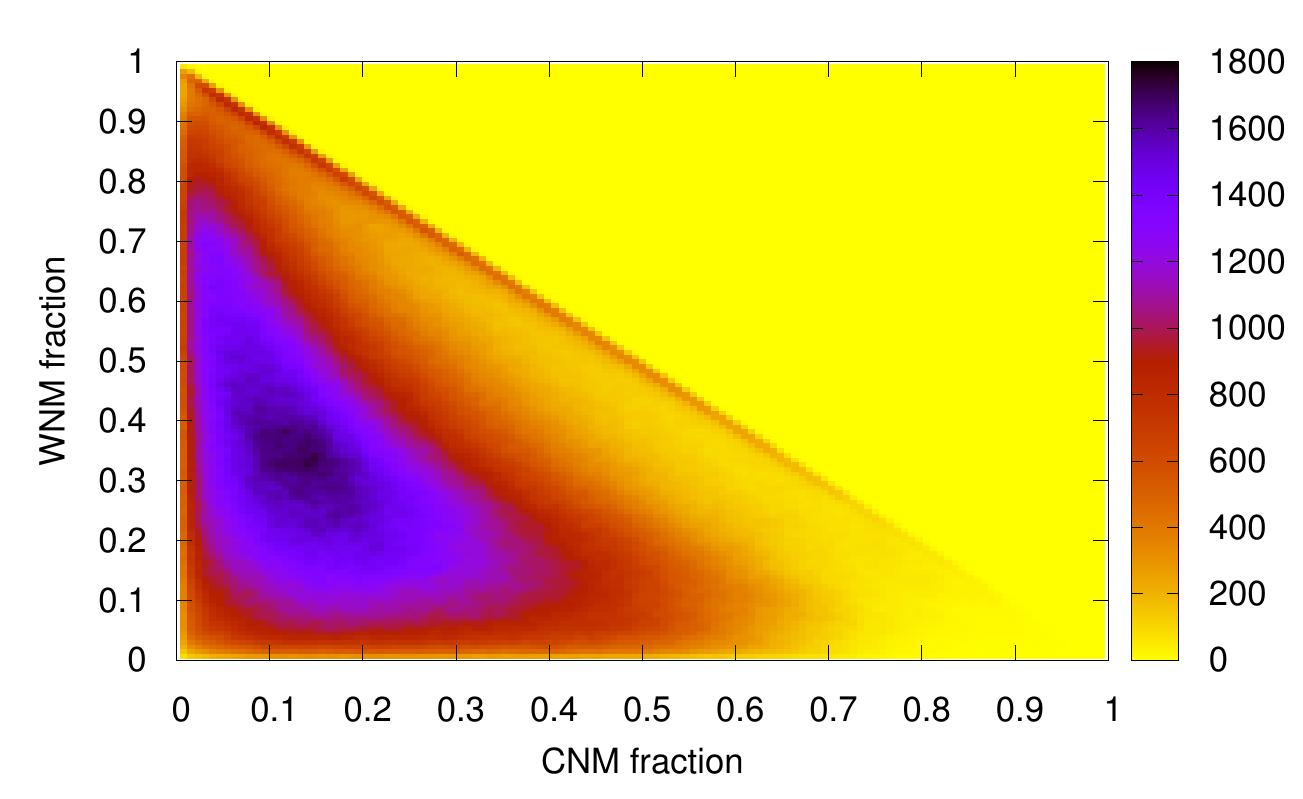}
   \includegraphics[width=4.4cm]{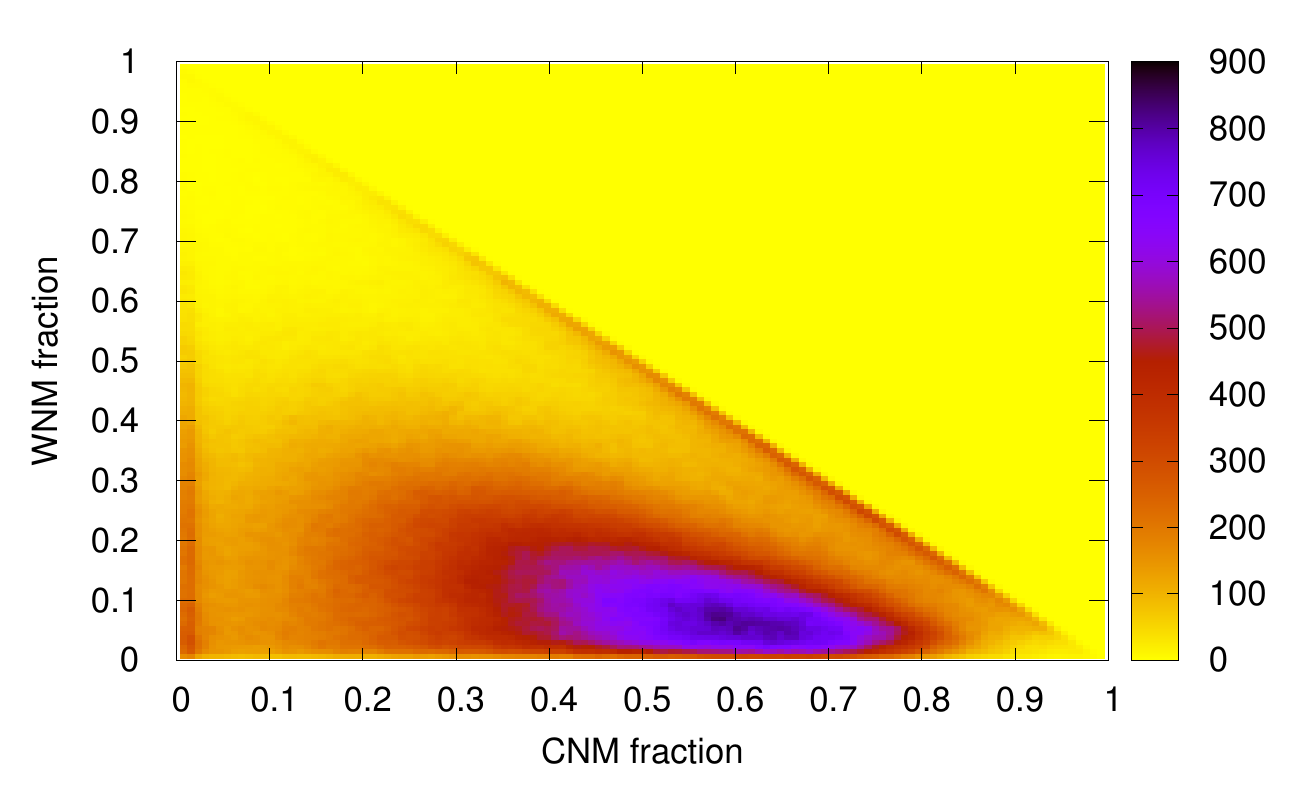}
   \caption{All sky 2D distribution functions, showing how frequent
     phase fractions of different phases are related to each other for $
     |b| > 20\degr$ in the velocity range $ -8 < v_{\mathrm{LSR}} < 8 $
     \kms. Top for CNM and LNM and bottom for CNM and WNM. Left:
     unconstrained \hi\ data; right: filamentary \hi\ structures only.}
   \label{Fig_correl_CLW}
\end{figure}

The median temperature for dust associated with filamentary \hi\ is
colder than the average temperature of $T_{\mathrm{dust}} = 19.41 \pm
1.54$ K \citep{PlanckXLVIII}.  The \hi\ filaments selected in the
current analysis are also slightly colder than the filaments studied in
\citet{Kalberla2016} with a median Doppler temperature of
$T_{\mathrm{D}} = 223 $ K. Prominent cold filaments reach $
f_{\mathrm{CNM}} \sim 0.65 $ (see Fig. \ref{Fig_correl_CLW}, also Sect.
\ref{Columntotal}). Typical Doppler temperatures are 170 K, with dust
temperatures of 19 K.  Such Doppler temperatures correspond to a FWHM
line width of 3.2 \kms\ and our sliding velocity window with $\delta
v_{\mathrm{LSR}} = 3 $ \kms\ matches well with this line width. We
conclude that the ISM in local filamentary structures is comparatively
cold.

The middle and lower plots in Fig. \ref{Fig_TS_USM} show that the
Doppler and dust temperatures in filaments are also correlated with the
component fraction $f_{\mathrm{LNM}}$ and $f_{\mathrm{WNM}}$ of the
associated warmer gas. For the LNM we fit for the
\hi\ $\mathrm{lg}(T_{\mathrm{D}}/\mathrm{[K]}) = 2.1789 \pm 0.0008 + (
0.3600 \pm 0.002) \cdot f_{\mathrm{LNM}}$; for the dust
$T_{\mathrm{dust}} = 19.165 \pm 0.002 - (0.012 \pm 0.004) \cdot
f_{\mathrm{LNM}}$ K.  For the WNM we obtain
$\mathrm{lg}(T_{\mathrm{D}}/\mathrm{[K]}) = 2.24195 \pm 0.0005 + (
0.50156 \pm 0.003) \cdot f_{\mathrm{WNM}}$ and $T_{\mathrm{dust}} =
18.994 \pm 0.002 + (1.070 \pm 0.006) \cdot f_{\mathrm{WNM}}$ K.

This implies that the Doppler temperatures of USM filaments tend to
increase moderately up to typical temperatures around $T_{\mathrm{D}} =
228$ K and $T_{\mathrm{dust}} = 19.17$ K for an LNM environment with a
component fraction of $ f_{\mathrm{LNM}} = 0.5 $.  Such a \hi\ Doppler
temperature differs only slightly from the median of $T_{\mathrm{D}} =
223$ K determined by \citet{Kalberla2016}.  For filaments embedded in
WNM with $f_{\mathrm{LNM}} = 0.5 $ we obtain a significantly higher
Doppler temperature $T_{\mathrm{D}} = 311$ K and correspondingly a
higher dust temperature $T_{\mathrm{dust}} = 19.53$ K. The temperature
differences would increase further for larger WNM component fractions,
however we observe little filaments associated with WNM in this state
(Fig. \ref{Fig_TS_USM} bottom). This deficiency may imply that we
observe here a transition to ionized gas.

Calculating average phase fractions for filaments we find that such
regions are dominated by the CNM: $f_{\mathrm{CNM}} \sim 0.46 $,
$f_{\mathrm{LNM}} \sim 0.37$, and $f_{\mathrm{WNM}} \sim 0.17$. Here
$f_{\mathrm{CNM}}$ is a lower limit only, our data are not corrected for
optical depth effects, for discussion see
\citet[][Sect. 5.9]{Kalberla2016}. Optical depth effects however are
mostly negligible for ${\mathrm{lg}}(N_{\mathrm{H}}/[{\mathrm{cm}}^{-2}])
\la 20.6 $ \citep{Lee2015}. Figures \ref{Fig_fcorrelNH} and
\ref{Fig_dust_CLW2} show that most of the CNM at high latitudes is below
this limit.

%=========================================================================

\section{\hi\ phases along the line of sight}
\label{PhaseRelations}

The results from the previous sections imply that the different
\hi\ phases are associated with each other; the observed gas composition
changes according to the environment. In Sect. \ref{coldspots},
considering the extension of CNM dominated regions in the plane of the
sky, we demonstrated that there is an anti-correlation between
$\overline{f_{\mathrm{CNM}}}$ and $\overline{f_{\mathrm{LNM}}}$.

To study further the relationship between CNM, LNM, and WNM we determine
the 2D frequency distributions of the observed phase fractions at
  the same position, hence along the line of sight. We use latitudes $
|b| > 20\degr$ and the velocity range $ -8 < v_{\mathrm{LSR}} < 8 $
\kms. Figure \ref{Fig_correl_CLW} shows how frequent the CNM with a
phase fraction $f_{\mathrm{CNM}}$ is associated with LNM or WNM and
phase fractions $f_{\mathrm{LNM}}$ or $f_{\mathrm{WNM}}$
respectively. To the left we display unconstrained data. In this case
the CNM contributes typically little \hi\ gas with $f_{\mathrm{CNM}} \la
0.15$. The distribution is bimodal. The associated warmer gas belongs
either to the LNM with typically $f_{\mathrm{LNM}} \sim 0.55$ or to the
WNM with $f_{\mathrm{WNM}} \sim 0.3$. Thus the LNM is dominant, in
agreement with Fig. \ref{Fig_dust_CLW1}.

The filamentary \hi\ gas has a very different composition. This is
demonstrated on the right hand side of Fig. \ref{Fig_correl_CLW}.  In
this case only filamentary \hi\ structures from the USM database have
been selected.  Clearly, filaments are outstanding with high phase
fractions $f_{\mathrm{CNM}}$. Further a sharp division between the
associated LNM and WNM phases is visible. The LNM typically occupies
regions with larger phase fractions $f_{\mathrm{LNM}}$ compared to the
WNM. The sharp diagonal stripe in the upper panel on the right hand side
of Fig. \ref{Fig_correl_CLW} confirms that also in filamentary
\hi\ structures CNM and LNM are closely anti-correlated. The
anti-correlation between CNM and WNM, shown in the panel below, has a
larger scatter.

\section{Discussion}
\label{Discussion}

Our most important result is the large average phase fraction
$\overline{f_{\mathrm{LNM}}} = 0.41$ and the close anti-correlation
between LNM and CNM. Most obvious are the phase relations from the
distribution of the different phases around cold spots
(Sect. \ref{coldspots}). The detailed anti-correlation between
$\overline{f_{\mathrm{CNM}}}$ and $\overline{f_{\mathrm{LNM}}}$ on small
and intermediate scales proves that the clumpy CNM is embedded in
LNM. The WNM surrounds both phases with an anti-correlation on larger
scales. The same kind of layered structure is observed around cold
filaments. These filaments are associated with cold dust. Doppler
temperatures as well as dust temperatures decrease with increasing
$f_{\mathrm{CNM}}$.

When distinguishing \hi\ associated with cold filamentary structures as
described in Sect. \ref{Filaments} from the unconstrained
\hi\ distribution it needs to be considered that filamentary structures
most probably are caused by sheets, seen edge-on
\citep{Heiles2005,Kalberla2016,Kalberla2017}. Similar structures with
different orientations, such that we do not have a favorable tangential
view (amplifying projected column densities), may exist but these parts
of the \hi\ distribution will not be observable with an outstanding
morphology. On the other hand, also \hi\ clouds that are warmer than the
filaments may contribute to the diffuse \hi\ phase.

From \hi\ observations alone it is difficult to decide whether the WNM
shown in Fig. \ref{Fig_correl_CLW} on the right hand side is physically
related to the CNM but we see in Fig. \ref{Fig_TS_USM} that a
significant part of WNM in filaments is associated with particular cold
dust that appears to be more smoothly distributed than the \hi. So we
are apparently left with a miracle; warm \hi\ gas associated with cold
dust along the line of sight. This phenomenon may be explainable in
context with the \citet{McKee1977} model. The prediction is that cold
gas must be associated with warmer gas, in addition there should even be
some ionized material at the same velocity. Indeed, \citet{Kalberla2017}
demonstrated that radio-polarimetric filaments can be associated with
cold \hi\ filaments. \hi\ Doppler temperatures were found to be
correlated with the CNM phase fractions and a re-inspection of the
Auriga and Horologium fields considered by \citet{Kalberla2017}
discloses that some of the USM filaments are also associated with cold
dust filaments.

Phase fractions discussed by \citet{Kalberla2017} were based on the
simplified assumption of a two-phase \hi\ medium. Nevertheless, it was
shown that increasing CNM phase fractions are correlated with decreasing
\hi\ Doppler temperatures and at the same time with anisotropies. In
addition steep spectral indices were found for the associated turbulence
power spectra that appear to be related to phase transitions. For a
bi-modal \hi\ distribution such a steepening is hardly explainable.
Considering local instabilities within a smoothly distributed WNM phase
we expect in case of phase transitions a cold gas distribution with
enhanced structures on small scales. This implies a local flattening of
the power spectrum, contrary to observations.

Our investigations prove that the CNM is associated with significant
amounts of LNM. The CNM is clumpy, the anti-correlation between
$f_{\mathrm{CNM}}$ and $f_{\mathrm{LNM}}$ implies therefore that the LNM
must be surrounding the colder gas on larger scales, up to scales of a
few degrees. For an average phase fraction $\overline{f_{\mathrm{LNM}}}
= 0.41 $ this results necessarily in excess LNM emission around the
CNM. Thus, the presence of LNM on larger scales leads to enhanced
emission on such scales, offering a natural explanation for the observed
steepening of local power spectra.

The Gaussian decomposition resulted in the following mean FWHM
line-widths of the CNM, LNM and WNM gas: 3.6, 9.6, and 23.3 \kms,
corresponding to Doppler temperatures of 283, 2014, and 11\,870 K. These
results can be used to estimate characteristic turbulent Mach numbers
for the \hi\ phases; $M = \sqrt{ 4.2 (T_\mathrm{ D}/T_\mathrm {kin} -
  1)}$ \citep[][]{Heiles2003a}. Using a typical kinetic temperature of
50 K for the CNM \citep{Heiles2003a} we obtain $M_\mathrm{CNM} =
4.4$. USM filaments have somewhat lower Doppler temperatures and
correspondingly $3.2 \la M_\mathrm{CNM} \la 3.7$ \citep{Kalberla2016}.
For the LNM it is hardly possible to define a characteristic kinetic
temperature. Using a lower limit of 500 K we derive a firm upper limit
$M_\mathrm{LNM} \la 3.6$ for the LNM Mach number. In case of the WNM we
derive $M_\mathrm{WNM} \sim 1.4$ for an kinetic temperature of
8000K. There is a systematic trend that Mach numbers decrease with
increasing Doppler temperatures for the different phases.

Our result that filamentary \hi\ gas is associated with cold dust might
be explainable by destruction of dust. Anisotropies in the observed
\hi\ distribution are probably caused by cold shells or sheets, seen
edge-on. As suggested by \citet{Heiles2003a}, these structures result
from shock waves, originating from supernova explosions, causing phase
transitions and rapid cooling of the \hi. Such shocks can destroy also
grains through sputtering. Large grains are ground down into smaller
grains \citep{Jones1996}. Low CNM temperatures in filaments may thus be
explainable by reduced photoelectric heating due to depletion of
polycyclic aromatic hydrocarbons (PAHs) \citep{Wolfire2003} but a direct
observational evidence for such a process is difficult for the dust
\citep{Tielens2008}.  \citet{Micelotta2010} model PAH processing in
interstellar shocks and find that PAHs do not survive shocks with
velocities greater than 100 \kms. We like to point out that we observe
low dust temperatures only for cold filamentary \hi\ gas. Without
Doppler temperatures from USM data we would not be able to quantify a
relation between dust and Doppler temperatures. The main evidence for
such a correlation comes from our USM data in direction to Loop I
\citep{Egger1995}. This is a prominent object and Loop I at an age of
several $10^6$ years is still expanding at a velocity of 17.3
\kms\ \citep{Frisch2018}. A depletion of PAHs appears plausible under
these conditions.

Dust variations in the diffuse interstellar medium have also been noted
by \citet{Ysard2015}. These authors use total \hi\ column densities from
the LAB survey at high Galactic latitudes for $ 10^{19} \la
N_{\mathrm{H}} \la 2.5~ 10^{20} $ cm$^{-2}$ with a sky coverage of about
12\%. They show that variations in the dust properties observed with
{\it Planck}-HFI cannot be explained by variations in the radiation
field intensity and gas density distribution. They conclude that small
variations in the dust properties can explain most of the variations in
the dust emission observed by {\it Planck}-HFI in the diffuse ISM.
Unfortunately, due to their column density limit, this analysis excludes
Loop I and most of the regions with cold \hi\ filaments.

One of the main drivers for the launch of {\it Plank} was the quest to
observe the cosmic microwave background B-mode signal induced by
primordial gravitational waves during cosmic inflation. However, there
are Galactic foregrounds and it is now established that these are a main
limiting factor for accurate cosmological parameters.  Polarized
Galactic foreground observations with {\it Planck} and \hi\ emission
data indicate that most of the filamentary dust structures are embedded
in filamentary CNM structures \citep{Clark2014,Kalberla2016}. Based on
this finding \citet{Ghosh2017} constructed a phenomenological dust
model, combining \hi\ data with an astrophysically motivated description
of the large-scale and turbulent Galactic magnetic field. They modeled
the large-scale polarized dust emission over the southern Galactic cap
by using GASS \hi\ data with selected velocity dispersions. They were
able to reproduce the {\it Planck} dust observations at 353 GHz for a
region of 3500 deg$^2$ at the southern Galactic cap. They tested their
model for \hi\ velocity dispersions $\sigma < 3$ \kms\ and $3 < \sigma <
7.5$ \kms\ and found no significant differences. These two ranges
correspond approximately to our CNM and LNM linewidth regimes. Thus the
results by \citet{Ghosh2017} are consistent with our conclusion that
cold CNM filaments and the surrounding LNM must be related to each
other.

\section{Summary and conclusion}
\label{Summary}

We used a Gaussian decomposition of the HI4PI survey to separate and
model the CNM, LNM, and WNM distributions. Our main results are:

Phase fractions depend on the selected velocity range. For
  the velocity range $ -8 < v_{\mathrm{LSR}} < 8 $ \kms\  41\% of the
  local \hi\ gas is LNM. Next important is the WNM with 34\%, the CNM
  contributes only 25\%.  
CNM or LNM are never observed as a single phases, they can only coexist
  with WNM.
The phase fractions for CNM and LNM are anti-correlated. This
  implies that the LNM surrounds the clumpy CNM. Both phases are
  embedded in the smoothly distributed WNM.
  
 CNM dominated regions have low dust temperatures,
  $T_{\mathrm{dust}} \sim 18.5$ K. The dust in WNM dominated regions is
  warmer; differences amount to about one K.
Dust temperatures as well as Doppler temperatures for CNM in
  filaments decrease with increasing $f_{\mathrm{CNM}}$.
Outside \hi\ filaments dust temperatures show a general trend to
  increase with LNM and WNM column densities. Here the presence of CNM
  (typically with $f_{\mathrm{CNM}} \la 0.2$ ) does not correlate with dust
  temperatures.

  The previous conclusions are based on high latitudes $ |b| >
  20\degr$. An extrapolation to lower latitudes suffers from large
  uncertainties but closer to the Galactic plane we did not get
  significantly different results.
It is difficult to define clear criteria for the separation of
  CNM, LNM, and WNM. The derived \hi\ phase fractions may be uncertain
  by 50\%. Such uncertainties do however not invalidate our general
  conclusions.

Previous studies have shown that cold filamentary \hi\ structures in the
diffuse interstellar medium are associated with dust ridges, aligned with
the magnetic field measured on the structures by {\it Planck} at 353 GHz
\citep{Clark2014,Kalberla2016}. These are also the structures where we
find high values for $f_{\mathrm{CNM}}$ and low dust temperatures. It is
becoming increasingly evident that the state of the ISM is significantly
affected by feedback processes.

%=========================================================================
\begin{acknowledgements}
  We thank the referee, Carl Heiles, for constructive criticism that
  helped to improve the paper, also J{\"u}rgen Kerp for valuable
  discussions and continuous support. U. H. acknowledges the support by
  the Estonian Research Council grant IUT26-2, and by the European
  Regional Development Fund (TK133). This research has made use of
  NASA's Astrophysics Data System.  EBHIS is based on observations with
  the 100-m telescope of the MPIfR (Max-Planck-Institut f\"ur
  Radioastronomie) at Effelsberg. The Parkes Radio Telescope is part of
  the Australia Telescope which is funded by the Commonwealth of
  Australia for operation as a National Facility managed by CSIRO. Some
  of the results in this paper have been derived using the HEALPix
  package.
   \end{acknowledgements}

{\it Note added in proof.}
After acceptance of the manuscript the 21-SPONGE \hi\ absorption line
survey against 57 background continuum sources became available
\citep{Murray2018}. These authors distinguish \hi\ phases for derived
spin temperatures with $T_{\mathrm{s}} < 250 $ K for the CNM, $250 <
T_{\mathrm{s}} < 1000 $ K for the unstable LNM, and $T_{\mathrm{s}} >
1000 $ K for the WNM. Accordingly they conclude that 28\% of the
\hi\ mass is associated with the CNM, 20\% belongs to the LNM, and 52\%
to the WNM. We determine for the region covered by this survey mostly
dust temperatures between 19 K and 19.5 K, independent of preferences in
phase fractions, and $\overline{f_{\mathrm{CNM}}} = 0.24 $,
$\overline{f_{\mathrm{LNM}}} = 0.45 $, and $\overline{f_{\mathrm{WNM}}}
= 0.31 $. \citet{Murray2018} do not detect a significant mass fraction
for \hi\ at $1000 \la T_{\mathrm{s}} \la 4000$ K. The LNM is a transient
phase. We explain discrepancies in the derived LNM and WNM phase
fractions as an indication that only about half of the LNM observed by
us is currently evolving to cold structures with recognizable optical
depths.

\begin{appendix} 
\label{app}

%=========================================================================
%=========================================================================
%=========================================================================

\section{On the XNM}
\label{XNM}

   In this paper we discuss CNM, LNM and WNM phases of \hi, but
   Figs.~\ref{Td3}, \ref{Td7} and \ref{Dis} contain also the fourth
   curve -- XNM. We mentioned that in many cases, but not always, the
   XNM Gaussians seem to be caused by the wide but weak wings of the
   lines from other phases. In a Gaussian decomposition such lines
   cannot be described by a single component as the Gaussians fall off
   too rapidly compared to the data. However, such lines are often
   successfully modeled with the Voigt profile \citep{Mitchell1971} -- a
   convolution of a Gaussian function with a Lorentzian, or, due to the
   computational expense of the convolution operation, the Voigt profile
   is approximated with a pseudo-Voigt one. Here we try to give a brief
   test, if such a more complicated fitting procedure may help us to
   avoid the XNM components in our decomposition.

\begin{figure}
%   \resizebox{\hsize}{!}{\includegraphics{Dif.pdf}}
   \resizebox{\hsize}{!}{\includegraphics{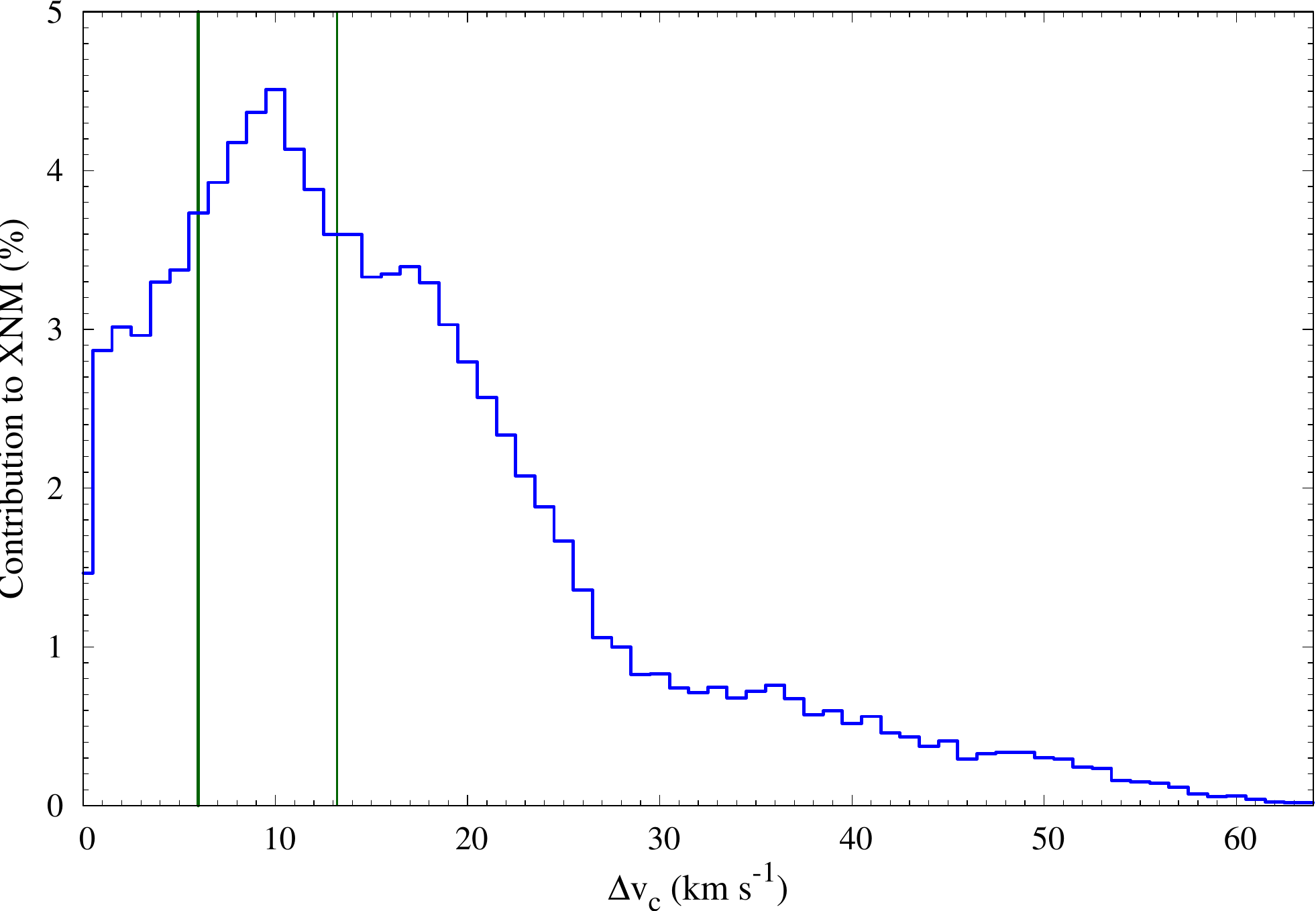}}
   \caption{Fractions of the total column density of the XNM from
     profiles with $N_{\mathrm{G}} = 2$ as a funcion of the velocity
     difference $\Delta v_{\mathrm{c}}$ between the centers of their
     Gaussian components. The shifts up to the thick green line are
     likely explainable with the uncertainties in the Gaussian
     decomposition and the thin green line marks the median of the
     distribution.  }
   \label{Dif}
\end{figure}

From the middle panel of Fig.~\ref{Td3}, we can see that the XNM is well
visible already for profiles, decomposed with only two low-velocity
Gaussians. With such profiles it is easy to estimate the presumable
improvements in the results, obtainable by using of Voigt
profiles. However, the Voigt profiles are only able to describe
symmetric non-Gaussian wings. Therefore, it is important to estimate,
how symmetric is the XNM component in each profile located relative to
another Gaussian of the same \hi\ profile.

To study the symmetry of the simple profiles, we selected from our
Gaussian database 51\,726 profiles with $N_\mathrm{G} = 2$, in which at
least one Gaussian had the width, typical for XNM ($\lg(\mathrm{FWHM}) >
1.64$ from Fig.~\ref{Td3}). For each profile we computed the positive
velocity difference $\Delta v_{\mathrm{c}}$ between the centers of the
two Gaussians. Fig.~\ref{Dif} gives the distribution of the
contributions (the percentage from the total column density of the XNM,
represented by the selected profiles), of the XNM components with
different shifts relative the main components of the profiles. We can
see that the maximum of the distribution corresponds to a considerable
shift between the centers of the Gaussians.  However, due to the
observational noise we cannot expect that the centers of the Gaussians,
which describe the main \hi\ peak and its wide wings coincide exactly
and the acceptable mismatch increases with the increasing width of the
XNM component.
   
\begin{figure}
%   \resizebox{\hsize}{!}{\includegraphics{Voi.pdf}}
   \resizebox{\hsize}{!}{\includegraphics{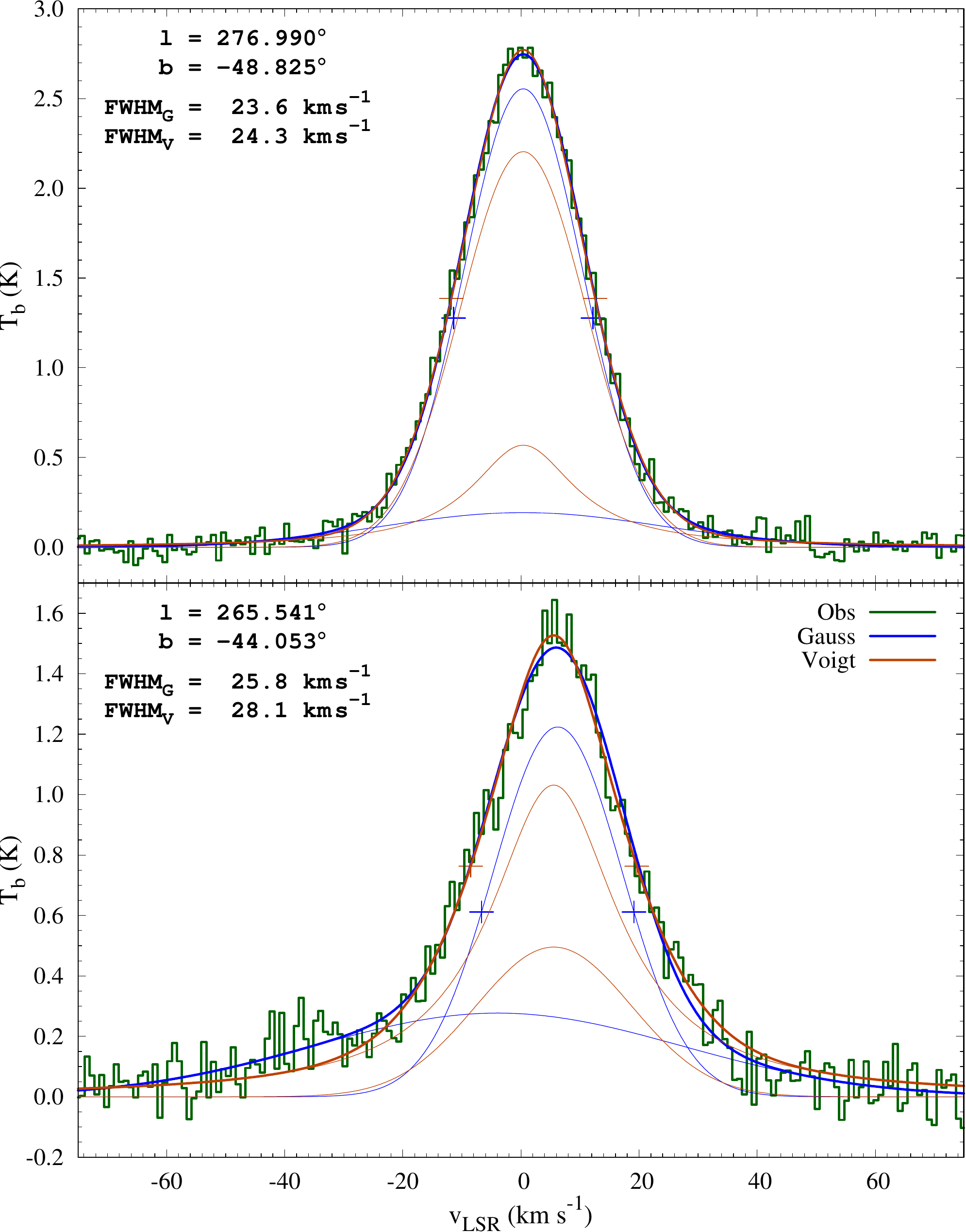}}
   \caption{Two example fits of the observations with the Gaussian
   components and with the pseudo-Voigt profile. The upper panel
   corresponds to the nearly ideal case for the fitting with the
   pseudo-Voigt profile and the lower panel represents the typical
   situation. Thick lines correspond to the observations and the models.
   Thin lines illustrate the Gaussian and the Lorentzian contributions
   to the final models. Crosses mark the points, which define
   $\mathrm{FWHM}_\mathrm{WMS}$ in two different models.}
   \label{Voi}
\end{figure}

We have studied the influence of the observational noise to the
uncertainties of the Gaussian parameters in \citet{Haud2000} and on the
basis of this study we expect that for our selected XNM Gaussians with
average $\mathrm{FWHM} = 56.5~\mathrm{km\,s}^{-1}$, the shifts up to
about $\Delta v_{\mathrm{c}} = 6~\mathrm{km\,s}^{-1}$ (broad green line
in Fig.~\ref{Dif}) between the centers of two components in each profile
may yet be acceptable for fitting both of these Gaussians with the
symmetric Voigt profile. From Fig.~\ref{Dif} we can see that most (about
79\%) of the XNM column density in studied simple profiles comes from
the Gaussians with larger displacements relative to the main component.
Corresponding profiles must be clearly asymmetric and for these we
cannot expect good fits with symmetric Voigt profiles. Moreover, the
Gaussians, which represent base-line problems or other asymmetric
features, may be centered anywhere on the velocity axis, but the wings,
describable by the Voigt profiles must be centered near our present main
component of the corresponding profile. Therefore, even at $\Delta
v_{\mathrm{c}} < 6~\mathrm{km\,s}^{-1}$ we may expect some contribution
from features for which modeling with the Voigt profiles may be
questionable and therefore $100-79=21$\% is an upper limit for the
fraction of simple profiles that can be modeled successfully
  with Voigt profiles.

\begin{figure}
%   \resizebox{\hsize}{!}{\includegraphics{Xnm.pdf}}
   \resizebox{\hsize}{!}{\includegraphics{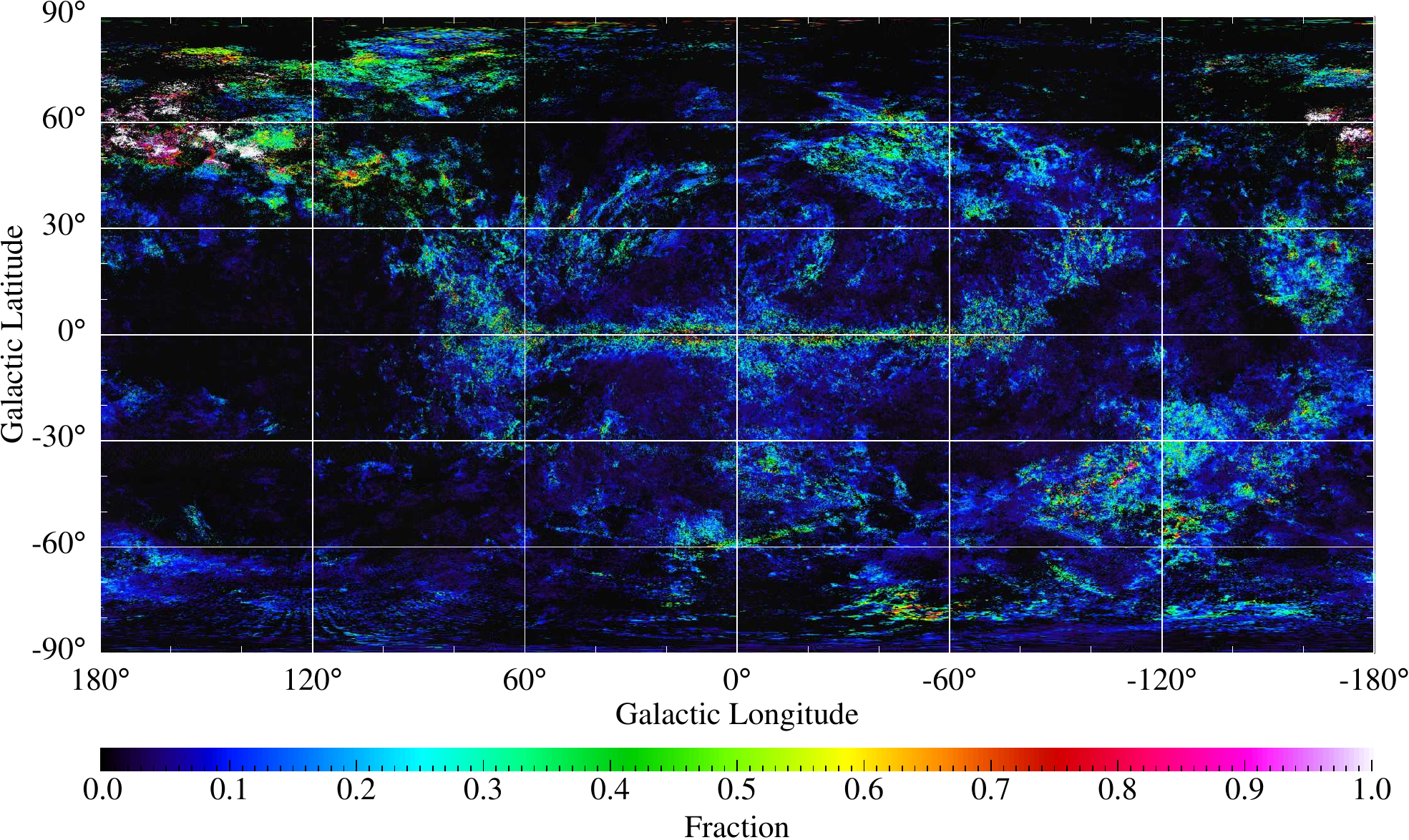}}
   \caption{All-sky distribution of the fractions $f_{\mathrm{XNM}}$,
   integrated over a velocity range $-8 < v_{\mathrm{LSR}} <
   8~\mathrm{km\,s}^{-1}$.}
   \label{Xnm}
\end{figure}

Of course, the estimate for the permissible displacement of the
Gaussians for successful fitting with the Voigt profile, is relatively
uncertain and it may be possible that contrary to our expectations even
at higher displacements satisfactory fits with Voigt profiles exist. To
test this, we give in Fig.~\ref{Voi} two examples. The upper panel
corresponds to nearly ideal case for fitting with the Voigt profile with
$\Delta v_{\mathrm{c}} = 0.06~\mathrm{km\,s}^{-1}$. The reduced $\chi^2$
values for fits with Gaussian components and with the pseudo-Voigt
profile are $(\chi^2/N_{\mathrm{dof}})_G = 0.981$ and
$(\chi^2/N_{\mathrm{dof}})_V = 0.975$, respectively. Both these values
satisfy our requirements for a good fit (see Sec. \ref{Gaussians}) and
as the fit with the pseudo-Voigt profile contains less free parameters
we would prefer this model. Such a choice removes from the decomposition
of this profile the XNM component and replaces the line width of the WNM
Gaussian $\mathrm{FWHM}_\mathrm{G} = 23.6~\mathrm{km\,s}^{-1}$ with
slightly higher value $\mathrm{FWHM}_\mathrm{V} =
24.3~\mathrm{km\,s}^{-1}$, but this change is considerably smaller than
other uncertainties in the separation of different gas phases.

For the lower panel of Fig.~\ref{Voi} we have chosen a more typical
example. $\Delta v_{\mathrm{c}} = 10.14~\mathrm{km\,s}^{-1}$ is close to
the maximum of the distribution in Fig.~\ref{Dif}, but still less than
the median of that distribution at $\Delta v_{\mathrm{c}} =
13.2~\mathrm{km\,s}^{-1}$ (thinner green line in Fig.~\ref{Dif}). The
contribution of the XNM component to the column density of this profile
($N_\mathrm{XNM}/N_\mathrm{Tot} = 0.39$) is also close to the median
($N_\mathrm{XNM}/N_\mathrm{Tot} = 0.40$) of the corresponding
distribution. As we can see, this profile is fitted well with Gaussians
($(\chi^2/N_{\mathrm{dof}})_G = 1.022$), but the fit with the pseudo-Voigt
profile is not any more a satisfactory replacement for the XNM component
($(\chi^2/N_{\mathrm{dof}})_V = 1.165$). The change of the line width is
also considerably larger (from $\mathrm{FWHM}_\mathrm{G} =
25.8~\mathrm{km\,s}^{-1}$ to $\mathrm{FWHM}_\mathrm{V} =
28.1~\mathrm{km\,s}^{-1}$), but as for a final fit we need at least one
more pseudo-Voigt profile, the interpretation of the changes becomes
already complicated.

Indirect indication that the modeling of the \hi\ profiles with more
general line shapes, other than Gaussians, cannot solve the problems
with the XNM, can be seen also from Fig.~\ref{Xnm}, which gives the
distribution of the XNM fraction, $f_{\mathrm{XNM}}$, in the sky. Except
the galactic plane, the $f_{\mathrm{XNM}}$ is often the highest in the
sky areas, where the total column density is low (see Fig. 2. in
\citet{Winkel2016c}). This is in particular true for the prominent
region $130 < l < 190\deg$ and $50 < b < 70\deg$. There are extended
regions with brightness temperatures $T_\mathrm{b} < 0.5 $ K at
$v_{\mathrm{LSR}} \sim 0$ \kms and for such a low emission remaining
instrumental problems (RFI, baseline problems and residual stray
radiation errors) can easily trigger Gaussian components that are
interpreted as XNM. Problems of this kind are recognizable as boxy
structures, discontinuities at the borders of individual observing
sessions in RA/DEC. About 0.5\% of all positions are affected.
Demanding for a better reliability of XNM components a more stringent
limit of $T_\mathrm{b} \ga 1 $ K, we find that 3\% of the XNM Gaussians
may possibly be affected by instrumental problems. The XNM fraction is
high also in the first and fourth quadrants near the galactic plane,
where the column density of the \hi\ is the highest, but there the
different features in the line profiles are so heavily blended with each
other that any decomposition of these profiles becomes unreliable,
yielding many components with very different properties. About 6\% of
all positions are affected.  In our study we have avoided making
conclusions on the basis of the observations near the galactic plane and
have used corresponding Gaussians only for tests of the limits, to which
the results, obtained from higher latitudes, are applicable.

Therefore, there are certainly cases, where the usage of the Voigt
profiles may give better results for the profile decomposition, but we
cannot expect that they will solve most of the problems with XNM.
Using Voigt profiles for the real symmetric wings of the \hi\ lines
improves (increases) somewhat our line-width estimates, but by fitting
asymmetric lines or some combinations of the real lines with more
questionable features may introduce additional uncertainties to the
line-width estimates and as we cannot in all cases automatically
distinguish the line wings and the more problematic features, the real
gain from the more complicated mathematics may remain rather
questionable. As a result, we consider the usage of the Voigt profiles
in the present work impractical.

We conclude that a Gaussian decomposition appears to provide a
reasonable parametrization of the XNM. About 3\% of the XNM may be
explainable as instrumental, additional 6\% in the Galactic plane as
affected by confusion. The rest appears to be associated with WNM
structures, mostly asymmetric. This is of course not a proof that such
structures must be real. The problem is more general, the question is
how to treat extended profile wings when processing instrumental
baselines of \hi\ survey observations. Extended asymmetric profile wings
may easily be fitted away as baseline errors but the approaches for
the LAB \citep{Kalberla2005}, GASS \citep{Kalberla2015}, and EBHIS
\citep{Winkel2016a} were conservative, taking care that such wings are
not eliminated. 

%=========================================================================
%=========================================================================
%=========================================================================

\section{Total column density distributions with uncertainties}
\label{Columntotal}

\begin{figure*}[!tbhp] %%  
  \centering
   \includegraphics[width=6cm]{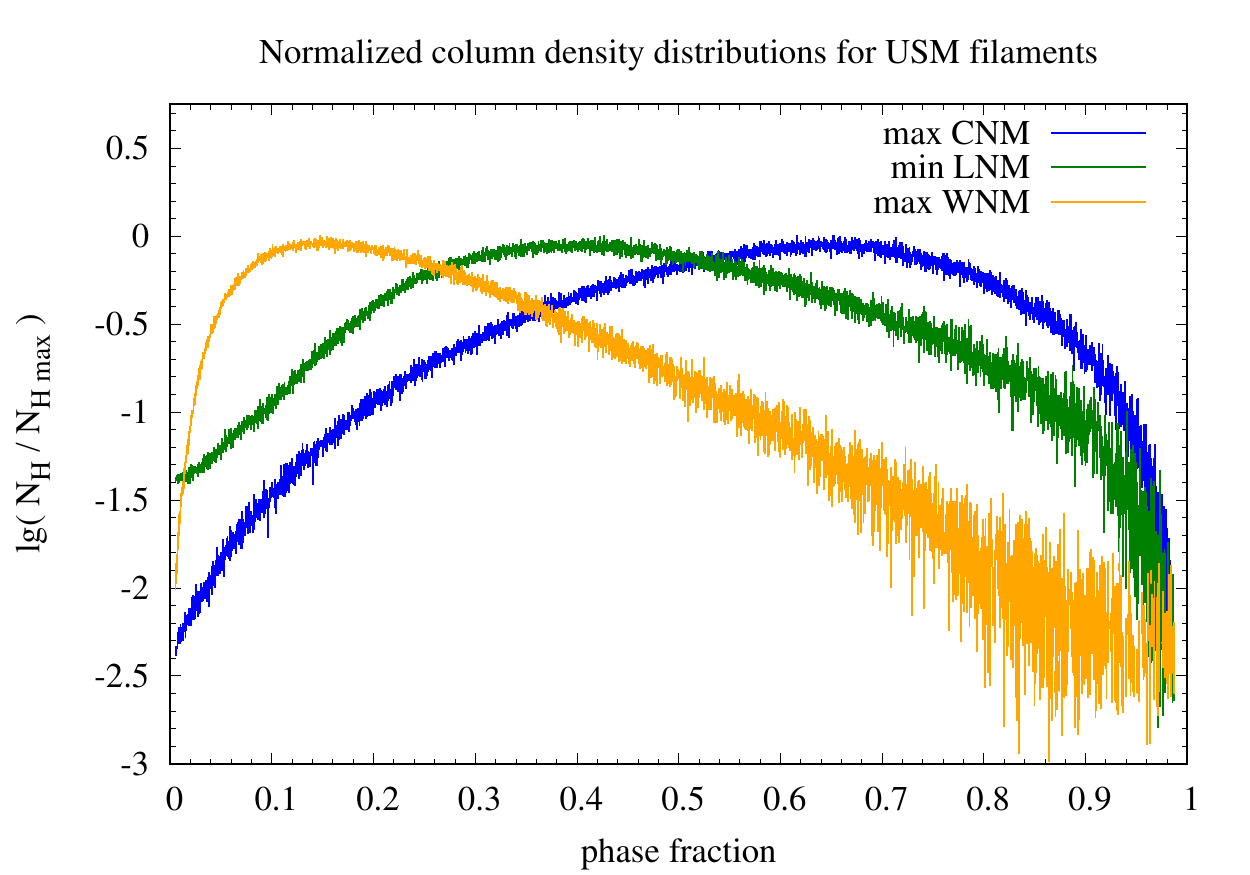}
   \includegraphics[width=6cm]{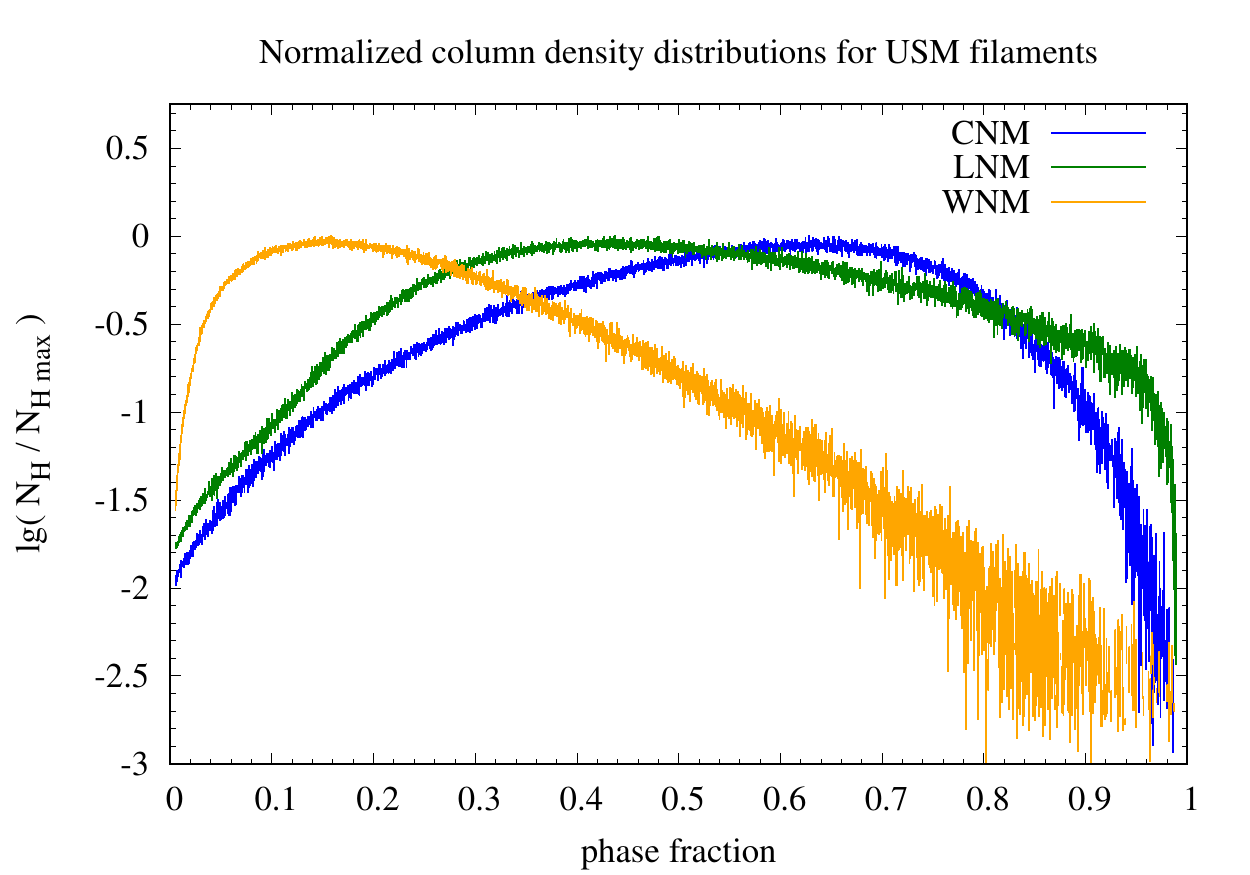}
   \includegraphics[width=6cm]{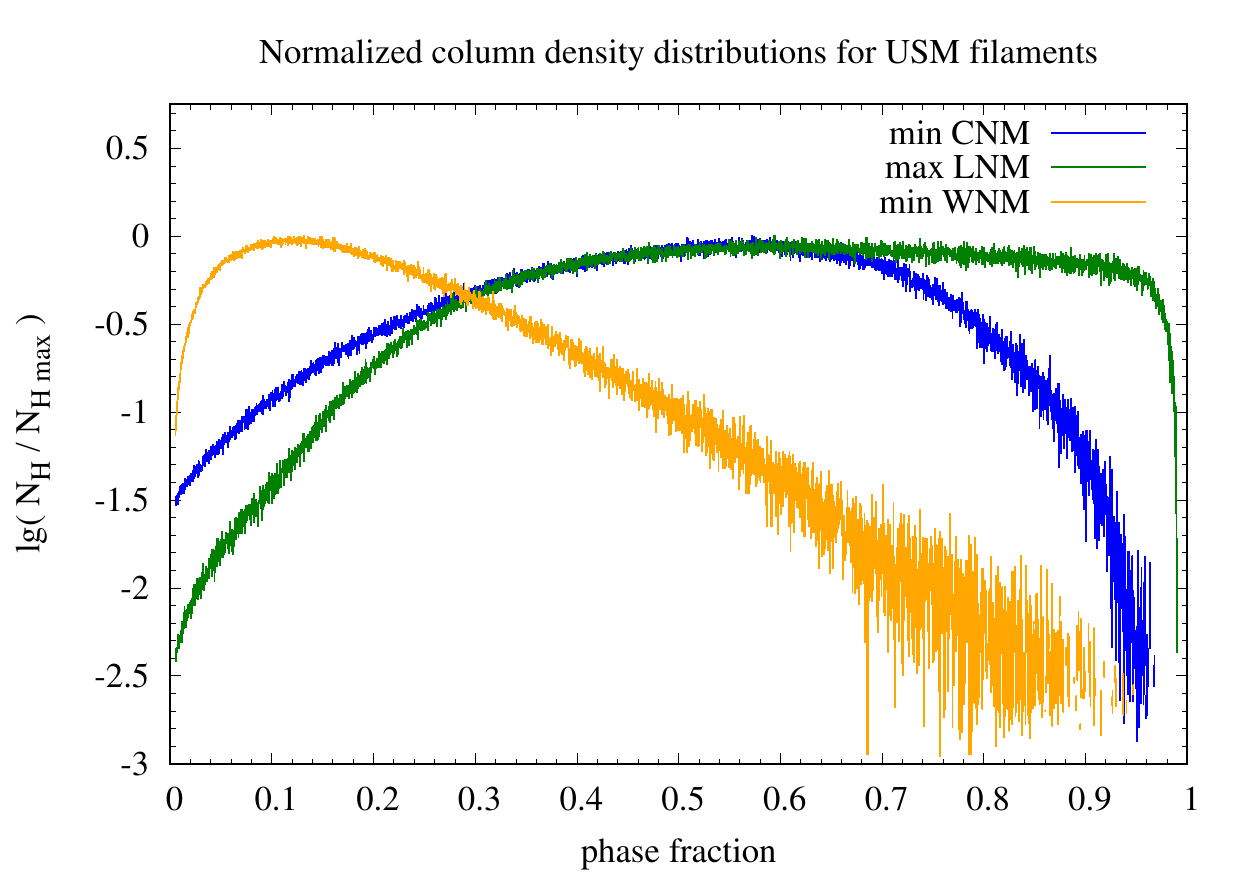}
   \includegraphics[width=6cm]{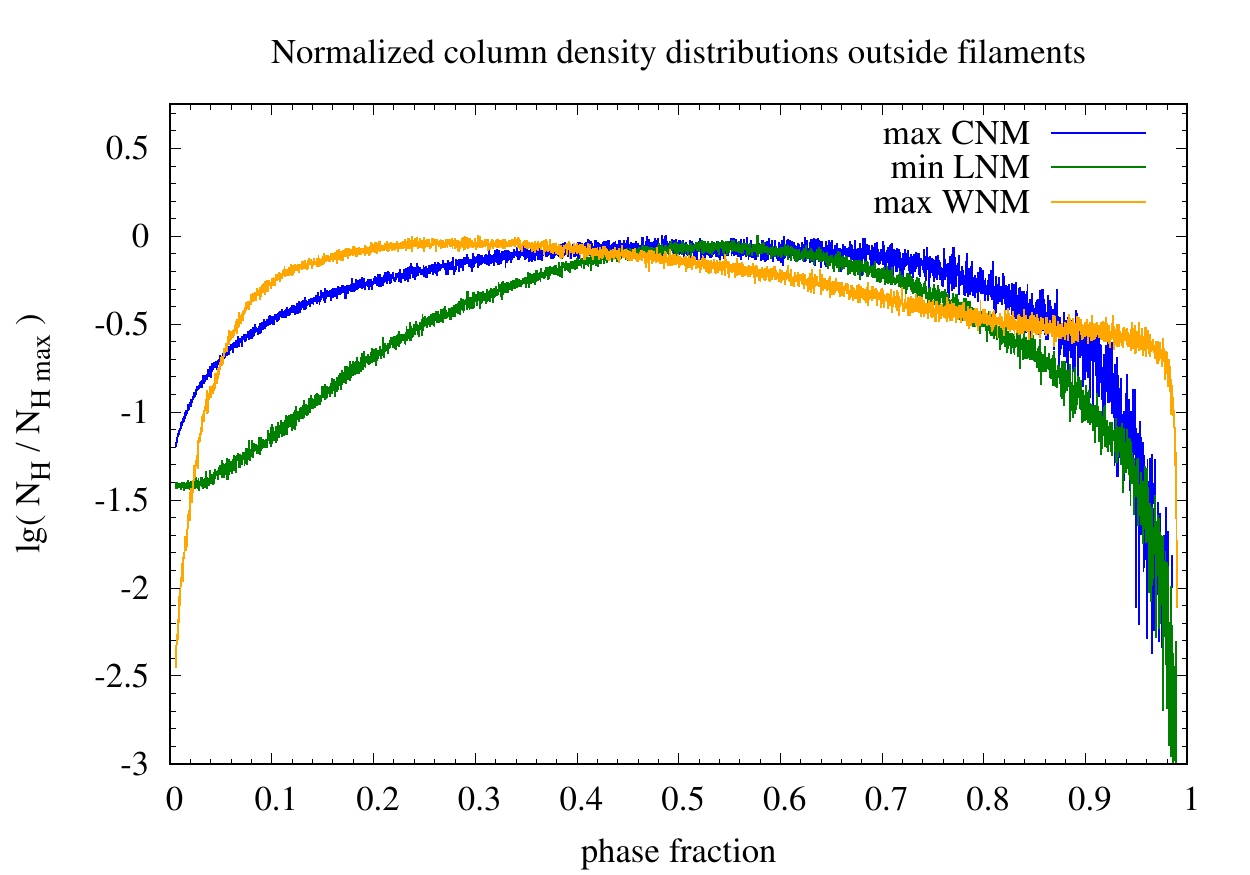}
   \includegraphics[width=6cm]{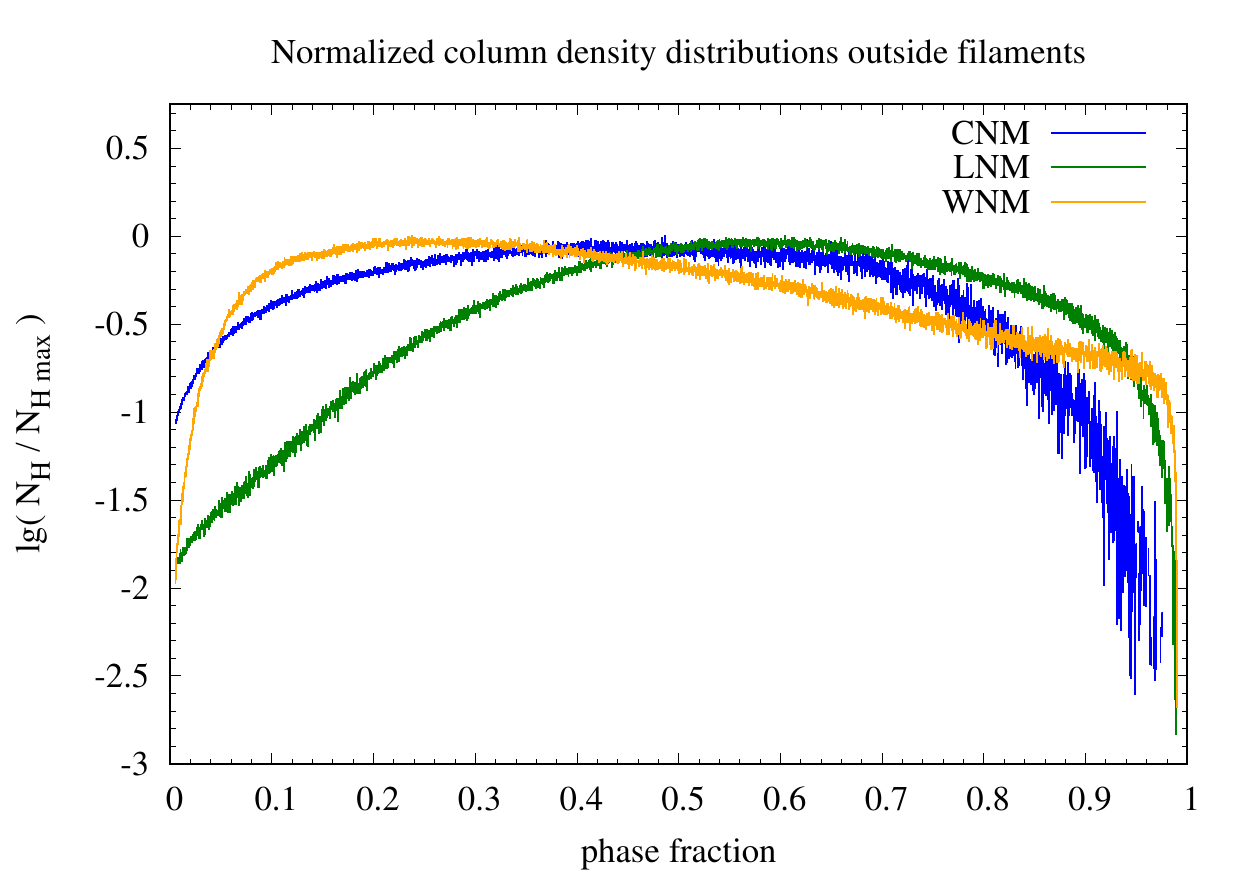}
   \includegraphics[width=6cm]{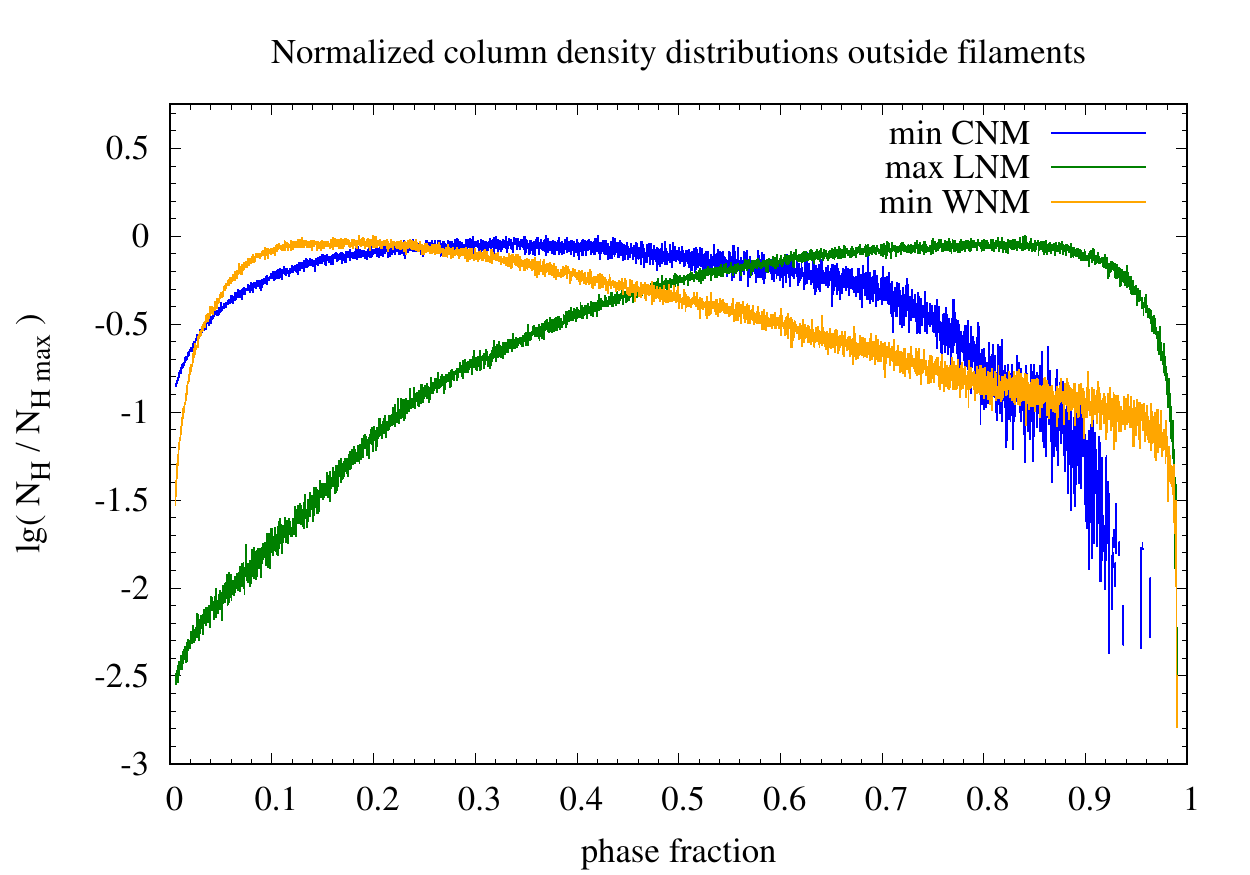}
   \caption{Normalized column density distribution functions for the
     phase fractions of the CNM, LNM, and WNM phases at high latitudes $
     |b| > 20\degr$.  Top: USM filaments in the velocity range $ -8 <
     v_{ \mathrm{LSR} } < 8 $ \kms. Bottom: \hi\ outside filaments at
     constant velocities $-1 < v_{\mathrm{LSR}} < 1 $ \kms. Middle: best
     fit results according to Fig.  \ref{Fig_VelDist},
     Sect. \ref{VelDep}. Left: minimum LNM, implying maximum CNM and
     WNM phase fractions. Right: maximum LNM with minimum CNM and WNM
     fractions respectively.  }
   \label{Fig_USM_HistoA}
\end{figure*}

With Fig. \ref{Fig_fcorrelNH} we presented 2D distributions for column
densities associated with individual phases. A more concise presentation
is possible if we summarize the column densities for all gas at a
particular phase fraction. For an optical thin medium, assuming further
that the distances of the three different observed gas phases do on
average not vary significantly, the total column density is proportional
to the total mass of the gas associated with a particular phase
fraction.

Figure \ref{Fig_USM_HistoA} shows for each of the phases how phase
fraction and contributing column densities are related to each other. We
distinguish between \hi\ associated with cold USM filaments (top) and
\hi\ observed outside such filamentary structures (bottom). As detailed
in Sect. \ref{SepPhases}, the separation of the CNM, LNM, and WNM is
burdened with rather large uncertainties (see also
Fig. \ref{Fig_VelDist}). To get an impression how far such uncertainties
affect the derived distributions, we compare several cases; the best fit
result (middle) together with distributions that were derived after
applying thresholds offset by the median standard deviation when
separating the individual gas phases. We distinguish two characteristic
cases, resulting in an upper limit for the CNM and WNM, implying a lower
limit for the LNM (max CNM, max WNM, and min LNM,
Fig. \ref{Fig_USM_HistoA} left) and a lower limit for the CNM and WNM,
implying an upper limit for the LNM (min CNM, min WNM, and max LNM
respectively, Fig. \ref{Fig_USM_HistoA} right).

The upper panels of Fig. \ref{Fig_USM_HistoA} show that CNM is dominant
in USM filaments with large phase fractions. A large amount of the gas
is concentrated in cold condensations, $f_{\mathrm{CNM}} \sim
0.65$. Most of the CNM is associated with LNM, correspondingly with
$f_{\mathrm{LNM}} \sim 0.4$. The WNM is unimportant with
$f_{\mathrm{WNM}} \sim 0.15$.  \hi\ gas that is not associated with
filaments (bottom) is characterized by a broad range of observed phase
fractions for all phases, in particular for the CNM. The LNM tends to
have predominantly $f_{\mathrm{LNM}} \ga 0.5$ while $f_{\mathrm{WNM}}
\la 0.5$.

Comparing the upper panels of Fig. \ref{Fig_USM_HistoA} with the lower
ones, confirms our previous finding that filamentary \hi\ is special in
the sense that the filamentary CNM is dominant for these
structures. Correspondingly the total CNM fraction is in this case twice
as high as for the non-filamentary sample. Filamentary \hi\ contains
little WNM, only half of the fraction derived for the non-filamentary
sample, see Fig. \ref{Fig_correl_CLW} for a complementary view.

The comparison plots (left and right hand side of
Fig. \ref{Fig_USM_HistoA}) show how far the variation of the thresholds
with phase contributions that we annotate as min or max affect
the derived CNM, LNM, and WNM distributions. In general the column
density distributions are shifted to higher phase fractions for the
max cases and opposite to lower for the min cases. However the
observed variations do not invalidate the conclusions drawn above that
filamentary \hi\ structures are enhancements in CNM, associated
predominantly with LNM envelopes that have significant phase fractions
$f_{\mathrm{LNM}}$. The WNM envelopes are in total less important.

\end{appendix}


\begin{thebibliography}{}

\bibitem[Audit \& Hennebelle(2005)]{Audit2005} Audit, E., \& Hennebelle,
   P.\ 2005, \aap, 433, 1

\bibitem[Clark et al.(2014)]{Clark2014} Clark, S.~E., Peek, J.~E.~G., \&
   Putman, M.~E.\ 2014, \apj, 789, 82

\bibitem[de Avillez \& Breitschwerdt(2005)]{Avillez2005} de Avillez,
   M.~A., \& Breitschwerdt, D.\ 2005, \aap, 436, 585

\bibitem[Dickey et al.(1977)]{Dickey1977} Dickey, J.~M., Salpeter,
   E.~E., \& Terzian, Y.\ 1977, \apjl, 211, L77

\bibitem[Dickey \& Lockman(1990)]{Dickey1990} Dickey, J. M., \& Lockman,
   F. J.\ 1990, \araa, 28, 215

\bibitem[Dickey et al.(2003)]{Dickey2003} Dickey, J. M.,
   McClure-Griffiths, N. M., Gaensler, B. M., \& Green, A. J.\ 2003,
   \apj, 585, 801

\bibitem[Egger \& Aschenbach(1995)]{Egger1995} Egger, R.~J., \&
   Aschenbach, B.\ 1995, \aap, 294, L25

\bibitem[Field, Goldsmith \& Habing(1969)]{Field1969} Field, G.~B.,
   Goldsmith, D.~W., \& Habing, H.~J.\ 1969, \apjl, 155, L149

\bibitem[Fitzpatrick \& Spitzer(1997)]{Fitzpatrick1997} Fitzpatrick,
   E.~L., \& Spitzer, L., Jr.\ 1997, \apj, 475, 623

\bibitem[Frisch \& Dwarkadas(2018)]{Frisch2018} Frisch, P., \&
   Dwarkadas, V.~V.\ 2018, arXiv:1801.06223

\bibitem[Gazol et al.(2001)]{Gazol2001} Gazol, A., V{\'a}zquez-Semadeni,
   E., S{\'a}nchez-Salcedo, F.~J., \& Scalo, J.\ 2001, \apjl, 557, L121

\bibitem[Ghosh et al.(2017)]{Ghosh2017} Ghosh, T., Boulanger, F.,
   Martin, P.~G., et al.\ 2017, \aap, 601, A71

\bibitem[G{\'o}rski et al.(2005)]{Gorski2005} G{\'o}rski, K.~M., Hivon,
   E., Banday, A.~J., et al.\ 2005, \apj, 622, 759

\bibitem[Haud(2000)]{Haud2000} Haud, U.\ 2000, \aap, 364, 83

\bibitem[Haud(2008)]{Haud2008} Haud, U.\ 2008, \aap, 483, 461

\bibitem[Haud \& Kalberla(2007)]{Haud2007} Haud, U., \& Kalberla, P. M.
   W.\ 2007, \aap, 466, 555

\bibitem[Heiles(2001)]{Heiles2001} Heiles, C.\ 2001, \apjl, 551, L105

\bibitem[Heiles \& Troland(2003a)]{Heiles2003a} Heiles, C., \& Troland,
   T.~H.\ 2003a, \apj, 586, 1067

\bibitem[Heiles \& Troland(2003b)]{Heiles2003b} Heiles, C., \& Troland,
   T.~H.\ 2003b, \apjs, 145, 329

\bibitem[Heiles \& Troland(2005)]{Heiles2005} Heiles, C., \& Troland,
   T.~H.\ 2005, \apj, 624, 773

\bibitem[Hennebelle \& Falgarone(2012)]{Hennebelle2012} Hennebelle, P.,
   \& Falgarone, E.\ 2012, \aapr, 20, 55

\bibitem[HI4PI Collaboration et al.(2016)]{Winkel2016c} HI4PI
   Collaboration, Ben Bekhti, N., Fl{\"o}er, L., et al.\ 2016, \aap,
   594, A116

\bibitem[Jones et al.(1996)]{Jones1996} Jones, A.~P., Tielens,
   A.~G.~G.~M., \& Hollenbach, D.~J.\ 1996, \apj, 469, 740

\bibitem[Kalberla et al.(1985)]{Kalberla1985} Kalberla, P.~M.~W.,
   Schwarz, U.~J., \& Goss, W.~M.\ 1985, \aap, 144, 27

\bibitem[Kalberla et al.(2005)]{Kalberla2005} Kalberla, P.~M.~W.,
   Burton, W.~B., Hartmann, D.\ et al. 2005, \aap, 440, 775

\bibitem[Kalberla et al.(2010)]{Kalberla2010} Kalberla, P.~M.~W.,
   McClure-Griffiths, N.~M., Pisano, D.~J., et al.\ 2010, \aap, 521,
   A17

\bibitem[Kalberla \& Haud(2015)]{Kalberla2015} Kalberla, P.~M.~W., \&
   Haud, U.\ 2015, \aap, 578, A78

\bibitem[Kalberla et al.(2016)]{Kalberla2016} Kalberla, P.~M.~W., Kerp,
   J., Haud, U., et al.\ 2016, \apj, 821, 117

\bibitem[Kalberla et al.(2017)]{Kalberla2017} Kalberla, P.~M.~W., Kerp,
   J., Haud, U., \& Haverkorn, M.\ 2017, \aap, 607, A15

\bibitem[Lee et al.(2015)]{Lee2015} Lee, M.-Y., Stanimirovi{\'c}, S.,
   Murray, C.~E., Heiles, C., \& Miller, J.\ 2015, \apj, 809, 56

\bibitem[McClure-Griffiths et al.(2009)]{Naomi2009} McClure-Griffiths,
   N.~M., Pisano, D.~J., Calabretta, M.~R., et al.\ 2009, \apjs, 181,
   398

\bibitem[McKee \& Ostriker(1977)]{McKee1977} McKee, C.~F., \& Ostriker,
   J.~P.\ 1977, \apj, 218, 148

\bibitem[Mebold(1972)]{Mebold1972} Mebold, U.\ 1972, \aap, 19, 13

\bibitem[Mebold et al.(1982)]{Mebold1982} Mebold, U., Winnberg, A.,
   Kalberla, P.~M.~W., \& Goss, W.~M.\ 1982, \aap, 115, 223

\bibitem[Micelotta et al.(2010)]{Micelotta2010} Micelotta, E.~R., Jones,
   A.~P., \& Tielens, A.~G.~G.~M.\ 2010, \aap, 510, A36

 \bibitem[Mitchell \& Zemansky(1971)]{Mitchell1971} Mitchell, A.~G.~G.,
   \& Zemansky, M.~W.\ 1971, Resonance radiation and excited atom.
   (Cambridge Univ. Press, Cambridge)

 \bibitem[Mohan et al.(2004)]{Mohan2004} Mohan, R., Dwarakanath, K. S.,
   \& Srinivasan, G.\ 2004, JA\&A, 25, 143

\bibitem[Murray et al.(2015)]{Murray2015} Murray, C.~E.,
   Stanimirovi{\'c}, S., Goss, W.~M., et al.\ 2015, \apj, 804, 89

\bibitem[Murray et al.(2017)]{Murray2017} Murray, C.~E.,
   Stanimirovi{\'c}, S., Kim, C.-G., et al.\ 2017, \apj, 837, 55

\bibitem[Murray et al.(2018)]{Murray2018} Murray, C.~E., Stanimirovic, S., Goss, W.~M., et al.\ 2018, arXiv:1806.06065 
   
\bibitem[Nidever et al.(2008)]{Nidever2008} Nidever, D.~L., Majewski,
   S.~R., \& Butler Burton, W.\ 2008, \apj, 679, 432-459

\bibitem[Planck Collaboration Int. XLVIII.(2016)]{PlanckXLVIII} Planck
   Collaboration Int. XLVIII. 2016, \aap, 596, A109

\bibitem[Roy et al.(2013)]{Roy2013} Roy, N., Kanekar, N., \& Chengalur,
   J.~N.\ 2013, \mnras, 436, 2366

\bibitem[Salpeter(1976)]{Salpeter1976} Salpeter, E.~E.\ 1976, \apj, 206,
   673

\bibitem[Saury et al.(2014)]{Saury2014} Saury, E.,
   Miville-Desch\^{e}nes, M.-A., Hennebelle, P., Audit, E., \& Schmidt,
   W.\ 2014, \aap, 567, A16

\bibitem[Spitzer \& Fitzpatrick(1995)]{Spitzer1995} Spitzer, L., Jr., \&
   Fitzpatrick, E.~L.\ 1995, \apj, 445, 196

\bibitem[Takakubo \& van Woerden(1966)]{Takakubo1966} Takakubo, K., \&
   van Woerden, H.\ 1966, \bain, 18, 488

\bibitem[Tielens(2008)]{Tielens2008} Tielens, A.~G.~G.~M.\ 2008, \araa,
   46, 289

\bibitem[V{\'a}zquez-Semadeni(2012)]{Vazquez-Semadeni2012}
   V{\'a}zquez-Semadeni, E.\ 2012, EAS Publications Series, 56, 39,
   arXiv:0902.0820

\bibitem[Winkel et al.(2016)]{Winkel2016a} Winkel, B., Kerp, J.,
   Fl{\"o}er, L., et al.\ 2016, \aap, 585, A41

\bibitem[Winkel, Lenz \& Fl{\"o}er(2016)]{Winkel2016b} Winkel, B., Lenz,
   D., \& Fl{\"o}er, L.\ 2016, \aap, 591, A12

\bibitem[Wolfire et al.(1995)]{Wolfire1995} Wolfire, M.~G., Hollenbach,
   D., McKee, C.~F., Tielens, A.~G.~G.~M., \& Bakes, E.~L.~O.\ 1995,
   \apj, 443, 152

\bibitem[Wolfire et al.(2003)]{Wolfire2003} Wolfire, M. G., McKee, C.
   F., Hollenbach, D., \& Tielens, A. G. G. M.\ 2003, \apj, 587, 278

\bibitem[Ysard et al.(2015)]{Ysard2015} Ysard, N., K{\"o}hler, M.,
   Jones, A., et al.\ 2015, \aap, 577, A110


\end{thebibliography}
\end{document}